# Semi-Supervised Classification of Social Media Posts: Identifying Sex-Industry Posts to Enable Better Support for Those Experiencing Sex-Trafficking

by

Ellie Louise Simonson

B.S. in Aerospace Engineering and
B.S. in Electrical Engineering and Computer Science
Massachusetts Institute of Technology, 2018

SUBMITTED TO THE DEPARTMENT OF ELECTRICAL ENGINEERING AND COMPUTER SCIENCE IN PARTIAL FULFILLMENT OF THE REQUIREMENTS FOR THE DEGREE OF

MASTER OF ENGINEERING IN ELECTRICAL ENGINEERING AND COMPUTER SCIENCE
AT THE
MASSACHUSETTS INSTITUTE OF TECHNOLOGY

FEBRUARY 2021

Signature of Author: _________________________________________________
Department of Electrical Engineering and Computer Science
September 7th, 2020

Certified by: ____________________________________________________
Richard Fletcher
Research Scientist
Mechanical Engineering
Thesis Supervisor

Accepted by: ____________________________________________________
Katrina LaCurts
Chair, Master of Engineering Thesis Committee

# Semi-Supervised Classification of Social Media Posts: Identifying Sex-Industry Posts to Enable Better Support for Those Experiencing Sex-Trafficking

by

Ellie Louise Simonson

## Abstract

Social media is both helpful and harmful to the work against sex trafficking. On one hand, social workers carefully use social media to support individuals experiencing sex trafficking. On the other hand, traffickers use social media to groom and recruit individuals into trafficking situations. Additionally, individuals experiencing sex trafficking can use social media as a means to meet sales quotas set by the traffickers [1]. There is the opportunity to use social media data to better provide support for people experiencing trafficking.

While Artificial Intelligence and Machine Learning have been used in work against sex trafficking, they predominantly focus on detecting Child Sexual Abuse Material. Work using social media data has not been done with the intention to provide community level support to individuals of all ages experiencing trafficking. Within this context, this thesis explores the use of semi-supervised classification to identify social media posts that are a part of the sex industry[1]. Instagram is one of several social media platforms used by individuals within the sex industry, and it was the primary source of data.

Several methods were explored for machine learning. However, the primary method used was semi-supervised learning, which has the benefit of providing automated classification with a limited set of labelled data. For this method, social media posts were embedded into low-dimensional vectors using FastText and Doc2Vec models. The data were then clustered using K-means clustering, and Cross-Validation was used to determine label propagation accuracy.

---

[1] Note that the sex industry consists of people in many circumstances. Some people freely choose to work in the sex industry, some people experience survival sex work, and some people experience sex trafficking. In most circumstances, it is impossible, even for experts, to tell whether an adult on social media is experiencing sex trafficking[2]. Therefore, this thesis only tries to identify sex industry posts as opposed to sex trafficking posts.

[2] Interview with Melinda Smith at Freedom Network USA [Online interview]. (2020, July 07).

The results of the semi-supervised algorithm were encouraging. The FastText CBOW model provided 98.6% accuracy to over 12,000 posts in clusters where label propagation was applied. The results of this thesis suggest that further semi-supervised learning, in conjunction with manual labeling, may allow for the entire dataset containing over 50,000 posts to be accurately labeled.

A fully labeled dataset could be used to develop a tool to identify an overview of where and when social media is used within the sex industry. Such a tool could be used to help determine where to provide public resources to support people experiencing trafficking, how to distribute resource funds, and where to provide awareness trainings.

Thesis Supervisor: Richard Fletcher
Title: Research Scientist

[1] Anthony, B., & et al. (2018, July). *On-Ramps, Intersections, and Exit Routes: A Roadmap for Systems and Industries to Prevent and Disrupt Human Trafficking SOCIAL MEDIA* (Rep.). Retrieved August 4, 2020, from Polaris website: https://polarisproject.org/wp-content/uploads/2018/08/A-Roadmap-for-Systems-and-Industries-to-Prevent-and-Disrupt-Human-Trafficking-Social-Media.pdf

# Table of Contents

# List of Figures

# List of Tables

# Acknowledgements

I would like to express my deepest gratitude to my thesis supervisor, Dr. Rich Fletcher. Dr. Fletcher provided invaluable support throughout my thesis. His multidisciplinary expertise enabled him to support this thesis from both a technical and nontechnical perspective, which was crucial to the success of this project. In addition, his guidance on how to best collaborate with and learn from nonprofits and access technical resources beyond those immediately at hand made this thesis a reality.

I could not have done this thesis without the opportunity to be a teaching assistant for the introductory Signal Processing course. This provided the necessary financial support and funding that made this project possible. I am grateful to Professor Dennis Freeman, Adam Hartz, Professor Elfar Adalsteinsson, Professor Jing Kong, and Professor Qing Hu for giving me the opportunity to be a part of the 6.003 - Signal Processing teaching staff.

I am grateful to Brandi Hardy and the Invisible Innocence Board of Directors who provided expertise and input from the view of an anti-human trafficking perspective. While Brandi provided expert advice and interdisciplinary connections, I am most grateful for Brandi's continued profound mentorship. Brandi provided emotional support throughout the thesis process, and she has been a strong role model who lives out her values of equality, determination, and compassion on a daily basis.

Without the expert input from Ashley Guevara, Assistant Clinical Director of My Life My Choice, this project could not have been realized. Ashley's technical support allowed the data in this project to be more accurately labeled. As researchers know, the ability to label data is exceptionally important.

I am grateful for the advice, expertise, and input that individuals from anti-human trafficking nonprofits across the USA have shared with me. As an individual with a predominantly technical background, I am extremely grateful for their advice. Without their input, technical researchers like me would not be able to effectively use our skills to reduce occurrences of human trafficking nor effectively support those experiencing human trafficking. In particular, I would like to thank Lisa Goldblatt, Melinda Smith, Melissa Do, and Hannah Darton.

The MIT students who chose to work alongside me on this overarching project helped make it a reality. I am grateful for the work of Bernardo Garcia Bulle Bueno who showed me how to create and use web servers, how to build and utilize databases, and how to effectively access these databases. I am grateful for the work of Helen Lu who helped implement a different semi-supervised model for this project. I am grateful for the work of Andrea Jessica David Jaba. Andrea worked alongside me on the originally proposed thesis project intended provide a more secure messaging system for those who have experienced human trafficking. They are all incredibly smart, humble, and encouraging fellow peers. I was fortunate to have the chance to work with them.

In addition, I would like to express my gratitude to Valene Mezmin, Felicity Slater, Tiffany Li, and Andrew Sellars for their valuable input that helped shape this thesis.

I would like to recognize Adam Hartz, Vera Sayzew, and Anne Hunter who were key in the completion of my thesis. They provided unending attention to academic and logistical detail as I navigated my Masters of Engineering at MIT. They quickly responded to emails, met in person, looked through administrative guidelines and provided emotional support to me throughout my journey at MIT.

Lastly, I would like to express my sincere appreciation for all of the love and support that my family, friends, and partner provided throughout this thesis.

# Semi-Supervised Classification of Social Media Posts: Identifying Sex-Industry Posts to Enable Better Support for Those Experiencing Sex-Trafficking

## Chapter 1: Introduction

Technology has long been used in ways that both help and harm society. Social media is no different. Social media is used by social workers and people experiencing sex trafficking in creative ways that provide support to those individuals experiencing sex trafficking [1]. On the other hand, social media is also used by traffickers to groom and recruit individuals into trafficking situations [1]. It is used by people experiencing sex trafficking as a means to connect with individuals who want to buy sex. In a similar vein, it is used by individuals looking to purchase sex as a means to connect with people in the sex industry [1]. Facebook, Instagram, Snapchat, messaging apps, YouTube, and Dating Sites have all been used in positive and negative ways within various types of human trafficking [1].

This thesis aims to determine whether or not semi-supervised classification can be used to classify social media posts as part of the sex industry. Semi-supervised learning uses some labeled data and a large quantity of unlabeled data. Posts are labeled as either "Yes, the post is in the sex industry", "Unclear, the post may be in the sex industry", or "No, the post is not in the sex industry." For the manually labeled data, I labeled several hundred posts, and computed the Inter-Rater Reliability (IRR) with the ratings of an expert in the field. This thesis uses the FastText and Doc2Vec models to embed posts in a 100-dimensional vector. It then uses Kmeans clustering to cluster the data. Finally, cross validation is used to determine the accuracy of label propagation among unlabeled posts.

It is important to note that not everyone in the sex industry is experiencing sex trafficking. People have a variety of different reasons why they are in the sex industry. Some may have freely chosen to work in the sex industry. Some may be experiencing survival sex work because societal discrimination and/or other factors have made sex work the best option available to them. Others may be experiencing sex trafficking. In many circumstances, it is impossible, even for experts in the field, to tell from an online post whether or not an individual offering to sell sexual services is experiencing sex trafficking [2]. Therefore, this research only claims to identify posts within the sex industry and not to identify sex-trafficking posts.

Throughout this thesis, I refer to various meetings with individuals from anti-trafficking nonprofits and documents presented from anti-trafficking organizations and sex-worker advocacy groups. These resources help to define ways in which appropriate future research on this project could provide a positive impact for individuals experiencing being in the sex industry. It also cautions against inappropriate use of this research that could have negative impacts for individuals experiencing being in sex industry.

# Chapter 2: Human Trafficking Overview and Terminology

## *2.1. Definition of Human Trafficking*

Human trafficking is a form of modern-day slavery, and it is illegal under international law, under federal law in the USA and within every state in the USA [3]. It was estimated in 2016 that 40 million individuals were enslaved in modern day slavery globally [4]. It is not only a global problem, but a domestic concern. People experiencing human trafficking are both US Citizens and non-US Citizens [6]. Cases of human trafficking typically fall under commercial sex trafficking, forced labor, and domestic servitude [6], but there are many other forms of human trafficking such as organ harvesting, forced marriage, illegal adoption, forced begging, child soldiers, and organized theft [7]. Human trafficking occurs in many ways, and it affects women, transgender, gender-nonbinary individuals, and men, as well as people from all racial and ethnic backgrounds. One place where sex trafficking is increasing in prevalence is social networking sites, such as Instagram [8].

## *2.2. Definition and Explanation of What Constitutes Sex Trafficking*

In the USA, sex trafficking occurs when "a commercial sex act is induced by force, fraud, or coercion, or in which the person induced to perform such act has not attained 18 years of age" [5, para. 11]. Sex trafficking also means "the recruitment, harboring, transportation, provision, obtaining, patronizing, or soliciting of a person for the purpose of a commercial sex act" [5, para. 11]. It is import to understand the definitions of force, fraud and coercion. The U.S. Office on Trafficking in Persons provides the following definitions:

> "**Force** includes physical restraint, physical harm, sexual assault, and beatings. Monitoring and confinement are often used to control victims, especially during early stages of victimization to break down the victim's resistance. **Fraud** includes false promises regarding employment, wages, working conditions, love, marriage, or better life. Over time, there may be unexpected changes in work conditions, compensation or debt agreements, or nature of relationship. **Coercion** includes threats of serious harm to or physical restraint against any person, psychological manipulation, document confiscation, and shame and fear-inducing threats to share information or pictures with others or report to authorities." [9, section 3 *How Victims Are Trafficked*]

Coercion is one of the three means sex traffickers use to induce someone into doing a commercial sex act. Since the term coercion is used less commonly in everyday language than force or fraud, it is defined again to help readers gain an even better understanding of what constitutes coercion. Legally, 22 U.S. Code § 7102 defines coercion as:

"(A) threats of serious harm to or physical restraint against any person;
(B) any scheme, plan, or pattern intended to cause a person to believe that failure to perform an act would result in serious harm to or physical restraint against any person; or
(C) the abuse or threatened abuse of the legal process" [5, para. 3]

### *2.3. The Seven Stages of Human Trafficking as Defined by the US Office on Trafficking in Persons*

Note that there are seven stages of human trafficking, as defined by the US Office on Trafficking in Persons. Theses stages are listed below:

"

1. **Recruiting** includes proactive targeting of vulnerability and grooming behaviors
2. **Harboring** includes isolation, confinement, monitoring
3. **Transporting** includes movement and arranging travel
4. **Providing** includes giving to another individual
5. **Obtaining** includes forcibly taking, exchanging something for ability to control
6. **Soliciting** includes offering something of value (only for sex trafficking)
7. **Patronizing** includes receiving something of value (only for sex trafficking)" [9, section 3 *How Victims Are Trafficked*]

These stages do not include the stages of exiting sex trafficking, and these next stages will be discussed later in this chapter.

### *2.4. Difference Between Sex Trafficking, Survival Sex Work, and Sex Work that was Freely Chosen*

Before discussing sex trafficking in further detail, it is crucial to differentiate between consensual sex work and sex trafficking. Not everyone who is involved in the commercial sex industry is experiencing sex trafficking [10]. In general, everyone involved in the sex industry has their own unique experience, but there are a few categories in which individuals may belong. People can often be a part of more than one of these categories, or may transition between categories.

1. **Freely chosen consensual sex work**: Some individuals freely choose to do sex work as their career, and could have chosen another viable career path if they wanted [10]. Some things that might support this community are the decriminalization of sex work and methods to help make sex work safer.
2. **Survival sex work:** Some individuals sell or trade sex for survival purposes because sex work is the only option or the best option out of many undesirable options for these individuals [2]. Survival sex work disproportionately affects communities that are discriminated against such as transgender individuals [11], as well as individuals who experience high vulnerability, such as homeless individuals [2], people of color, or trans people of color [12]. Individuals in this category may prefer to have another option besides sex work, but choose themselves to do sex work since it is the best option that they have available to them. Some things that might serve this community, is to addresses systemic discrimination and individual discrimination against these communities, as well as to provide effective support to vulnerable populations.
3. **Sex trafficking**: Some individuals are experiencing sex trafficking, and are in the sex industry because of force, fraud or coercion. Sex trafficking happens in a wide variety of ways, and there is no one way someone experiences sex trafficking. According to Polaris' 2017 data, the top risk factors for vulnerable populations of sex trafficking include being homeless or runway youth, mental health challenges, being in the child welfare system, using substances, and recently migrating or relocating [13]. Likewise, there are many different things that might serve this community. Some ways that individuals could help serve people experiencing sex trafficking is through providing social work, providing legal case management, or by working to reduce or address the vulnerabilities that increase the risk of being trafficking. This is by no means an exhaustive list of ways to provide support.

4. **Labor trafficking:** Although people often associate sex trafficking with the sex industry, labor trafficking can happen in the sex industry as well. Labor trafficking within the sex industry can happen in massage parlors, camera work, or strip clubs. [14].

The previous list is by no means exhaustive in the types of experiences people have, nor in the ways to provide support to the communities. It is meant to draw attention to the fact that people selling commercial sex have a wide variety of experiences, not all of it is trafficking, and different individuals in the sex industry may have different needs or wants. Many of the needs and wants are shared among many of the individuals in the sex industry with diverse experiences.

Regardless of the reason for entering the sex industry, prostitution is a dangerous career for the individuals providing commercial sex. Potterat et al. report overall and cause-specific mortality rates among female prostitutes in *Mortality in a Long-term Open Cohort of Prostitute Women* using the mortality rate per 100,000 person-years and the standardized mortality ratio (SMR). The SMR is adjusted for age and race and is computed taking the expected number of deaths and dividing that by the total observed number of deaths. The SMR in this report represents how much more likely a prostitute is to experience a specific type of death [15]. Potterat calculates a mortality rate of active prostitutes of 459 per 100,000 person-years, with a SMR of 5.9. Adjusting for age and race, prostitutes were 5.9 times more likely to die in a given year than non-prostitute individuals [15]. Potterat found that the leading causes of death were homicide, at 19%, and drugs, at 18%. Other significant causes of death were accidents and alcohol related deaths at 12% and 9% respectively [15]. In terms of homicide, the calculated homicide mortality rate for active prostitutes is 229 per 100,000 person-years, and a SMR of 17.7. Adjusting for age and race, this means that a prostitute is 17.7 times more likely to be murdered in a given year than non-prostitute individuals [15]. This shows that being a prostitute is indeed a dangerous career regardless of the reason for entering the sex industry. This is not being said to discredit the validity for those who choose this as a career path. However, it is important to understand that people practicing consensual or survival sex work and those experiencing sex trafficking may want to leave the sex industry at some point due to the dangers they face that other professions do not face.

## 2.5. The Life Cycle of a Trafficking Case

This section discusses the stages of exiting prostitution. In *Exiting Prostitution: An Integrated Model*, Lynda M. Baker, Rochelle L. Dalla, and Celia Williamson discuss a model of exiting prostitution [16]. It is important to note that consensual or survival prostitution is different from sex trafficking, but some of the same reasons to leave the situation may be shared. For example, fear of violent customers, (bad 'dates'), fear of a violent pimp (or traffickers in the case of people experiencing trafficking), or the overall increased risks that come with providing commercial sex may be shared reasons for wanting to leave the situation. The ways in which people leave prostitution or a sex trafficking situation may be different, but they may also share many similar aspects, depending on the individual situations. Likewise, many of the barriers that prevent people from leaving prostitution and leaving sex trafficking are shared, while others are different. Baker, Dalla, and Williamson discuss numerous barriers that prostitutes who would like to exit prostitution may experience. These barriers are categorized into four categories including individual factors, relational factors, structural factors, and societal factors [16]. It goes without saying that not every person experiences all of these factors, and people may have factors that are not included on this list. The barriers listed below in Table 1 affect individuals exiting prostitution at every stage of the process, and they make it harder for individuals to exit prostitution.

*Table 1 - Factors that make exiting prostitution more challenging [16]*

| | |
|---|---|
| **Individual Factors** | • Psychological trauma/injury from violence<br>• Chronic psychological stress<br>• Self-esteem/shame/guilt<br>• Physical health problems<br>• Lack of knowledge of services<br>• Self-destructive behaviors<br>• Substance abuse<br>• Mental health problems<br>• Effects of trauma from adverse childhood |
| **Relational Factors** | • Limited conventional formal & informal support<br>• Strained family relations<br>• Pimps<br>• Drug dealers<br>• Social isolation |
| **Structural factors** | • Employment, job skills, limited employment options<br>• Basic needs (e.g. housing, homelessness, poverty, economic self-sufficiency)<br>• Education<br>• Criminal record<br>• Inadequate services |
| **Societal factors** | • Discrimination/stigma |

In their article, Baker et. al. also present a model of exiting street prostitution that is shown below in Figure 1. Some aspects of this model may be similar to those who exit sex trafficking, which will be discussed later.

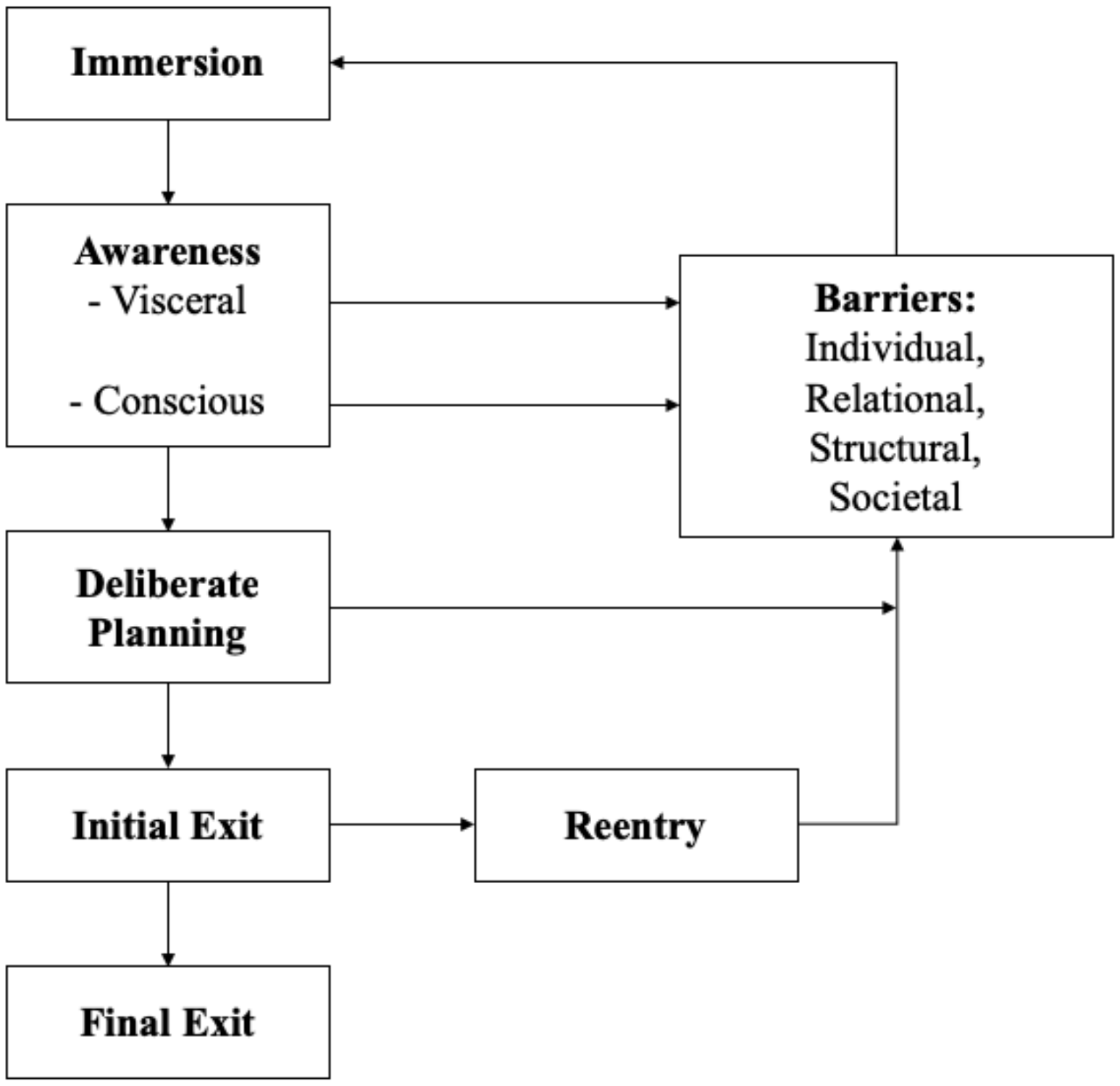

*Figure 1 - A model that represents the process of exiting street prostitution, [16, pg. 16].*

In this model *Immersion* is similar to the precontemplation stage in the Stages of Change model. Here there are no thoughts of leaving prostitution. *Awareness* consists of both visceral awareness and conscious awareness. *Visceral awareness* are unconscious feelings of wanting to leave prostitution. Visceral awareness may become *conscious awareness* when an individual acknowledges and begins to processes the feelings of wanting to leave. In the *deliberate planning* stage individuals may asses support resources available to them including formal and informal resources, and they may contact support providers or communicate with family or friends who are not in the sex industry [16]. The *initial exit* stage is when an individual actively uses these support resources, for example, meeting with a service provider, moving in with a family member, or attending counseling, etc. Some communities have extensive resources, while others do not, and access to informal or formal support is critical during the initial exit stage [16]. Separately, it is important for the resources in the communities to be trauma informed so that they can best provide support to people who have experienced trafficking. In prostitution, as well as in sex-trafficking, many people re-enter into the sex industry (or into sex trafficking). The stage of *reentry* is when an individual re-enters the sex industry and again engages in commercial sex work. Note that the average individual who experiences sex trafficking, returns to 'the life' (of sex trafficking) about 7 times after their initial exit of 'the life' [17, time 16:22]. The *final exit* stage occurs most often after a series of exiting and reentering the sex industry. Due to the complicated nature, the *final exit* is not defined in *Exiting Prostitution: An Integrated Model*. It may look different for each individual, and it cannot be explained for all individuals by a certain amount of time passing since when they were a prostitute, nor can it be explained for all individuals as having new careers or housing or life plans. Throughout all of these steps, many barriers are present that make it harder to reach the stage of *final exit* for someone in the sex industry. These barriers contribute to the reason why many people re-enter into the sex industry, or why many people may never able to reach the stage of *final exit* [16].

### *2.6. People-First Terminology*

In this thesis, we will refer to individuals who have been trafficked and are no longer being trafficked as 'individuals who experienced trafficking'. In addition, I shall refer to individuals who are currently being trafficked as 'individuals experiencing trafficking'. The reasoning behind this is given below.

Different organizations use different terminology to refer to someone being trafficked. Some organizations use the word 'victim' to explicitly state that these individuals did not choose to be trafficked nor did they choose to be in the situation they are in. However, some people who experienced trafficking do not want to be called a victim because they may not want pity or they may not choose to identify as a victim. Therefore, some organizations use the word 'survivor' to explicitly state that at all times of being trafficked, the individuals were doing everything they could to do in order to survive the situation they were in. The term survivor also has a positive connotation while victim has a negative connotation. However, some people who experienced trafficking do not want to be called survivor because they believe their identity is much more than a survivor of human trafficking and using ‘survivor’ minimizes the rest of their identity. Therefore, some organizations use the terminology ‘people with lived experience’ in order to use a people-first language and because people who have been trafficked have much more to their identity than being a survivor of human trafficking. However, ‘people with lived experience’ does not describe what kind of experience the people lived. Therefore, in order to avoid confusion in this paper, and in an attempt to use a people-first terminology, we will refer to individuals who have been trafficked and are no longer being trafficked as ‘individuals who were trafficked’ and individuals who are currently being trafficking as ‘individuals experiencing trafficked’.

# Chapter 3: The Role of Technology in Human Trafficking

Social media platforms have had significant impacts on many aspects of our society, so it is not a surprise that social media and human trafficking are connected. Social media is used by traffickers, buyers of sex, and people experiencing trafficking in the cycle of sex trafficking [1]. It is also used by people experiencing trafficking, families of people who are experiencing trafficking, leaders who have experienced trafficking in the past, social workers, and others to reduce sex trafficking [1].

## *3.1 Social Media Used to Facilitate Human Trafficking*

Social media is used to facilitate trafficking by traffickers and buyers. Traffickers also force or require people experiencing human trafficking to use social media to facilitate their own trafficking [1].

Social media is used to facilitate human trafficking by traffickers in various ways. Traffickers use social media to isolate and control people they traffic. For example, traffickers isolate people experiencing trafficking by monitoring and controlling their social media accounts. Traffickers also may impersonate or stalk the people they are trafficking on social media or isolate them from support systems. They may post intimate images or spread rumors around the people they traffic [1]. Traffickers use social media for recruitment. For example, traffickers may also use social media to recruit individuals by posting fake or deceptive job opportunities, as well as creating online 'relationships' to groom individuals [1]. Traffickers use social media as a means to communicate or provide services with potential buyers. For example, traffickers may use social media to communicate prices, location and contact information, and it may include, (possibly through a link), explicit live streams or photos of adults or children [1]. While traffickers may use the trafficker's own social media account, they often use the account of the people they traffic [1].

Traffickers are not the only individuals who use social media to facilitate trafficking. People experiencing human trafficking may be forced or required to use social media to assist in their own trafficking. Polaris surveyed people who had experienced trafficking and found that 26% of participants were exploited by their trafficker on their personal social media accounts [1].

Facebook is the most commonly used social media platform used by people experiencing trafficking. Instagram is the second most commonly used platform used by people experiencing trafficking [1].

Polaris published a report on how industries can prevent and disrupt human trafficking. In their report for Social Media industries, they provide a table of social media platforms that have been used to support trafficking. This is shown in Table 2 below [1, p. 16].

*Table 2 - Overview of some intersections between social media platforms have with different types of trafficking. [1, p. 16]*

| | Types of Social Media Platforms | | | | | |
|---|---|---|---|---|---|---|
| Types of Trafficking | **Facebook** | **Instagram** | **Snap-chat** | **Chat apps (Kik, Kakao Talk, WeChat, WhatsApp)** | **Dating Sites & Apps** | **YouTube** |
| Agriculture & Animal Husbandry | X | | | X | | X |
| Arts, Sports & Entertainment | X | X | | | | |
| Bars, Strip Clubs & Cantinas | X | X | | | | X |
| Domestic Work | X | | | X | | |
| Escort Services | X | X | X | X | X | |
| Illicit Massage Businesses | X | X | | X | | |
| Outdoor Solicitation | X | X | X | | | |
| Personal Sexual Servitude | X | X | X | X | X | |
| Pornography | X | X | | X | | X |
| Remote Interactive Sexual Acts | | | X | X | X | |
| Restaurants & Food Services | X | | | X | | |
| Traveling Sales Crews | X | X | X | | | X |

Polaris's report *On-Ramps, Intersections, and Exit Routes: A Roadmap for Systems and Industries to Prevent and Disrupt Human Trafficking SOCIAL MEDIA* written by Brittany Anthony is a good resource for any individual wanting to work on the intersection of human trafficking and social media. It is too long to be summarized in this thesis, but would provide useful information beyond what is reported in this thesis for future researchers.

### *3.2. Social Media Being Used to Reduce Human Trafficking*

Social media is not only being used to facilitate human trafficking, but it is also being used to reduce human trafficking. Service providers may use social media to connect with people still experiencing trafficking [1]. Twenty percent of people who had experienced trafficking communicated via private messages with service providers [1]. In addition, people experiencing trafficking may be able to use their social media to reach out to family or friends who are not being trafficked for support [1].

Some aspects of certain social media platforms can make it safer for a survivor to use their platform to communicate with service providers or loved ones. These include disappearing messages, which can be seen in Snapchat, Instagram and Facebook Mobile. This reduces the chance that traffickers, who may be monitoring their accounts, will know they are communicating with service providers or loved ones [1]. Other social media services, such as location discovery settings, the ability to "check-in" and post one's location can help people experiencing trafficking. People who have experienced trafficking in the past are often very tech-savvy and can use social media features like blocking, disabling location services, and reducing profile visibility to help keep themselves safe [1].

Another way that entities use social media to reduce human trafficking is through photo detection tools, like PhotoDNA. These tools can be used to identify child sexual abuse material (CSAM) when it is uploaded. Tools like this can then send the information to the National Center for Missing and Exploited Children [1]. Overall, 19% of people who had experienced trafficking said that social media was a part of their exit [1].

### 3.3. Organizations Using Technology to Identify Child Exploitation or Child Sexual Abuse Material

Naturally, there are organizations outside of social media, service providers, and individuals who are using technology and the internet to reduce human trafficking. Tech Against Trafficking mapped out how 300 different entities and organizations are using technology to reduce human trafficking. This thesis will specifically discuss organizations used to identify child exploitation or child sexual abuse material (CSAM) and organizations using advertisements for sex to reduce human trafficking. There are several organizations used to identify child exploitation or child sexual abuse material (CSAM) and organizations using advertisements for sex for anti-human trafficking purposes. The organizations listed on Tech Against Trafficking in this space include Safer, Spotlight, Content Safety API, Analyze, Childsafe.ai, Project Arachnid, IdTraffickers, CEASE.ai, Crisp, Deepdive, and WAT – traffic bot [18]. A table can be found below listing the entity names and descriptions, as well as the websites of these organizations.

*Table 3 lists several organizations that identify child exploitation or child sexual abuse material (CSAM) for anti-human trafficking purposes, as well as organizations that use advertisements for sex for anti-human trafficking purposes.*

| **Entity Name** | **Description from Tech Against Trafficking Interactive Map** | **Website** |
|---|---|---|
| Safer | Safer was created by Thorn to identify, report, and remove CSAM. It is end-to-end encrypted [18] | https://getsafer.io/ |
| Spotlight | Spotlight was created by Thorn to aid law enforcement in fighting human trafficking. It archives over 100,000 new sex ads per day, and allows law enforcement to search for specific information like a phone number, as well as identifying ads that might have children [18] | https://spotlight.thorn.org/ (Only available for law enforcement.) |
| Content Safety API | Google developed Content Safety API. It is an API that allows fewer people to be exposed to CSAM when reviewing material. It uses deep neural networks to process the images, and then it prioritizes material for review [18]. | https://www.blog.google/around-the-globe/google-europe/using-ai-help-organizations-detect- |
| | | and-report-child-sexual-abuse-material-online/ |
| Analyze | Analyze is a software that can detect nudity (via skin tones) and explicit content (via body motions) to assist in digital investigations of CSAM [18] | https://www.griffeye.com/ |
| Childsafe.ai | Childsafe.ai is used by law enforcement in the US, and it uses artificial intelligence to model child exploitation risk [18]. | https://childsafe.ai/ |
| Project Arachnid | Project Arachnid searches the web for child exploitation imagery. It compares imagery with information submitted in police reports around the world, and has identified 40,000 images of child exploitation [18]. | https://projectarachnid.ca/en/ |
| IdTraffickers | IdTraffickers uses biometrics to compare images of people reported missing to images of "escort ads" and human trafficking databases. This information is then submitted to law enforcement to help identify and provide aid to people experiencing trafficking, as well as to identify and track traffickers [18]. | https://idtraffickers.com/ |
| CEASE.ai | CEASE.ai uses artificial intelligence to identify new CSAM. By identifying new CSAM material, this can help law enforcement decide which CSAM material requires immediate action. It also can help social networks identify CSAM before the CSAM is posted [18]. | https://www.twohat.com/cease-ai/ |
| Crisp | Crisp's goal is to provide a safer environment for youth who use online games and social networks. They crawl public and dark websites to deliver safer social media [18]. | https://www.crispthinking.com/about-us/ |
| Deepdive | Deepdive scrapes CSAM from web advertisements | http://deepdive.stanford |
| | and creates a searchable database of these advertisements to aid in interventions [18]. | .edu/showcase/memex |
| WAT – Traffic bot | WAT – Traffic bot looks to connect images, words, and data from sexual advertisements and other sources including social media [18]. | https://www.cs.cmu.edu/~hovy/papers/14dgo-WAT.pdf |

This being said, I felt that a lot of work was being done by technology companies in identifying Child Sexual Abuse Material (CSAM), and a lot of this work or other work on identifying sex trafficking was being done on websites on the dark web, or sites specific to the sex industry. Only a small number of them, comparatively, work on identifying things on social media sites [18]. However, it was unclear which social media site(s) could have the most impact.

### *3.4. Instagram is Used by Youth More than Facebook or Twitter*

Historically, individuals experiencing trafficking were forced or coerced into soliciting sex on the streets, in brothels, or in hotels and motels. Solicitation of sex on the streets is more 'dangerous' for traffickers (also called pimps) and for buyers of sex because law enforcement could see the buyer and solicitor in person during solicitation and patronization. After the internet became widely accessible, solicitation of sex has since transitioned to become more and more prevalent online. Solicitation of sex occurs on online platforms such as websites dedicated to advertising sexual services, craigslist, and social network platforms such as Instagram, Facebook, Twitter, and Tinder [8]. During an interview with Lisa Goldblatt Grace, the co-founder and director of My Life My Choice, we learned that in her experience, youth today are more likely to use Snapchat or Instagram than Facebook or Twitter [19]. It was important that research only be done on publicly available data accessible without an account. Based upon the interview with Goldblatt, the data in the Polaris report, and which social media networks have publicly available data, Instagram was a good fit for this research project.

It is important to note, that because social media is so tied to human trafficking, in both exploitation and in providing services, it is very important that organizations using social media to fight against human trafficking and social media companies work in a trauma informed manner. It is important that they collaborate and consult with people who have experienced trafficking in addition to NGO professionals [1].

### *3.5 Further Ways in Which Organizations and Companies Use Technology Against Trafficking*

From Tech Against Trafficking's Interactive Map, it is clear that technology is being used to fight human trafficking in many ways [18]. Organizations use technology for anti-human trafficking purposes ranging from victim or trafficker identification, supply chain management, worker engagement, awareness raising or education, data trends, ethical shopping, to victim case management [18]. These tools are used to reduce multiple forms of trafficking, including labor trafficking, sex trafficking, and other forms of human trafficking [18]. The technology in the majority of the platforms are web/cloud based or on a mobile platform, but they are also based upon network platforms, blockchain, big data, artificial intelligence, satellite technology, as well as many other types of technology [18]. The target users spread across a wide variety of individuals including NGOs, community or family members, businesses, potential victims, governments, law enforcement, victims or consumers [18]. The majority of the technologies target cross-sector work, but some focus specifically in the public or informal sector, or in particular industry sectors such as tourism/hospitality, transportation, textiles, or the fishing industry [18].

It is clear that technology has begun to be widely used in a variety of ways for anti-human trafficking purposes. This is not to say by any means that humanity is close to ending human trafficking. Human trafficking is widely prevalent around the world. However, given Tech Against Trafficking's report, researchers and tech professionals can find a variety of areas to do anti-human trafficking work.

### *3.6 Opportunities and Existing Needs for Technology Tools*

Throughout the work of this thesis, several opportunities and existing needs for technology became apparent. Before proceeding with any of these ideas, it is extremely important to get feedback and work with individuals from NGOs and people who have experienced trafficking (i.e. survivors who are in a leadership position willing to give feedback). Without that feedback, the projects could result in unintentional harm. These are several ideas that could provide a positive impact in the anti-trafficking work. They include the following:

1. **Peer Messaging Platform** - Creating a safer messaging platform for people who have experienced trafficking. Currently, people who have experienced sex trafficking return to 'the life' (of sex trafficking) about 7 times after their initial exit [17, time 16:22]. In order to reduce this, it is important for survivors to have a strong connection with other individuals. Support groups can be a great way to form these connections, but current methods of communication that are available, such as texting, WhatsApp, Facebook Messenger, etc. are not well suited for survivors of sex trafficking within the context of a support group. People who have experienced trafficking and who are participating in support groups may still be interacting with pimps, bottoms, recruiters or other victims of sex trafficking who are still in 'the life'. As mentioned earlier, survivors often return to 'the life' multiple times before they have fully exited 'the life'. In order to address the unique situations and needs of these survivors, one need for technology is to developing a messaging app that includes additional security features that make communication between people in the support group safer and more secure.
2. **False Advertisement Detection** - Another need for technology is to identify advertisements on social media that are actually false advertisements. These false advertisements may be posted by traffickers or traffickers may force people experiencing trafficking to post the false advertisements. This may occur in many different ways. A sex trafficker could post about modeling opportunities [1]. A labor trafficker or the traveling sales crew being made to work for the labor trafficker may post advertisements boasting lots of cash earned [1]. False job opportunities for other countries may be posted in areas where there are few economic opportunities. The individuals are then trafficked out of the countries and into the new country where they may experience labor trafficking. Technology could be used in this situation to identify potential labor traffickers or to provide resources to the communities being targeted by these false advertisements.
3. **Automated Detection of Traffickers** - One area technology could be used is in comparing the social media accounts of known and convicted traffickers to a random sampling of the public. This could be used to identify potential traffickers. Alternatively, analyzing the social media accounts of known and convicted traffickers could lead to detecting patterns that traffickers may exhibit in their social media accounts.

4. **Disseminating Information about Resources** - Social media could be used to disseminate information about educational and economic resources to communities who have vulnerabilities known to increase the risk of human trafficking.
5. **Providing Feedback to Local Governments** - Identifying communities with higher sex industry social media posts could allow cities to provide public education and awareness of national and local resources that people experiencing trafficking can take advantage of.
6. **Measuring Effects of Policy** - SESTA-FOSTA was a law that was recently passed in 2018. The goal of the law was to decrease sex trafficking online. There have been some unintended negative consequences that have made people in the sex-industry have to move to more dangerous platforms because of it. Data science is needed to better measure the positive and/or negative effects of such laws and new policies on people experiencing sex trafficking and people who are doing consensual or survival sex work. Analyzing how SESTA-FOSTA changed the online sex industry would be very important for informing future laws [14].
7. **Improving Data Quality and Metrics** - Good data has always been one of the biggest challenges for individuals working in anti-trafficking work. Data is often very poor in this field and estimates of trafficking have wide ranges because of it. In addition, individuals have not accurately measured or determined how sex trafficking has changed due to different laws and policies and programs. There can be many factors that affect this: individuals may think trafficking is increasing in their area after an awareness program. It may likely be an increased awareness that alerts individuals to trafficking and allows them to report it, rather than an increase in trafficking itself. Or for example, once prostitution is legalized, the way data is collected might change, meaning that confounding factors make it hard or impossible to tell whether trafficking has increased or decreased. In addition, many statistics come from reports by concerned community members or people experiencing trafficking. That itself could be a confounding factor. What about the individuals who are unable to reach out to resources? What if community members cannot see these individuals? What data does that result in us missing? There is a lot to be done in improving data collection in the anti-trafficking area. Professor Roberto Rigobon at the MIT Sloan and Jessica Brunner at Stanford's Handa Center were working on subsets of this topic. Figuring out ways to collect good data for labor, sex, and other types of trafficking and implementing those data collecting techniques would be extremely valuable to the fight against human trafficking.

8. **Exploring Trafficking in Correctional Facilities** - Labor Trafficking is prevalent in the U.S. Prison Complex. It would be interesting to see how technology could be used to combat labor trafficking in the U.S. Prison Complex since this is an area that is not commonly looked at by technology organizations.
9. **Underrepresented Areas** - Looking solely at organizational areas that are underrepresented in the Tech Against Trafficking Interactive Map, three or less organizations worked in areas providing Victim Case Management and Payment Security. Three or less organizations worked in industries of Food & Beverage Processing, Auto & Technical Manufacturing, Domestic Work/Servitude, Wholesaling & Retail, or Healthcare & Pharmaceutical Industries. Ironically, of the target users, people who have experienced trafficking and exited trafficking have relatively few organizations serving them. The only other target users with fewer organizations serving them are consumers, 'other', and standards & certification bodies [18].
10. **Identifying Sales Crews Participating in Labor Trafficking** - One other side note is that social media is not only used by individuals connected with sex trafficking. Social media is also used by labor traffickers and those experiencing labor trafficking. For example, sales crews use social media to post false advertisements. And there may be a trail of reviews from disappointed customers who have not received items they bought. This information could be found on websites like Reddit, RipOffReport.com, Better Business Bureau, or Complaintsboard.com. This could be a very useful way to use social media or websites to try to identify sales crews participating in labor trafficking [1].

In addition to the technology opportunities listed above, various categories of technology tools are also being explored for these applications. These include: blockchain, big data, artificial intelligence and machine learning, mobile and network platforms, radio frequency, satellite technology, IoT, deep/dark web technology, and RFID [18].

I hope that these ideas can provide some encouragement to readers who are interested in working in the field of anti-human trafficking. Certainly, there are many different types of technology and technical skills that can be used to combat human trafficking.

# Chapter 4: Scope of this Thesis and Problem Definition

## *4.1 Discussion of Possible Options and Trade-offs*

In July 2018 before I began my thesis, I knew that I wanted to use technology to work in the anti-human trafficking space. It is a topic that I care deeply about, and I wanted to make sure I did my work in an ethical, well-thought out way, that was well informed. At the time, I could not find any professors or researchers at MIT working in this space, so I reached out to Dr. Fletcher since I knew his lab's research is focused on having positive social impact in the health sector. Dr. Fletcher was open to me working in his lab on this topic, and we began the process of determining what an impactful research project in this area might look like. This is relevant because the context surrounding this thesis had a strong impact on the discussion of possible options, as well as what trade-offs existed among them.

The first project that we explored was a safer communication platform for individuals in a support group who had experienced human trafficking. The main trade-off for this project was ensuring I had the skillset to execute it within the MEng timeframe. Dr. Fletcher and I felt that partnering with a nonprofit would give us the best chance at developing a tool or research project that could have a tangible positive impact in this space. After reaching out to several individuals, we settled on working with the non-profit Invisible Innocence in Bismarck, ND. The founder Brandi Hardy showed a strong ability to communicate well, create local connections, and showed excellent mentorship skills. The board of Invisible Innocence also included a survivor of human trafficking and that individual was able to help guide us throughout the problem definition to ensure that the voice of people who have experienced human trafficking and the voice of professionals at NGOs are taken into account. We developed the idea of a project that would create a safer messaging platform for people who have experienced trafficking. This is the first project listed in Opportunities and Existing Needs for Technology within Chapter 3. We worked with Invisible Innocence to design a messaging app that individuals in Invisible Innocence's support group could use to more safely communicate with each other. This involved several safety features including having an escape button so people could quickly exit the app if in danger, an inconspicuous app logo, like a calculator, where the messaging portion of it could only be accessed via a specific code, as well as a monitoring function that allowed Invisible Innocence staff to remove an individual if they tried to recruit someone back into 'the life' (of sex trafficking). There were several other features, but these were all things that Invisible Innocence believed could help the individuals they were serving have a safer way of communicating with each other. Ultimately, after about a year of project planning and work, we had to end this project on our end. My background is predominantly at the intersection of Aerospace Engineering and Electrical Engineering and Computer Science, so I have more experience with artificial intelligence and machine learning algorithms, signal and systems, and data science. I did not have any background in Android or iPhone app development, server development, systems engineering, database management, nor frontend or backend software engineering. While I was comfortable learning new information and topics for my thesis, the amount of information I needed to learn to do this project was not feasible to complete in the MEng timeframe. This project still has value and is a project future researchers or individuals could implement in collaboration with a nonprofit to have a positive impact in this space. Ultimately, the skillset required to complete this project did not align with my skillset, and so we chose to switch to a new project that would fit better within my skillset. This was a trade-off between choosing a project that was known to be needed by a nonprofit and choosing a project within my skillset.

In order for the new project to fall closer to my background, we decided the project should be related to machine learning, artificial intelligence, signals & systems, or data science. The thesis needed to go from problem formulation, to data acquisition and storage, data cleaning, and analysis within a year, so it was important that data was easy to generate or get given the limited time frame. Social media plays a large role in both reducing and facilitating human trafficking, and the data is relatively easy to access for certain social media organizations from a technical standpoint. Twitter even has an API that allows individuals to collect and analyze data. Therefore, as a team, Dr. Fletcher, Bernardo Garcia Bulle Bueno, who worked with us for several months on this project, and I generated several ideas related to social media. Throughout this time, we met with nonprofits to determine what might be helpful within this space. Some of the ideas that we generated during this time were ideas #2 - False Advertisement Detection, #3 - Automated Detection of Traffickers, #4 - Disseminating Information about Resources, and #5 - Providing Feedback to Local Governments that are discussed in Opportunities and Existing Needs for Technology within Chapter 3.

When deciding among these projects, there were several aspects and trade-offs that needed to be considered. Several factors that played major roles were:

a. How accessible is the relevant data?
b. What would the labeling process look like?
c. What legal and ethical concerns are there?
d. What do nonprofits believe to be useful?
e. Does this project fit within our skillset?

There were several reasons why we chose not to do some of these projects. For example, we were unsure how we might be able to find and label false advertisement posts. (In retrospect, Polaris may have had more PDFs and documents allowing us to identify a method to find and identify these false advertisement posts.) Finding patterns in social media accounts of known traffickers would also have been a challenging data collection task, because we would need to sift through legal court cases to identify known, convicted human traffickers. Finding their personal social media accounts would prove to be challenging work, and it would likely result in a small database that required a lot of manual work to generate. That wouldn't have been feasible within our timeframe.

### *4.2 Monitoring and Detection of Sex Industry Posts in Social Media and Potential Uses of this Tool*

Ultimately, we decided to work in designing a machine learning model that could identify posts in the sex industry. The overarching goal of this project was to detect sex industry posts in social media. This involves determining which social media platform to use, having data for both posts in the sex industry and not in the sex industry, and storing them securely on a remote server in a database. Then the data must be cleaned and labeled as sex industry posts or not sex industry posts. After a large database has been labeled, supervised machine learning can be used to detect sex industry posts.

Before discussing the potential use cases for this, it is important to note that the term 'sex industry' is specifically being used in various parts of this thesis as opposed to 'sex trafficking'. It was already discussed that there are many different individuals with different experiences in the sex industry. Some choose to work in it consensually, some choose to work in it for survival needs, and some are trafficked within the sex industry. It is impossible, or very challenging, to tell whether a post in the sex industry is sex trafficking or consensual. Rather than labeling all sex work as trafficking or all sex work as consensual, we choose to use the term ‘sex industry’ which encompasses the many different scenarios. We do not believe we have the expertise to classify a post as sex trafficking or not. Even experts in the field can find it very challenging or impossible [2]. Therefore, the term ‘sex industry’ is used for situations where most professionals in this field or I would not be able to tell whether a post is representative of sex trafficking or not.

One potential use case for this includes identifying communities with high rates of sex industry posts and using this knowledge to publicly post information on services and resources in the community. It could also be used to identifying temporal and geographical trends in the sex industry on social media. Due to the advice from Melinda Smith at the Freedom Network USA, it was determined that it is best to only use this data for aggregate data analysis rather than using it to attempt to provide resources to people potentially experiencing trafficking [2]. This will be discussed further in the future work section.

### *4.3 Analysis of Images vs Text*

When deciding what features of posts to analyze, there are a number of possible features that were considered. The algorithm could use the text or images in the post or the time or location of the post. However, we did not want confounding factors if a temporal or spatial analysis was to be done later, so we chose not to use temporal or spatial information when labeling. In regards to images, they could easily cause unintended bias within our model. Transgender and people of color make up a disproportionate percentage of people in the sex industry due to oppression and discrimination in society [11], [12]. A machine learning model may end using these attributes in images to classify a post as in the sex industry resulting in further discrimination and bias against these individuals. In fact, this is already a problem in anti-trafficking machine learning models that have looked at images. People who are women of color or transgender individuals have been falsely accused of posting content in the sex industry and have been banned from sites [14]. In addition, many people post content that is intended to be a sexual image that is not related to the sex industry at all. Individuals may simply enjoy posting ‘sexy photos’ for compliments or a feeling of empowerment. Therefore, due to the potential harm that using images may cause, this thesis does not include any images of photos in the database nor in analysis of the data. Images may prove useful for classifying this information, but it would require additional time to ensure that these biases are not inherent in the classification or database. Therefore, we choose to only analyze text from social media posts in this thesis.

### *4.4 Focus of Thesis: Development of a Semi-Supervised Model for Classification*

Initially, the problem definition of the thesis proposal was to develop a supervised learning model that monitored and classified posts as part of the sex industry or not. However, due to the fact that supervised learning models need large databases of labeled data, we needed to have a way to label over 50,000 posts. This cannot reasonably be done manually, given that these posts require someone with knowledge of the field to label them. Therefore, the end goal of this thesis is to develop a semi-supervised model that can take some manually labeled data, and apply these labels to other posts. In order to build this model, the thesis needs to have text from posts that are in the sex industry as well as posts that are not in the sex industry to reduce bias and future errors. The data must be stored securely on a remote sever in a secure database and backed up on a separate machine. This data then must be cleaned and vectorized into a mathematical format rather than text itself. Finally, this data can be used to create a semi-supervised learning model that will result in all posts being labeled. The next step, not covered in this thesis, is to test various supervised machine learning models on this data to build a model that can monitor social media to detect sex industry posts. This last step would make the use cases discussed above viable.

# Chapter 5: Related Work

### *5.1 Semi-Supervised Learning to Detect Potential Sex Trafficking vs Consensual Sex Work Posts on Backpage.com*

There has been work done to try to determine which online posts on are sex trafficking, as opposed to posts by individuals who chose to work in the sex industry. Alvari, Snyder, and Shakarian created a semi-supervised learner to identify sex trafficking posts on Backpage.com. They used hand-labeled data labeled by experts - both a person who was trafficked and by an individual in law enforcement, and they used unlabeled data. They developed a selection of features that could be used to identify human trafficking from literature and expert advice. These were Advertisement Language Pattern, Words and Phrases of Interest, Countries of Interest, Multiple Victims Advertised, Victim Weight, and Reference to Website of Spa Massage Therapy. They use these to collect 999 posts that may be human trafficking, and then use the 2D t-SNE transformation so that their samples were easily visualized in two clusters using k-means clustering. Then Latent Dirichlet Allocation (LDA) modeling was used to determine the 25 most representative topics. 150 of these posts were labeled by the experts. From that, they use Python's scikit-learn package to do label propagation semi-supervised models of the data. They used the radial basis function (RBF) kernel and the K-nearest neighbor (KNN) to do label propagation. Their results are 90.42% accuracy for the KNN kernel, and 92.41% accuracy for the RBF kernel. Their work suggest that semi-supervised learners can be successful at identifying human trafficking posts on Backpage.com [8].

### *5.2 Using Multimodal Analysis on Instagram to Detect Posts About Drug Use*

There has also been work done to try to determine Instagram posts that have illicit drug abuse or drug dealing on Instagram. While this is not detecting sex trafficking, the process is very similar. They indeed state in this paper that the approach is generalizable to other problems such as human trafficking. Yang and Luo used a multimodal analysis to track illicit drug dealing and abuse on Instagram. Yang and Luo proposed a scheme to identify drug-related posts, as well as detect drug dealer accounts. They used a multimodal data that included images, text, relational information, and temporal patterns that had 88% accuracy in detecting drug-related posts. Given a list of drug-related terms, they collect potentially drug-related posts based upon hashtag-based searches on Instagram. They used image-based and text-based classifiers to filter potential drug-related posts. Their image-based classifier uses images from a search engine database. There are 4,819 potentially drug-related Instagram posts they downloaded, and these are manually annotated as drug-related or not by experts in the field. They train the image and text-based classifiers for the drug-related posts separately, and the outputs of these classifiers are then weighted to give an overall weighted average. They use a multi-task learning method to combine search engine augmented datasets with the images from the Instagram posts. In their text-based classifier, only the top 200 frequent words are drawn. They then use uni-gram for hashtags and bi-gram for captions and keep the top 1,000 features, resulting in a 2,000-dimensional vector for each post. A Naïve Bayes classifier is used to classify the posts. The image and text-based classifiers are then fused at the decision-level. Overall, they achieve an 88.1% accuracy and a 0.75 F1-score, while text-only classification was around 81.7% accuracy [20].

### *5.3 Organizations Using Social Media Posts or Sex-Industry Website Advertisements for Anti-Human Trafficking Purposes*

There are a good number of other organizations that use technology for anti-human trafficking purposes. Tech Against Trafficking is an organization that mapped out how 300 different organizations and entities were using technology for anti-human trafficking purposes [18]. The organizations that are loosely related to this thesis are listed in the table below. They are loosely related because they use machine learning or artificial intelligence, or they are at the intersection of human trafficking and social media. Table 4 includes the name of the entity, a description for that entity, and the entity's website. Some of these overlap with Table 3.

*Table 4 - Entities using Artificial Intelligence, Machine Learning, or looking at the intersection of human trafficking and social media for anti-human trafficking purposes [18]. Some of these overlap with Table 3.*

| **Entity Name** | **Description from Tech Against Trafficking Interactive Map** | **Website** |
|---|---|---|
| Safer | Safer was created by Thorn to identify, report, and remove CSAM. It is end-to-end encrypted [18] | https://getsafer.io/ |
| Spotlight | Spotlight was created by Thorn to aid law enforcement in fighting human trafficking. It archives over 100,000 new | https://spotlight.thorn.org/ |
| | sex ads per day, and allows law enforcement to search for specific information like a phone number, as well as identifying ads that might have children [18] | (Only available for law enforcement.) |
| Content Safety API | Google developed Content Safety API. It is an API that allows fewer people to be exposed to CSAM when reviewing material. It uses deep neural networks to process the images, and then it prioritizes material for review [18]. | https://www.blog.google/around-the-globe/google-europe/using-ai-help-organizations-detect-and-report-child-sexual-abuse-material-online/ |
| Analyze | Analyze is a software that can detect nudity (via skin tones) and explicit content (via body motions) to assist in digital investigations of CSAM [18]. | https://www.griffeye.com/ |
| Childsafe.ai | Childsafe.ai is used by law enforcement in the US, and it uses artificial intelligence to model child exploitation risk [18]. | https://childsafe.ai/ |
| Project Arachnid | Project Arachnid searches the web for child exploitation imagery. It compares imagery with information submitted in police reports around the world, and has identified 40,000 images of child exploitation [18]. | https://projectarachnid.ca/en/ |
| IdTraffickers | IdTraffickers uses biometrics to compare images of people reported missing to images of "escort ads" and human trafficking databases. This information is then submitted to law enforcement to help identify and provide aid to people experiencing trafficking, as well as to identify and track traffickers [18]. | https://idtraffickers.com/ |
| Human Trafficking Ad Classifier | Human Trafficking Ad Classifier is a machine learning model that does text classification. It is a character-level and a deep-learning model [18]. | https://github.com/nasa-jpl-memex/HT-ad- |
| | | classifiers |
| CEASE.ai | CEASE.ai uses artificial intelligence to identify new Child Sexual Abuse Material (CSAM). By identifying new CSAM material, this can help law enforcement decide which CSAM material requires immediate action. It also can help social networks identify CSAM before the CSAM is posted [18]. | https://www.twohat.com/cease-ai/ |
| Crisp | Crisp's goal is to provide a safer environment for youth who use online games and social networks. They crawl public and dark websites to deliver safer social media [18]. | https://www.crispthinking.com/about-us/ |
| Deepdive | Deepdive scrapes CSAM from web advertisements and creates a searchable database of these advertisements to aid in interventions [18]. | http://deepdive.stanford.edu/showcase/memex |
| Freedom Signal | Freedom Signal is a collection of technology that is used to connect with people experiencing trafficking, or interrupt people buying sex. It is used to determine the amount of sex industry online, and provide data and analytics. Freedom signal was created by Seattle Against Slavery [18]. | https://www.seattleagainstslavery.org/technology/ |
| Prewave | Prewave is used to identify risks that can affect supply chains. It works real time, in multiple languages, and captures information on social media [18]. | http://www.prewave.ai/ |
| LegisGATE | LegisGATE runs General Architecture Text Engineering on different document resources. It works on information from court appeals, social media, and the news. LegisGATE is a part of MemexGATE 18]. | https://github.com/nasa-jpl-memex/memex-gate |
| WAT – Traffic bot | WAT – Traffic bot looks to connect images, words, and data from sexual advertisements and other sources including social media [18]. | https://www.cs.cmu.edu/~hovy/papers/14dgo-WAT.pdf |

These entities in the table are predominantly listed as a resource for readers who are interested in related work. Tech Against Trafficking has descriptions, like the ones above, for all 300 entities researched and the entities can be searched for and filtered at: https://techagainsttrafficking.org/interactive-map/ [18]. This could be very useful for graduate students, who use research for anti-human trafficking purposes, to identify organizations or entities working in their specific areas of interest.

# Chapter 6: Practical Challenges for Conducting Research in this Field

Anytime work is done in this space, there are many ethical concerns to be thought through. The anti-human-trafficking space is very complicated and has many interweaving aspects that affect each other. For example, legislature might make a company act in a way that harms survivors but helps the company, while a lack of that legislature may make other companies intentionally exploit human trafficking. Law enforcement may have positive and negative effects, while some officers provide some people with resources, and other officers cause trauma and harm to other individuals [2]. Nonprofits trying to gain funding for their organizations to provide aid to people who have experienced trafficking may exploit and 'tokenize' people who have experienced human trafficking. They may not cite research correctly and spread misinformation among the community about what human trafficking looks like [2]. Some red flag identification techniques lead to racism that discriminate against bi-racial families [2]. Anti-trafficking information sent to people who've experienced trafficking could result in their being punished and abused by their traffickers if the trafficker finds them in possession of that information [2]. Tools are used for both good and bad in trafficking, and each person has their own unique experiences that are not always clear cut. Sex workers, people experiencing survival sex, and people experiencing sex trafficking have different experiences and may want different services. Other times they may want the same services. On top of that, whenever working within this field, one must consider the many related legal aspects. All of these things can and do interact together, and they create a truly complex environment where legal, ethical, moral, and past outcomes must be considered.

## *6.1 Ethical Concerns and Tactics to Reduce Harm*

### 6.1.1 Two Camps of Anti – Sex Trafficking Groups

One of the first things to note, is that there are generally two camps within the anti-sex trafficking organizations. One group considers all sex work to be exploitative, while another group considers people to be masters of their own choice [2].

From my experience, viewpoints presented by nonprofits and sex-worker groups vary widely, and some of the viewpoints are quite similar to two groups of feminists, the abolitionist feminists and sexual liberation feminists. Jessica Swanson discusses the argument between abolitionist feminists and sexual liberation feminists on whether to legalize or criminalize prostitution in the paper *Sexual Liberation or Violence Against Women? The Debate on the Legalization of Prostitution and the Relationship to Human Trafficking* [21]. Sexual liberation feminists believe that sex work is a choice, a valid career, and believes that it gives agency, autonomy, and choice to a woman [21]. Abolitionist feminists believe that society is inherently patriarchal and oppressive of women. They view prostitution as violence against women and as a tool to support the patriarchal global economy [21]. Swanson looks at the Netherlands, which fully legalized prostitution, the United Kingdom, which legalized selling and buying of sex, but criminalizes owning a brothel or pimping, and the United States, which decides on a state-by-state basis, but in general criminalizes both selling and buying sex. Swanson also discusses Sweden as a country where buying sex is criminalized, but selling sex is legal, and cites the decreased levels of street prostitution in Sweden and a decrease in men buying sex in Sweden. Swanson claims that there are no current measures that indicate one of these policies results higher rates of human trafficking than others. She claims that the efforts and arguments should be put into other areas of focus. Swanson acknowledges that there are challenges since the laws are quite new. The majority of anti-trafficking laws were less than 15 years old when this paper was published in 2016 [21].

Swanson argues that human trafficking happens not only domestically, but at a global scale. This means that individuals experiencing human trafficking transnationally have their own culture and experiences that they bring to countries where they are trafficked. Likewise, they experience the new culture in the countries they are trafficked within. Countries have different views of women, and both the sexual liberation feminist arguments and abolitionist feminist arguments are rooted in Western beliefs that not all countries share. In addition, Swanson argues that vulnerabilities are what ultimately increase the risk of someone experiencing human trafficking. These vulnerabilities include, but are not limited to, a lack of economic opportunities, a need to provide financial support for family, oppression, desperation, or a need to travel to another country [21]. In addition, the Polaris Project identifies some significant risk factors for human trafficking which include mental health concerns, being a homeless or runaway youth, having recently migrated, being involved with the child welfare system, and substance use [22]. Swanson argues that while the arguments presented by the Abolitionist Feminists and Sexual Liberation Feminists are valid, both arguments fail to view the entire complex issue and generalize in a way that does not include all individuals [21]. There are individuals who freely choose sex work, there are those who choose it out of desperation or survival, and there are those who are experiencing sex trafficking and are forced, coerced, or experiencing fraud. Each person within the sex industry is unique, and every person has their own background, culture, experience, and viewpoint [2].

Some things that Swanson argues may help reduce human trafficking include understanding the level of desperation potential ‘victims’ experience, and truly addressing those vulnerabilities including economic disparity, gender inequality, racial and religious discrimination, violence, homelessness, support for those experiencing drug addictions, etc. Swanson believes it would be better for laws to punish the buyer rather than the victims as well as providing assistance to victims where they are not required to ‘snitch’ about their trafficker [21]. Swanson also argues that criminal penalties for traffickers should reflect the seriousness of the crime. For example, the majority of traffickers convicted in the Netherlands experienced less than 1 year in jail. The relaxed punishment for traffickers means that there is little risk to the trafficker and people who experienced human trafficking are less likely to want to testify against their trafficker for fear of retaliation. Swanson also argues that the laws should be more straightforward to increase convictions [21].

During an interview with Melinda Smith MSW, who is the Director of Partnerships at Freedom Network USA, Smith presents another argument that disagrees with criminalizing the buyer of sex. Criminalizing buyers of sex brings law enforcement into scenarios with prostitutes and can lead to traumatization of people, either experiencing trafficking or participating in consensual sex work. This is because law enforcement is not trauma-informed care, and some law enforcement officers are known to have raped, harmed, punished, returned individuals experiencing trafficking to their traffickers, and been clients to individuals selling sex [2]. In addition, people have been sexually assaulted and raped by law enforcement as either a proof that they were selling sex to arrest them or as a threat of “do this, or I will arrest you” mentality. All of this spreads distrust and fear of law enforcement among people in the sex industry [2]. In addition, buyers of sex who are publicly shamed or criminally prosecuted sometimes take out their anger at individuals in the sex industry, which creates a dangerous situation for people in the sex industry [2]. This prevents many in people in the sex industry, who experiencing rape or assault, from coming voluntarily to the police. When police are not trained in trauma-informed response, like social workers are, they can cause further harm to people experiencing human trafficking. Smith argues that criminalizing or decriminalizing sex work affects not only the pimp, but the person selling sex as well. Smith believes decriminalizing the buying and selling of sex work makes it safest for people in the sex industry [2].

All in all, the issue of human trafficking is incredibly complex. Good data in the field of anti-human trafficking work is hard to access and this can mean that it is hard for countries to determine if and when human trafficking is increasing or decreasing. In addition, the voices of people who have experienced human trafficking or are consensual sex workers, are often left out of the conversation which means that laws intended to do good can cause unintended harm, or may do both good as well as causing harm. One example of such a law is the SESTA-FOSTA law that the US passed recently, which will be discussed later on in this chapter and relates directly to this project.

### 6.1.2 Potential Harm Caused by Reaching Out to Individuals

Initially, this thesis planned to build a machine learning algorithm that could identify posts in the sex industry. One of the applications of this was that an expert in the nonprofit or social work setting could identify potential sex trafficking cases. They could then reach out and offer support, resources, or information to individuals who may be experiencing sex trafficking. These ideas were generally supported by the organizations and individuals whose philosophy may align more closely with the Abolitionist Feminist philosophy that prostitution is inherently exploitative and serves a patriarchal global economy. After reading resources on several sex worker sites, it became apparent that this had the opportunity to cause unintended harm. I reached out to ten sex-worker advocacy groups, transgender advocacy groups, and nonprofits who believe that not all sex work is trafficking or exploitative. These different viewpoints allowed me to identify potentially harmful outcomes from the initial use case of our machine learning algorithms.

Note that only one of the ten organizations was willing or able to meet with me. This was the nonprofit Freedom Network USA. Other groups either did not respond, responded initially but later declined to respond, or informed me that meeting with them was out of the budget of this project. There may be several reasons for this. Sex workers are, generally speaking, not paid a salary wage, so any time they spend meeting with me directly takes away from their income as compared to a nonprofit staff member. In addition, sex work is generally illegal in the USA so there is considerable risk to a sex worker meeting with an unknown individual despite the fact that I claim to be a graduate student seeking their input. Another reason is that sex workers are generally not valued by the society as a whole, but they do recognize that they deserve to be valued individuals and therefore, request money for using their time to support another project. Lastly, I am not creating a project aimed to support sex workers, but rather to fight human trafficking. Anti-trafficking work has often not considered their voices sometimes resulting in unintended harm. There may be a distrust already of anti-trafficking work, resulting in hesitancy to have discussions. However, if is important for future work to hear these voices, so I would recommend future research work in this space to set aside as least $300 for a consultation with a sex worker consultant from a sex-work advocacy group. This was not in our budget, because we were not aware that we would need it.

The tools developed in this thesis could cause unintended harm if a social worker or nonprofit reached out to someone experiencing sex trafficking on social media. Traffickers can control the social media accounts of those experiencing sex trafficking. If a trafficker found evidence that the person experiencing trafficking was communicating with or contacted by a social worker, they could be punished violently by their trafficker [2], [1]. Separately, there are some ways social workers contact people experiencing trafficking, including disappearing messages, but that would be inappropriate for the context of this project since there is no known time when a social worker could send a disappearing message and ensure the trafficker could not view it [1].

Unintended harm could also occur if a nonprofit with an Abolitionist Feminist Philosophy reached out to a consensual sex worker. The language that the nonprofit might use could suggest that the career the sex worker choose was not morally acceptable, or not a valid choice. This could result in further distrust between consensual sex workers and nonprofits [2]. This would not only be harmful to the sex worker, but sex workers are one of the best resources to reach out to people experiencing sex trafficking. They may be able to identify which people in the sex industry are being trafficked and have contact with people experiencing sex trafficking that social workers or other professionals cannot have contact with. [12].

Another challenge is that, in general, no professional social worker, nonprofit staff member, or law enforcement official can identify a social media post that is posted by someone experiencing sex trafficking vs someone participating in consensual or survival sex work. The only time it is clear, is when the individual is clearly or explicitly under 18 years old, at which point sex work is legally considered sex trafficking [2].

Another area for potential harm is if law enforcement had access to the machine learning algorithm and could identify people in the sex industry. Then there may be a rise of people in the sex industry on social media experiencing trauma at the hands of the police [2]. As mentioned earlier, some law enforcement officers abuse their powers and sexually assault people in the sex industry. Traffickers of foreign people experiencing trafficking may also use deportation by police as a means of control. Immigrant sex workers may also fear deportation and fear the police. These concerns suggest providing the identifying Machine Learning algorithm to law enforcement may cause unintended harm.

Therefore, based upon these reasons, we believe that this algorithm should only be used for aggregate data analysis. For example, one might use this Machine Learning algorithm to identify cities with large volumes of sex work on Instagram and post a list of resources in public areas within that city. It would not target any specific individual, but might create a larger awareness in that area of what resources exist should someone experiencing sex trafficking decide to seek assistance. In addition, if there were communities with a high amount of sex industry on social media, local communities could look into what vulnerabilities may be creating an increased amount of survival sex work. Then those communities could address those vulnerabilities so individuals engaging in survival sex work may have the option to choose another career if they want.

### 6.1.3 SESTA-FOSTA/Legal Aspects that Affect Use of this Project

One example that highlights the complexity of determining how to morally do this project is highlighted by the SESTA-FOSTA law. SESTA stands for Stop Enabling Sex Traffickers Act, and FOSTA stands for Allow States and Victims to Fight Online Sex Trafficking Act. This became law on April 11, 2018. One part of this law changes the Communications Decency Act to state that web companies who knowingly, actively facilitate sex trafficking can be prosecuted by the state and sued civilly. This law came into place because Backpage.com was knowingly altering online advertisements where the individual being advertised was clearly a minor. They knowingly altered them to be less suspicious to law enforcement. Families of the individuals who were trafficked and advertised on Backpage.com were told in court that all web companies are protected from the content created by others on their site. This prompted the SESTA-FOSTA law [1].

Another aspect of the SESTA-FOSTA law is that it prohibits using the internet to facilitate or promote the prostitution another individual. This meant that many websites that facilitated prostitution voluntarily shut down. While this law did send a strong message to traffickers, it also created fear among individuals in the sex industry. Without the help of these popular sites to meet sale quotas, people experiencing trafficking could encounter serious 'punishment' at the hands of traffickers if they do not make enough money [1]. It also means that social media companies now actively try to remove profiles that are soliciting sex because facilitating prostitution of another person is prohibited by law. This could have serious implications if someone tries to partner with Instagram, Facebook, or other social media companies to try to reduce human trafficking. The company most likely will remove the sex industry profiles rather than finding a way, aggregate or not, to support those individuals. Removing profiles related to the sex industry does not help people experiencing trafficking in general. People experiencing trafficking may have a harder time filling the sales quota, or they may have to resort to more dangerous methods of selling sex such as on the streets. The law was intended to stop traffickers and those who knowingly supported sex trafficking, like the executives at Backpage.com [2]. However, the law now indirectly negatively affects anyone in the sex industry including people experiencing sex trafficking [2]. This means that using social media to detect trafficking in partnership with a social media company is much more challenging, if not impossible. This makes data acquisition much more difficult for many anti-sex trafficking projects in the social media area.

### *6.2. Legal Aspects for Researchers to Consider*

This project not only has many ethical and legal aspects to consider. This next section simply addresses some legal information that anti-sex-trafficking researchers in the future may find useful to be aware of. It is, of course, not intended to be legal advice of any kind. This does not cover everything that a researcher could run into, because each research project is different and would have different aspects to it. But it may raise awareness among researchers of some important considerations. If researchers don't have a dataset to use, some researchers may want to scrape data. This chapter discusses several legal concerns researchers may have if they wanted to web-scrape social media data. This chapter also discusses legal risks associated with accidentally, or intentionally, scraping, storing, or accessing content that might be related to or comprise of a minor engaged in sexual conduct, also known as child pornography.

### 6.2.1 Laws Regarding Child Pornography

Researchers who work in the anti-sex trafficking space should be aware of laws regarding child pornography, since they may accidentally come across images that are child pornography. The federal law that criminalizes child pornography related offenses defines child pornography as a visual depiction where:

"(A) the producing of such visual depiction involves the use of a minor engaging in sexually explicit conduct; and
(B) such visual depiction is of such conduct" (18 U.S.C. 2252)

Researchers should note that accessing, downloading, or storing images of child pornography or storing URLs of the child pornography could make a researcher liable under child pornography related offenses. If a researcher does come across child pornography, it's extremely important that the researchers report any child pornography or known sex trafficking of minors to the National Center for Missing & Exploited Children and remove that content from their dataset/computer. This is important for the sake of the minor being exploited and to help reduce the probability that the child pornography appears on other sites in the future.

### 6.2.2 Laws Regarding Collecting Data Automatically

Researchers likely will want to work in collaboration with the respective companies if they would like to use the publicly available data on social media. If not, a researcher would want to consider legal risk associated with scraping public social media data without permission. There are several laws to consider if someone chose to collect data from a social media site whose Terms of Service do not permit social media scraping. They are the Computer Fraud and Abuse Act (CFAA), the California Penal Code §502(c)(2), and the trespass to chattels. In addition, social media data is copyrighted by the original poster, so Copyright laws would apply. There may be other laws as well, depending on the social media company or where the lawsuit would take place. The best way to avoid these laws is to not scrape data from social media sites.

#### *6.2.2.1 Breach of Contract Claim*

Researchers may want to know about the Breach of Contract Claim. First, by going to a social media site, even without an account, the user may be accepting the Terms of Service. Sites in California can sue for a breach of contract if four things are proven. I will not list them all, but one of the things that must be proven is damages to their site [23], [24], [25]. Researchers should not automatically collect data in any way that would or could cause damages or harm to a site, for both moral and legal reasons.

#### *6.2.2.2 CFAA Violation Claim and the California Penal Code §502(c)(2)*

If a website brings a Breach of Contract Claim, they can also bring a CFAA claim and a California Penal Code §502(c)(2) claim. The CFAA is a criminal law and therefore, it could result in heavy fines and/or imprisonment. The CFAA can be applied, among other situations, if a person "intentionally accesses a computer without authorization or exceeds authorized access, and thereby obtains… information from any protected computer" (18 U.S.C. § 1030(a)(2)(c)). The CFAA also defines a protected computer as, among other things, a computer "which is used in or affecting interstate or foreign commerce or communication, including a computer located outside the United States that is used in a manner that affects interstate or foreign commerce or communication of the United States" (18 U.S.C. § 1030(e)(2)(b)). I do not have a law degree, but many social media companies seem to work at an international scale. I suspect that they may be used in foreign communication or commerce, depending on how our government chooses to define those terms. Regardless, the CFAA claim and California Penal Code §502(c)(2) claim are claims that researchers in this space may want to be aware of.

#### *6.2.2.3 Trespass to Chattels Claim*

A trespass to chattels claim can result when an individual interferes with someone else's property causing injury. Researchers may want to consider the trespass to chattels and should make sure that their research is not interfering with other people's or company's property in a way that causes harm.

#### *6.2.2.4 Copyright Laws*

Also, researchers should know that social media posters own the copyright to the text, videos, images, and any medium of expression of their social media posts. By scraping social media posts, this would mean a researcher could be liable under the Copyright Act. However, a researcher may be protected because the Copyright Act has a fair use exception. Columbia University created a fair use checklist that can help individuals know whether their use of copyrighted material is favored by the fair use exception. This checklist can be viewed at: https://copyright.columbia.edu/content/dam/copyright/Precedent%20Docs/fairusechecklist.pdf [26].

Overall, there are several things researchers need to think about when working on an anti-trafficking project that involves social media. Hopefully this information can raise awareness among researchers of some important legal considerations.

# Chapter 7: Overview of Solution and Technical Design

The goal of this thesis is to classify over 50,000 Instagram posts as sex industry or not sex industry using the text from the post. The general solution that this thesis puts forth is to classify them using a semi-supervised machine learning algorithm. This project can then be broken down into smaller subcategories including data acquisition/management, manual labeling, and semi-supervised classification, which is shown in Figure 2.

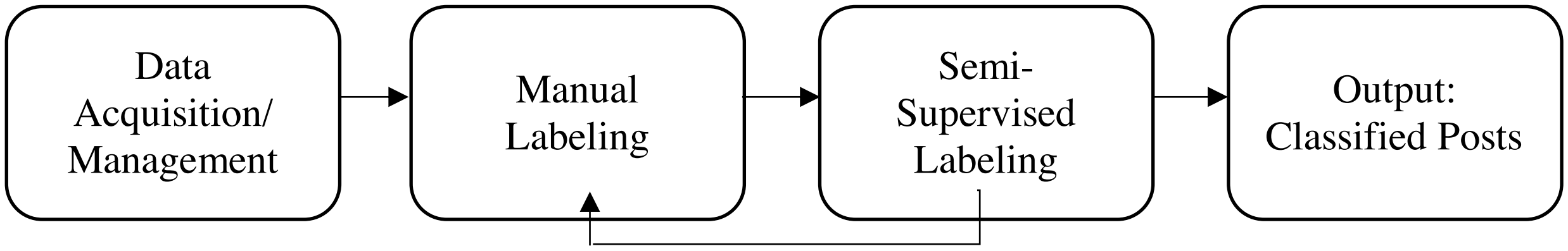

*Figure 2 – Thesis work flow chart*

Data acquisition and management includes multiple components. In regards to data acquisition, there is a dataset with the text from over 50,000 Instagram posts. I also worked to ensure secure storage of this data in a MySQL database on a remote Amazon Web Services (AWS) server. Several security measures are taken to ensure that the data is stored securely. For example, I require a unique username and password for anyone working on this project, and those users can only access the data on their specific IP addresses. For example, even with my username and password, I could not access the database on another team member's Wi-Fi, and vice versa. Neither of us would be able to access the database from any IP address that I did not explicitly approve. There are several other means of keeping the data secure that are discussed later. The process of data acquisition and management will be discussed further in Chapter 8: Data Acquisition and Management.

The MySQL database is then used in the manual labeling of the posts. Having read anti-trafficking reports, watched anti-trafficking documentaries, audited a modern-day slavery class, and met with anti-trafficking nonprofits over the past 3 years, I felt I was well equipped, compared to most technical graduate students, to manually label the data. Given that I am not an expert in the anti-human trafficking field, I choose to have an expert in this field manually label data as well. I then use this to determine the Inter-Rater Reliability (IRR). After she labeled the data, she met with me to explain how she labeled the data, so I could use this to improve my own labeling of the data. Manual data labeling is discussed in further detail in Chapter 9 – Manual Data Labeling.

The next step of the thesis is the semi-supervised labeling. This includes several subparts including data cleaning, clustering, and using the existing manual labels to label the clusters, and propagating the labels to the other data points in the clusters. Figure 3 depicts the flow of the semi-supervised learning.

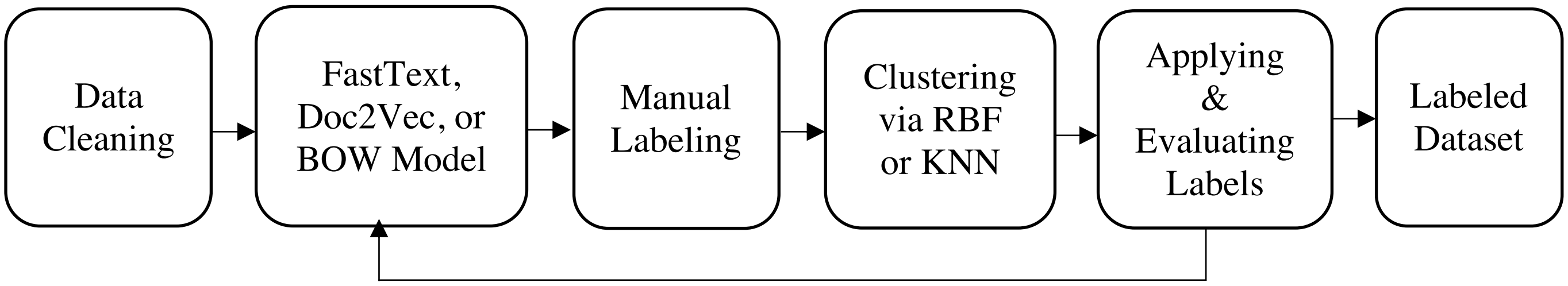

*Figure 3 – Semi-Supervised Learning Flow Chart*

The semi-supervised model should only apply labels in clusters that are clearly in the sex industry or not in the sex industry. Clusters of posts that are not clearly in or not in the sex industry need to be manually labeled. However, this semi-supervised learning model allows the amount of manual labelling to be significantly reduced. Posts can be represented mathematically via Word2Vec, Doc2Vec, FastText, or Bag of Words. One reason FastText is a good way to model Instagram posts is because FastText also looks at character level meanings, so even if a new word is discovered after the model is built, the FastText model can still process that word [27]. This is important for social media sites like Instagram that use hashtags that are constantly changing. Word2Vec and Bag of Words cannot handle seeing words outside of the original model’s vocabulary. Both Word2Vec and FastText embed contextual meaning within a word. For example, star and moon may have a high similarity, even though on a character basis they are quite different. Bag of Words does not encode this contextual meaning. Once the posts are vectorized, clusters can be formed using Kmeans Clustering, K-Nearest Neighbors (KNN) or Radial Basis Function (RBF). This will be discussed in further detail in Chapter 10 – Data Cleaning and Chapter 11 – Semi-Supervised Learning.

This thesis is different from past work because many of the tech anti-trafficking organizations and DARPA funded Memex research projects specifically look at websites designed to facilitate prostitution or look at CSAM [18]. This project instead is in an area that seeks to identify sex industry on social media independent of age in an effort to reduce human trafficking. Social media posts related to sex industry are different in many ways from sites intended to advertise prostitution. Sites intended to sell prostitution have advertisements that can be much more explicit, while individuals in the sex industry on social media have to ‘filter’ their messages so that the social media sites do not kick them off of social media. As was mentioned earlier in this thesis, SESTA-FOSTA has resulted in many websites blatantly removing any profile associated with the sex industry, regardless of whether or not it might be sex trafficking. This is because, as mentioned before, it is illegal now for websites to facilitate the prostitution of another person [1]. Research has been done to identify posts that may be sex trafficking on sites designed to sell prostitution [8], and research has been done to identify posts that are related to drug usage on Instagram [20], but, as far as I am aware, public research has not yet been done to identify posts that are in the sex industry on Instagram or on social media. Instagram may itself be doing their own company research in this area in attempts to adhere to SESTA-FOSTA regulations or for their own company reasons, but this is unknown to me. Doing this will allow machine learning models to be developed to detect posts in the sex industry on Instagram, both of buyers and sellers of sex. This could allow public resources to be disseminated in communities with higher rates of sex industry.

# Chapter 8: Data Acquisition and Management

## *8.1 Dataset Information*

In order to do this research, we needed to have a dataset of posts from Instagram that included both posts related to the sex industry and posts that are not related to the sex industry to reduce bias in our algorithms. We have a dataset that has text from over 50,000 posts from posts both related to the sex industry and posts not related to the sex industry. Currently, this database is fixed and cannot have more posts added to it. This unfortunately means that some data that would be very helpful for this thesis cannot be obtained, included knowing whether a comment or reply was posted by the original poster or another user.

## *8.2 AWS Server Security*

I stored the data with the best security that I knew how to implement on an AWS server. The data is currently stored in a MySQL database on an AWS server. The AWS server is encrypted. The AWS server only allows specific IP addresses with a public/private key pair to access the server itself via SSH over the TCP Protocol. I am the only individual with SSH access to this server now and in the past. The MySQL database must be accessed via the MySQL/Aurora Port 3306, over the TCP Protocol. While writing this thesis, I am the only individual with access over Port 3306 or with SSH access. However, throughout the past summer and spring 2020, the AWS server could be accessed by team members who were working on this project at that time, as well as by other AWS servers that I owned that needed access to the database. While the server does not require a password to connect over Port 3306 on the TCP protocol, the MySQL database itself does require that.

## *8.3 MySQL Database Security*

In order to access the MySQL database, each team member has their own individual username and password coupled with their IP address. This means that in order to access the database, an individual had to not only know a correct username and password combination, but they had to be accessing it from the corresponding IP address(es) approved for that username/password combination. The root username and password could only be accessed via the actual root login on the AWS server hosting the MySQL database, which required a private/public key pair login coming from my IP address. Passwords for this database were random long strings of number and letters both capitalized and lower case. They were all well over 40 characters, meaning there were much more 4.96e71 combinations for a hacker, who got onto our IPs and knew our unique usernames, to try before they could access the database. These procedures can help the database protect from attacks from other individuals who may be breaking into MySQL databases to steal information. I recognize that there are other ways that attacks can be made. For example, someone within a company could have bad intentions and use their inside knowledge and access to complete an attack. I helped mitigate this by reducing the number of individuals who had access to the database. There was only one team member who needed regular access to this database besides me. By severely limiting who has access to the database, this allows the data to be stored more securely with much lower risk. This was done to protect the data of the individuals whose posts are in the dataset we are using. All of the data that is in the database was publicly available data that anyone, even without an Instagram account, could access. Regardless, it is extremely important to protect the data for ethical and legal reasons.

### *8.4 MySQL Database Structure*

The structure of the MySQL Database is shown below in Figure 4. There are various tables in the MySQL Database that consist of over 50,000 posts. Since this thesis only looks at classifying posts as sex industry or not sex industry based upon the text in the post, each post may consist of the post caption, the post comments, and the post replies. The aspects of the database that are used in terms of the semi-supervised learning model are *italicized*. The other attributes of the database are used for other things, including labeling the posts or reporting posts to the National Center for Missing and Exploited Children when relevant. Most of the structure of this database is the same as a database Bernardo Garcia Bulle Bueno made while working on this project. I followed the same structure as his database and added tables as necessary. This is shown in Figure 4.

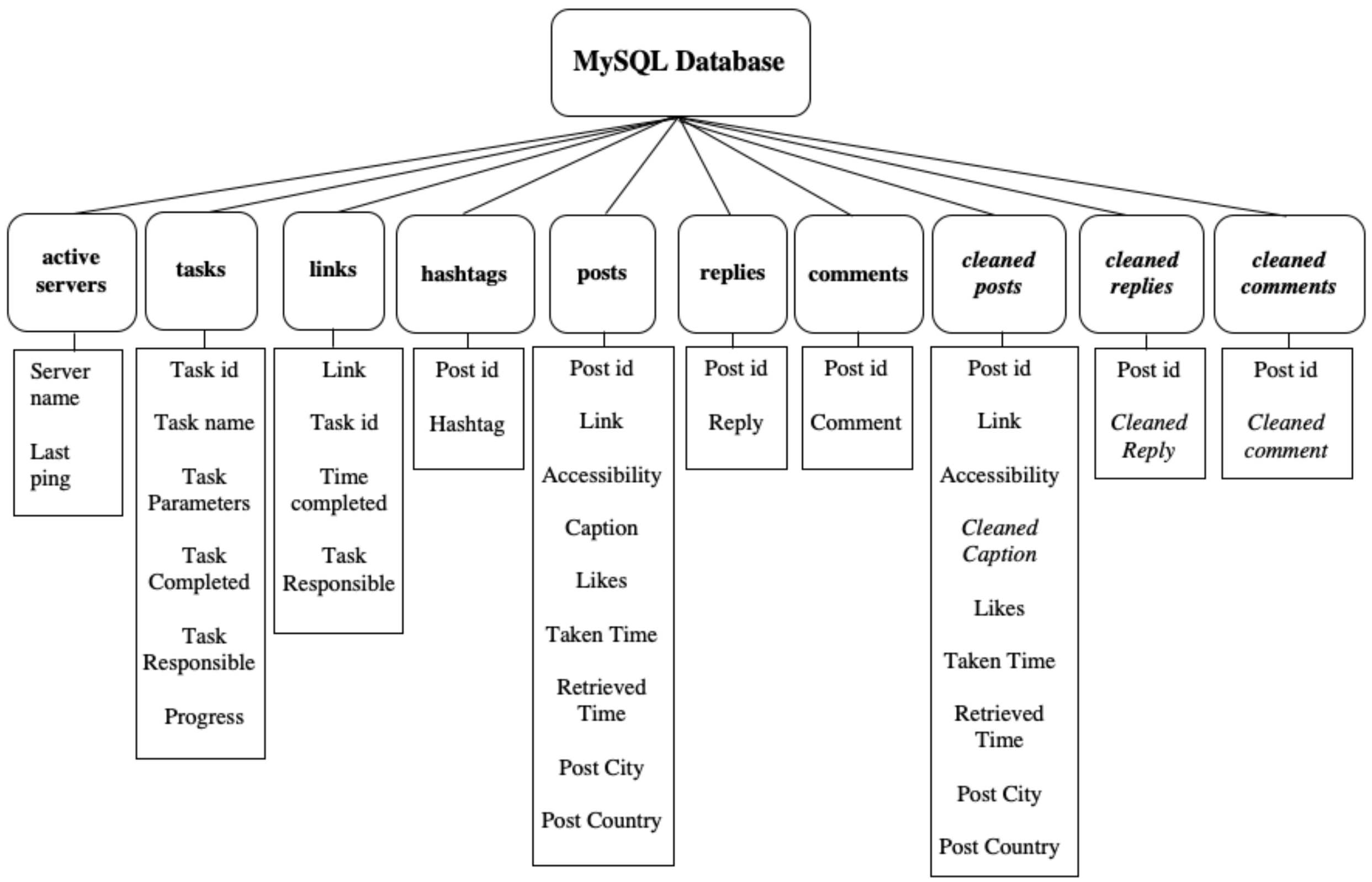

*Figure 4 - MySQL Database Structure*

# Chapter 9: Manual Data Labeling

## *9.1 Manual Labeling with Expert Ashley Guevara LICSW*

In order to create a semi-supervised model that labels all 50,000+ posts, there needed to be a smaller subset of manually labeled data. For some projects, anyone can manually label data. For example, deciding whether a picture has a cat or not could be labeled by almost anyone. For this project, we need to classify the post as either in the sex industry or not in the sex industry, and someone with professional or trained experience is needed. There is a lot of room for bias in the classification. For example, someone might classify sexy photos of a woman as part of the sex industry when that post may have nothing to do with the sex industry. Therefore, we proposed having two individuals label the data. Since this is my thesis, and I have been learning about human trafficking for the past three years, I was one of the individuals to label the data. However, I recognize that I am not an expert in the field, especially compared to social workers who work with people who have experienced sex trafficking. The second individual who did manual labeling is Ashley Guevara LICSW, a licensed social worker who works for My Life My Choice. Guevara labeled 201 posts so that I could do an Inter-Rater Reliability to decide if I was labeling the data well.

## *9.2 Definition of the Sex Industry for this Thesis*

When talking to Guevara, I asked her to define her definition of the sex industry, I stated that my goal was to identify posts related to the specific sex industry areas that might involve exploitation or might be areas that vulnerable populations are providing the services. I explained that we are looking at posts on both the demand (buyers, for example) and the supply side. I gave this explanation on what the sex industry means for this project and explained possible use case scenarios for this research. With that information, Guevara gave a definition for the sex industry that she used to label the data. Guevara explained that she viewed the sex industry as anyone using any part of their body for the sexual pleasure of someone else in exchange for something of value: money, shelter, presents, etc. [28].

### *9.3 Labels Permitted When Labeling the Posts*

Using this definition, Guevara labeled 201 posts, and I labeled 500 posts. When labeling the posts, there were four options for labeling.

- Removed, or 'removed': Some posts had been removed, and those were labeled 'removed'. From a technical standpoint, if a post is labeled as removed, it is as if the post was never labeled in the first place.
- Part of the Sex Industry or 'yes': The second option is when the post is in the sex industry, and it is labeled as 'yes' as based upon Guevara's definition of the sex industry.
- Unclear if Part of the Sex Industry or 'unclear': The post could be labeled as 'unclear' if it is not clear whether the post is related to the sex industry or not.
- Not Part of the Sex Industry or 'no': If a post is not related to the sex industry, it is labeled as 'no'.

In addition to using the publicly available data from the post to label the post, we sometimes need to use the profile to label the post if the post itself cannot be clearly labeled. When we need to use the profile to label the post, there is a separate label that describes whether or not the profile was needed to label the post. It is labeled as 'yes' if the profile page was needed to label the post, and 'no' if the profile page was not needed to label the post. This is important, because knowing this may indicate whether having Natural Language Processing (NLP) on the Post Text is sufficient or whether analysis of the profile may be required in some cases for the machine learning to adequately label the data.

### *9.4 Post Attributes that Help Label the Data*

**Notice to reader: In the next paragraph, some readers might find hashtags related to the sex industry to be psychologically harmful, triggering or upsetting. In this case, the reader may skip to 9.5.**

When labeling the data, there were several things that helped me identify whether a post was part of the sex industry or not. Usually, there was terminology, as part of a profile describe, hashtags, a post caption, or comments that indicate there was some exchange occurring in the sex industry. After discussing with Guevara, she stated that she similarly found these things helpful for identifying posts as part of the sex industry or not. In particular, she found hashtags helpful [28].

This may be because a user can search for posts by hashtag. Therefore, it is advantageous for both the person trying to sell and trying to buy within the sex industry to use hashtags that allow the other individual to find their post. For example, someone selling sex may use the hashtag #paypig or #findom to indicate that they are in the sex industry, and someone trying to buy sex may use the hashtag #sugarbabywanted or #sugarbabyneeded.

**End of paragraph including potentially psychologically harmful, triggering or upsetting hashtags.**

### *9.5 Post Types: Control Posts and 'SIH' Posts*

Guevara labeled 201 posts, separated into two groups, which allowed me to calculate the Inter-Rater Reliability (IRR) between her labels and my labels. Before discussing the IRR, it is important to note the categories that the posts that were separated into. The dataset has both control posts, and posts that are more likely to be in the sex industry. The control posts are in the dataset because they have one of 500 of the most commonly used hashtags on Instagram. These posts are unlikely to be part of the sex industry, and are more likely to include pictures of food, dogs, or people on vacation. These posts are very rarely labeled as part of the sex industry, although it has occurred. These posts are referred to as Control Posts.

There are posts that are more likely to be in the sex industry in the database because the post had hashtags that were indicative of the sex industry. These posts are referred to as posts that are more likely to be a part of the Sex Industry because of their Hashtags, or SIH Posts. It does not mean the posts *are* part of the sex industry, just that they are more *likely* to be part of the sex industry.

A Random Sample of both Control Posts and SIH posts are chosen to be labeled by the expert. The Random Sample of posts represents the dataset as a whole and is a sample of both Control Posts and SIH Posts.

In addition, a SIH Sample of only SIH posts are chosen to be labeled by the expert. The SIH posts represent the subset of the dataset that is more likely to be a part of the sex industry.

These four categories, Control Posts, SIH Posts, Random Sample, and SIH Sample are referenced throughout this chapter in the IRR Tables.

### 9.6 Inter-Rater Reliability (IRR): Post Samples Chosen

Of the 500 posts I labeled, I created two samples, both of size 100 posts, and I made sure none of these two 100-samples had overlapping posts. The first sample had 100 posts from all the posts I labeled randomly chosen to be labeled by the expert. This sample is referred to as the Random Sample, and it represents on average how well I labeled the posts compared to the expert in the field. 100 of these 500 posts were chosen to be labeled by Guevara. Then I chose 100 randomly chosen SIH posts from the remaining 400 posts. As a reminder, these SIH Posts included hashtags indicative of the sex industry and were more likely to be a part of the sex industry. This allowed me to see how well I was labeling the SIH Posts. Lastly, not as a part of either of the groups, I wanted to see how Guevara labeled a control post that appeared to be part of the sex industry, so I included one extra post for the purposes of learning, bringing the total posts labeled by Guevara to 201.

Most of the Control Posts are easy to label because they are clearly not part of the sex industry and contain posts about food, pets, motivational quotes, sports games, etc. However, the SIH Posts are much more challenging to label. The labeler has to determine the answers to questions like, what parts of the foot fetish community may be in the sex industry? If a user asks the original poster to be their sugar baby, but the original poster does not have that in mind, is the post part of the sex industry? Are offerings to sell images and videos, but no meet-ups in person, part of the sex industry? After Guevara labeled the posts, Guevara offered to discuss with me how she labeled the data so I could label the data more accurately in the future. This is discussed later in this chapter.

### 9.7 Initial IRR before Feedback Call

Table 5 below shows the Initial IRR accuracy. Remember that sometimes posts were removed, meaning that they could not be labeled. Of the 200 posts I asked Guevara to label, 48 had been removed by the time she labeled them. Therefore, there were 152 posts that could be compared during the Initial IRR. Of these 152 posts, there are a few categories that need to be defined. The Random Sample category includes 84 of the original 100 posts in the Random Sample because 16 of them had been removed. The Random Sample category represents an average of how well I am labeling the data compared to Guevara. It contains both Control Posts and SIH Posts. The IRR for the Random Sample was 83.3% correct, so on average, I was labeling my posts the same as Guevara would label them 83.3% of the time. The SIH Category originally contained 100 posts, but 32 posts had been removed by the time Guevara labeled the data. This category of posts had hashtags indicative of the sex industry. The IRR for the SIH Sample was 67.6%, so of posts that had hashtags indicative of the sex industry, I was labeling them correctly about 2 out of 3 times. I then combined all of the posts Guevara labeled, including the one control post that I believed to be part of the sex industry described earlier. I separated these posts into Control Posts and SIH Posts. The 53 control posts had a 98.1% IRR accuracy. There was one post that we differed on. Later during the Feedback Call with Guevara, she expressed that she didn't feel this post was 'unclear' but rather a 'no'. This brought the IRR for the Control Posts to 100%. When looking at all SIH Posts from both the Random Sample and the SIH Sample, there was an IRR accuracy of about 65%. The incorrect labels consist of subparts that are partially or completely incorrect. Completely incorrect labels are when Guevara and I labeled the posts 'no' and 'yes' or vice versa. Partially incorrect labels occur when one person labels the post as 'unclear' and the other labels it as 'no' or 'yes'. This indicates that one of us was unsure or unclear, rather than having completely different viewpoints. Of the posts that are labeled incorrectly, they are roughly half partially incorrect and half completely incorrect. See Table 5 on the following page.

*Table 5 - Initial Inter-Rater Reliability between expert social worker, Ashley Guevara LICSW, and Ellie Simonson. *Later on, during the Feedback Call, Guevara expressed that she didn't feel the one Control post we labeled different was an 'unclear' but rather a 'no' post, which brought the IRR for the Control Posts to 100%.*

| Initial IRR | Random Sample (Included 84 posts) | SIH Sample (68 posts) | Portion of All Posts that were Control Posts (53 Posts) | Portion of All Posts that were SIH Posts (100 Posts) |
|---|---|---|---|---|
| Percent Correct / Same labels (%) | 83.3% | 67.6% | 98.1%* | 65% |
| Percent Incorrect / Different labels (%) | 16.7% | 32.4% | 1.9% | 35% |
| Subpart of Incorrect Labels that are Completely Incorrect (yes vs no) (%) | 7.1% | 16.2% | 0% | 17% |
| Subpart of Incorrect Labels that are Partially Incorrect (unclear vs no, or unclear vs yes). | 9.5% | 16.2% | 1.9% | 18% |

This clearly shows that almost all of the difference in our labeling was in the SIH posts that had hashtags indicative of the sex industry. My labeling matched only about 2 out of 3 SIH posts with Guevara's labeling, and I was labeling roughly 1 out of 6 SIH posts completely incorrectly. This was not acceptable for my labeling. I clearly needed to learn better how Guevara was labeling the posts to understand what I was doing differently.

### 9.8 Feedback Calls with Expert

Because my labels did not match well with the expert's labels, Guevara and I had two feedback calls to help me better understand how she labeled the data. There were several things that I learned from these calls.

First, I learned that even an expert in the field can mislabel the posts due to some of the key information not being in plain sight. There were 8 posts that Guevara changed her label on when I specifically asked her about those posts and showed her information indicative of the sex industry. Sometimes, this information had been hidden in the comments, and one had to expand all the comments and search for the original poster's name in order to see this information. Other times, the information indicative of the sex industry was on different posts by the same user, and this wasn't clear unless the labeler happened to look at one of those posts. When I asked her about those posts, curious what I had done incorrectly, she clarified that she had not seen that information and wanted to change her label of those posts. This same thing has happened to me in the past as well. The 8 posts Guevara wished to change her labels of represented just over 5% of the 153 posts she labeled. Note that I didn't go over every post with Guevara, so this number may have been higher. There may have been posts with additional information that was not seen by either of us.

Second, I learned that Guevara considers many of the posts in the foot fetish community to be a part of the sex industry. Initially, when labeling, I felt that the foot fetish community was not a part of the sex industry. This was because I had thought that someone selling photos or videos of their feet may be less likely to be being exploited. In addition, feet are not genitalia or parts of the body in the USA that are typically considered private. Therefore, I would generally label posts in the foot fetish community as not part of the sex industry. However, when I had the Feedback Call with Guevara, I learned two things. First, Guevara defines the sex industry to be when someone is using their body for the sexual pleasure of someone else in exchange for money or something valuable like shelter, presents, or other valuables. Taking photos or videos of one's feet and selling them to people with a sexual foot fetish, would fall under using one's body to provide sexual pleasure for someone else in exchange for something of value. Second, Guevara explained that individuals buying foot fetish photos or videos from someone may use that very information to exploit or pressure that individual into providing more. For example, the person selling could be intending to only sell foot photos and videos. However, the person buying the photos and videos may threaten to release the photos to the public, show them to the boss or family of the person selling, or publicly humiliate that individual unless the person selling provides more sexual services to the person buying. Then this cycle could continue and the person selling may feel more and more trapped as they are coerced into providing more sexual services. This was the area where I mislabeled most of the posts. There were at least 10 posts in the foot fetish category out of the 153 posts that I originally labeled as no and Guevara labeled as yes. However, I also learned that there are posts within the foot fetish category that are not within the sex industry as well. Not all foot fetish posts are looking to exchange photos, videos, or experiences for something of value. For example, some users may repost images of others' feet that they find attractive [28].

**Notice to reader: In the next paragraph and table, some readers might find hashtags related to the sex industry to be psychologically harmful, triggering or upsetting. In this case, the reader may skip to the end of Table 6.**

During this first call, I was also able to ask which hashtags were truly indicative of the sex industry. I saw many hashtags that were on posts in the sex industry, but it was unclear to me which hashtags were indicative of a sex industry post (such as #findom), and which had much broader meanings (such as #paypal). From a list of over 80 hashtags, I asked Guevara which hashtags were indicative that the post was in the sex industry. She labeled each hashtag as either

- 'Yes' - the hashtag is indicative of the sex industry,
- 'Maybe' – the hashtag may be indicative of the sex industry, but is not always indicative of the sex industry,
- 'No' – the hashtag is not indicative of the sex industry, and
- 'Unknown' – Guevara is not sure whether this hashtag is indicative of the sex industry or not.

Table 6 shows the labels of the hashtags. Please note that if this information from this thesis becomes generally known to social media companies, they may block these hashtags, and then different hashtags may take their place.

*Table 6 - Classification of hashtags in regards to the sex industry [28].*

| Yes – Hashtags indicative of Sex Industry | Maybe – Hashtags may be indicative of the sex industry, but not always | No – Hashtags not indicative of the sex industry | Unknown – Guevara is unsure whether these are indicative of the sex industry |
|---|---|---|---|
| #bunnyranch<br>#cashslave<br>#escortmassage<br>#financialdominatrix<br>#financialdominatrix<br>#financialslavery<br>#financialsub<br>#findom<br>#findomaus<br>#findomfetish<br>#findomgoddess<br>#findomlatina<br>#findomme<br>#findommegoddess<br>#findomslave<br>#findomsub<br>#finsub<br>#flymetoyou<br>#girlfriendexperience<br>#humanatm<br>#humanatmwanted<br>#luxurycompanion<br>#moneydom<br>#moneyslave<br>#nudesforsale<br>#nurumassage<br>#payforplay<br>#paypig<br>#paypiggies<br>#paypiggypay<br>#paypigs<br>#paypigswanted | #cashcow<br>#dominatrix<br>#dommelife<br>#dømme<br>#elitecompanion *<br>#feet<br>#feetfetishnation<br>#feetporn<br>#feetworship<br>#femdomm<br>#footfetishcommunity<br>#footfetishgroup<br>#footfetishnation<br>#footmodel<br>#footworshipping<br>#humiliatrix<br>#prettyfeet<br>#spoiled<br>#spoiledlife<br>#spoilher<br>#spoilme<br><br>* #elitecompanion was classified as: 'Yes' or 'Maybe' by Guevara | #cashapp<br>#fuckyoupayme<br>#paypal | #cashfag<br>#cashmeet<br>#cashqueen<br>#walletdrain<br>#walletdraining<br>#walletrape<br>#walletrinse |
| #paypigswelcomed<br>#paypigwanted<br>#payslave<br>#sellingnudes<br>#sexworkiswork<br>#sugarbabie<br>#sugarbabieswantedcollege<br>#sugarbabycollege<br>#sugarbabyneededuk<br>#sugarbabyus<br>#sugarbabywanted<br>#sugardad<br>#sugardaddieswanted<br>#sugardaddy<br>#sugardaddyneeded<br>#sugardaddyneededasap<br>#sugardaddyseason<br>#sugardaddywanted<br>#sugardating<br>#sugardoll<br>#sugarlover<br>#sugarmomma<br>#vipmaidens<br>#egirl | | | |

**End of paragraph and table including potentially psychologically harmful, triggering or upsetting hashtags.**

Knowing which hashtags are indicative of the sex industry, is very helpful for labeling and can clarify whether a post is actually within the sex industry or not. There are also other considerations when labeling a post as part of the sex industry or not. During a second call with Guevara, I wanted to ask about different scenarios. Guevara explained what she would label those scenarios to help give me a guiding rubric while labeling the posts. For each scenario, I asked Guevara what should would label the post.

The rubric rules are listed below. Reminder: ‘yes’ label means in the sex industry, ‘unclear’ is unclear whether it’s in the sex industry, and ‘no’ means that the post is not in the sex industry.

**Rubric Rule 1: Sex Industry Hashtags**

If the original poster uses a hashtag in the ‘Yes/Sex Industry’ Category in their post, the post is labeled **yes** [29].

**Rubric Rule 2: Feet Fetish Posts**

- Case 1: Feet fetish posts and/or profiles where the original poster explicitly state they are selling photos or videos, offering online or in person meetups, or currently financially supporting someone are labeled as **yes.**
- Case 2: Feet fetish profiles where someone collects posts of women and reposts them are labeled as **unclear**.
- Case 3: Feet fetish posts and/or profiles that do not explicitly state anything beyond feet fetish hashtags are labeled as **unclear**.
- Case 4: Feet fetish posts that explicitly state they are not interested in financial transactions, sugar babies, etc. are labeled as **no** [29].

**Rubric Rule 3: Posts where the user profile or other posts indicates that the user is in the sex industry, but the user does not clearly indicate on the post itself whether or not it is in the sex industry.**

**When we are considering the user’s profile in addition to the post:**

- Case 1: When the original poster doesn’t explicitly use sex industry hashtags, **but their profile page/info** makes it clear that they are in the sex industry, and other users **do** make sex industry comments, then this post is labeled as **yes**.
- Case 2: When the original poster doesn’t explicitly use sex industry hashtags, **but profile page/info** makes it clear, but other users **do not** make sex industry comments, then this is labeled as **yes**.
- Case 3: When the original poster doesn’t explicitly use sex industry hashtags, but their **other posts** are clearly in sex industry and other users **do** make sex industry comments, then this post is labeled as **yes.**

- Case 4: When the original poster doesn't explicitly use sex industry hashtags, but their **other posts** are clearly in sex industry and other users **do not** make sex industry comments, then this is labeled as **yes** [29].

**On the contrary, if the research's goal is to identify solely the post,** then all of those cases would be labeled as either no or unclear, because one can't be 100% sure that that post itself is part of the sex industry [29]. **(See the description for these two labeling options beneath this rubric.)**

**Rubric Rule 4: Posts where the original poster gives no indication of sex industry in their post(s) or profile, but other users make sex industry comments on their post.**

These posts are labeled as **no** [29].

**Rubric Rule 5: Foreign Language Posts:**

The posts in the dataset were chosen because they had specific hashtags. The hashtags that were chosen only included ASCII characters, and the majority, if not all of the hashtags, were in English. However, there are still some posts where the bulk of the comments are not in English.

- Case 1: Posts written in a foreign language that are clearly non-sexual content are labeled as **no.**
- Case 2: Posts written in a foreign language that include sexual content, must be done on a case by case analysis. They may be labeled as **unclear.**
- Case 3: Posts written in a foreign language that use hashtags used from the Hashtags indicative of Sex Industry List are labeled as **yes** [29].

In rubric rule 3, Guevara explained that depending on the use case scenario of the research, there are two ways to label the posts. In one case the user's profile and other posts can affect whether the post is labeled as part of the sex industry. In this case, the research is related to whether or not someone is at risk for being exploited or whether an individual is interested in buying products from the sex industry. In this case, it is relevant whether the user is posting in the sex industry in general. Sometimes this can be helpful because Instagram may not allow all of the comments to be shown publicly. This means that it may not be possible to tell if the original poster included hashtags indicative of the sex industry because not all comments are visible. For example, if the original poster's profile indicates that they are in the sex industry, and the post being labeled had users commenting about allowances for a sugar baby, then in this scenario it may make sense to label the post as part of the sex industry. On the other hand, if the goal of the research is to solely identify posts in the sex industry and it separates the user profiles from the post itself, then the scenario described above may have the post labeled as a no or as an unclear. If this research was being used in a way to 'punish' people for being part of the sex industry, such as if Instagram took down people's posts, then it would be very important for this research to look at the post only. In this case, false positives would be worse than false negatives, since people would be unfairly punished. If this research was being used in a way to provide communities or individuals with more access to resources, then it would be better to label these posts as 'yes' since false positives would mean a community may get more resources or put up advertise local resources more effectively. In this case, false positives are better than false negatives, because false negatives may result in a lack of resources within a community.

I think it is important to note that between the two calls, Guevara's decision changed on Rubric Rule 4. Initially, Guevara labeled these posts as unclear, but later changed her mind to label them as no. Since the first call, she had realized that there are some individuals who will post asking for a sugar baby on posts that have photos of beautiful people, even if the original poster has no intention of participating in the sex industry. Therefore, she felt it was not accurate to label those posts as unclear, but instead to label them as no. I say this because by following the rubric above, there will be a number of posts I would label as no or yes instead of unclear that Guevara would have labeled as unclear when she first labeled the posts. Therefore, we might expect a higher level of posts where Guevara's labels and my second labels disagree because Guevara's label is unclear and mine is yes or no.

### *9.9. Updated IRR for Labels Identifying Solely the Post*

When I relabeled the posts the second time, I created two sets of labels because of Guevara's comment for Rubric Rule #3. The first IRR I will compare is the label for labeling solely the post. This is specifically for Rubric Rule #3. I expect these labels to be less accurate than the IRR for the labels which consider the profile and other post(s) by the user. This is expected because Guevara stated that since the intended use cases of this research are to enable more resources to be provided, then it would be better to have false positives than false negatives. This is equivalent to saying that it is better to provide resources in situations where someone might not need them than to not provide resources in a situation where someone might need those resources. Therefore, I believe that when Guevara labeled the posts, she took into consideration the profile information. When I labeled the posts using the post only for Rubric Rule 3, I did indeed increase my accuracy. Specifically, the SIH Sample and All SIH Posts increased in accuracy from 67.6% to 70.9% and 65% to 72.5%, respectively. And, as mentioned earlier, the majority of incorrect cases fall within the 'Partially Incorrect' category. This is likely because when labeling Posts Solely, I labeled several posts unclear that Guevara may have labeled as yes because of Rubric Rule #3. These IRR Ratings can be seen in Table 7.

*Table 7 - IRR After Feedback Calls with Expert. Posts were labeled only taking into consideration the post in regards to Rubric Rule #3.*

| IRR After Calls with Guevara, for Labeling Post Solely | Random Sample (Included 76 posts) | SIH Sample (55 posts) | Portion of All Posts that were Control Posts (52 Posts) | Portion of All Posts that were SIH Posts (80 Posts) |
|---|---|---|---|---|
| Percent Correct / Same labels (%) | 90.8% | 70.9% | 98.1% | 72.5% |
| Percent Incorrect / Different labels (%) | 9.2% | 29.1% | 1.9% | 27.5% |
| Subpart of Incorrect Labels that are Completely Incorrect (yes vs no) (%) | 1.3% | 7.3% | 0% | 6.2% |
| Subpart of Incorrect Labels that are Partially Incorrect (unclear vs no, or unclear vs yes). | 7.9% | 21.8% | 1.9% | 21.2% |

## *9.10 Updated IRR for Labels Identifying Post Using Profile or User's Other Post(s)*

The second set of labels that I applied to the posts when relabeling used not only the post, but the profile and user's other posts as well. This distinction is specifically for Rubric Rule #3. As expected, the IRR between my labels and Guevara's labels increased. The Portion of All Posts that were Control Posts had an IRR of 100%. The SIH Sample had an IRR of 80%, and the Portion of All Posts that were SIH Posts had an IRR of 80% as well.

The posts that were Completely Incorrect (yes vs no) were only 3.6% for the SIH Sample and 5.0% for Portion of All Posts that were SIH Posts. These IRR results can be seen in Table 8.

*Table 8 – Updated IRR Ratings after Feedback Call with Expert Guevara. Posts were labeled including information from the profile and other posts made by the user in regards to Rubric Rule #3. *See paragraph below to explain why this number may high.*

| IRR After Calls with Guevara, for Post or Profile | Random Sample (Included 76 posts) | SIH Sample (55 posts) | Portion of All Posts that were Control Posts (52 Posts) | Portion of All Posts that were SIH Posts (80 Posts) |
|---|---|---|---|---|
| Percent Correct / Same labels (%) | 93.4% | 80.0% | 100% | 80% |
| Percent Incorrect / Different labels (%) | 6.6% | 20.0% | 0% | 20% |
| Subpart of Incorrect Labels that are Completely Incorrect (yes vs no) (%) | 2.6% | 3.6% | 0% | 5.0% |
| Subpart of Incorrect Labels that are Partially Incorrect (unclear vs no, or unclear vs yes). | 3.9% | 16.4%* | 0% | 15.0%* |

*The posts that are Partially Incorrect (unclear vs no, or unclear vs yes) is much higher than the Completely Incorrect Labels (yes vs. no). I suspect this is because Guevara told me during the second feedback call that in retrospect, she would have labeled less posts as unclear and instead labeled them as no. This is in regards to Rubric Rule #4. This may account for the relatively

large percentage of partially incorrect posts that I labeled as no and Guevara may have labeled as unclear, but in retrospect she would have labeled as no.

The overall IRR on the Random Sample is 93.4%. This is much better than the initial IRR of the Random Sample, which was 83.3%. The IRR of the Updated Portion of All Posts that were SIH Posts is 80%. This is much better than the initial IRR of All Posts that were SIH Posts, which was 65%.

### *9.11 Relabel Manually Labeled Data*

Ultimately, the anticipated use cases of this research are to identify where, when, and how the sex industry is being used on social media and to use this aggregate data to increase availability of resources to individuals who may be experiencing sex trafficking. Therefore, I relabeled all the 201 posts labeled by Guevara. Of the original 500 posts that I had labeled, I only kept the labels from the Controls Posts. This is because the IRR for the Control Posts was consistently at or near 100%. When relabeling, I specifically choose to label the posts taking into account the user's profile and user's other post(s) in regards to Rubric Rule #3. I did this because of several reasons. First, if the use case of this thesis is to use the aggregate data analysis to increase resource availability, then I believe it is better to have accidental false positives as opposed to accidental false negatives in the labels. Second, I believe that in many of the posts, the individuals have used their other posts or profile to indicate that they are in the sex industry. Therefore, many other users who follow their posts already know that. As a result, the original poster may not need to use hashtags that explicitly indicate the post is within the sex industry in order to alert other users to that knowledge. I recognize that this may cause some challenges for the semi-supervised learning machine algorithm. If it does not perform well, then future researchers may want to have a database that includes the user's profile and a sampling of their other posts when analyzing the data. Unfortunately, our dataset does not have that additional information about the profile and profile's other posts, so we cannot include that in our analysis.

# Chapter 10: Data Cleaning

In order to do applied NLP research, one must clean the data. This is especially important in this thesis because Instagram posts do not follow the same rules that other types of text, like books or newspapers, follow. In addition, I wanted to remove or alter information that might be linked specifically to an individual, like their phone number or username. Therefore, I cleaned the data in several ways before applying the semi-supervised learning.

## *10.1 Remove Usernames*

First, I remove usernames that start with an @. In replies on Instagram, the person posting the reply automatically has an @username for the individual they are replying to. These usernames do not give any meaning to the text. In addition, they are personal private information that should not be associated as either in the sex industry or not in the sex industry. They do not generalize well to new posts that may be seen, and all in all, must be remove. Therefore, I identify usernames as word that starts with @, and remove all such usernames.

## *10.2 Clean Emojis*

After removing usernames, I edit emojis to be words representing what the emoji is. The emojis in the posts are originally represented as Unicode-Escape, and these need to be converted to something with character level meaning that represents something in the model. For example, \ud83e\udd40 represents the wilted_flower emoji that looks like a rose drooping. The text \ud83e\udd40 at a character level does not mean anything in the English language, whereas wilted_flower would. Wilt is related to wilted, and wilted makes sense alongside flower in the English language. By converting emojis to English word descriptors, this can help the FastText models have a better understanding of the meaning and context of these emojis.

## *10.3 Clean Apostrophes*

Apostrophes are represented in many ways in the original post data. These include ’ and ' and & x27 and '. Apostrophes are important in this context because they can indicate a contraction. Contractions are expanded at a later stage. However, it is not practical for these apostrophes to have so many representations, so all apostrophes are changed to be represented by '.

### 10.4 Remove Accents

Most of the posts in the database are in English, but some posts may include other languages with accents or English posts with accents in words such as café. In other languages, accents do indeed serve the purpose of differentiating different words. And with the accent, a word could have a completely different meaning. However, in social media, users may or may not use accents, even if the language they are writing in requires accents. Since most of the posts are in English, and those that are not in English are not guaranteed to use the accents correctly, accents are removed from the posts. They are replaced with their most similar non-accent correspondent. For example, é would become e, and ñ would become n.

### 10.5 Remove Punctuation

Many forms of punctuation do not add to the sentiment or meaning of the text. For example, #dog, “dog?” or “dog!” does not have a significantly different meaning from dog. The punctuation that is removed from the text is ! and : and ; and = and ? and . and \ and #. Punctuation including – and + are useful in identifying phone numbers and $ is useful for identifying a potential exchange of money, and these do have conceptual meaning in these posts. For example, people trying to connect with someone providing services in the sex industry will often provide their WhatsApp number or they indicate the amount of money they are willing to pay on a weekly or monthly basis. For this reason, numbers are kept as well since they contribute to this conceptual meaning. However, all phone numbers are edited for the privacy of the data, which is discussed later.

### 10.6 Make Text All Lower Case

Upper- and lower-case letters generally do not relate to meaning of a word, but an uppercase character is distinguished from its lowercase counterpart. Dog and dog would be separate words. Sometimes, this does correspond to a different meaning because of proper pronouns. For example, Daisy is a name and daisy is a flower. Apple is a company and apple is a fruit. However, for the vast majority of upper- and lower-case letters in these posts, the upper- or lower-case letter does not represent a conceptual difference. Distinguishing between the two would be counterproductive.

### *10.7 Remove Duplicate Letters*

Sometimes in social media posts, people use duplicate letters to show emphasis. People might saw “awwwwwwwwwww” or “I loooooooooove it!” These words would not have any significant change in a conceptual meaning if they had been spelled correctly “aww” or “I love it!”. This means that “awwwwww”, “awwww”, and “aww” would all be changed to the same word “aww” instead of being three separate words. I do this naively so that any set of 3 or more of the same letters are changed to be just two of those letters. English words, in general, don’t have more than two letters duplicated. In our examples above, this would become “aww” and “I loove it!”. While there are some downsides to this, since love and loove would be separate words, it still reduces the number of redundant letters. It would also be impractical to scan every word with 3 or more letters in a row to see if the correct spelling of that word should have two letters or one letter. This is because people often do not spell things correctly on social media, and so it would be impractical to check. Note that I do not implement a spell check on these posts because that would be very challenging given the number of hashtags, abbreviations, slang or other non-dictionary ‘words’ that exist in these posts.

### *10.8 Remove Post Username*

Instagram includes the username of the individual who posted the Instagram post in the post’s caption. Again, usernames do not provide conceptual context to the meaning of the post, and we do not want any algorithms to associate certain usernames with the sex industry. Therefore, we remove the section of the post caption that includes who posted the Instagram post.

### *10.9 Expand Contractions*

Contractions such as don’t or I’m have expanded equivalents. Do not and I am in this case. While it may not be necessary to expand contractions, it does not make sense to waste the model’s resources on contractions that could be expanded. Then instead of the model identifying both don’t and do not, it only needs to model do not. Since expanding contractions in the English language is a nontrivial task, I used Gensim’s glove-twitter-25 model to expand the contractions.

This model was built using 2 billion tweets and has 1.2 million words in its vocabulary [30]. I use the 25-dimensional model for expanding the contractions. I need to make the text lowercase again after doing this since some contractions include upper case words, like I.

### *10.10 Remove Newline*

Ultimately removing the newline affects whether or not the entire comment or reply should be considered 'one sentence'. I felt that simply by looking through the posts and manually labeling them, most times an entire reply or comment stayed within the same context. While I wasn't entirely sure what the best option was, I ultimately felt that it would be more accurate to keep the entire reply or comment together as one 'sentence' for training because the meaning often didn't change from one line to the next.

### *10.11 Alter Phone Numbers to a Fake Phone Number*

People who wanted to buy services in the sex industry sometimes post their phone number or WhatsApp number on the publicly available post. Therefore, I believe that a phone number does have a conceptual meaning for some of these posts. However, we clearly do not want someone's phone number associated with the sex industry. Therefore, phone numbers are detected and changed to 555-555-0123. I change it to this because this is a fictitious phone number that does not belong to anyone. (Note that all phone numbers XXX-555-0100 to XXX-555-0199 are fictitious.

# Chapter 11: Semi-Supervised Learning

## *11.1 Different Ways to Vectorize Text: Bag-Of Words, Word2Vec, FastText, and Paragraph Vector (Doc2Vec)*

Once the dataset was fully cleaned, then I was able to use different Natural Language Processing (NLP) tools to convert the text from the posts into mathematical vector representations. This allows for semi-supervised machine learning to be done. There are various NLP algorithms that allow for text to be converted to a numerical vector. The four methods that I considered were Word2Vec, FastText, Bag-Of-Words (BOW), and Doc2Vec. Each of these methods have their own pros and cons, and ultimately, I choose to look at FastText and Doc2Vec for the reasons discussed within the next several paragraphs.

The BOW model is a one-hot-encoding model. Each document is a long vector, and each element within the vector represents the number of times a word occurs within a document. The length of the vector is the length of all words trained in the model. Unfortunately, the ordering of the words and the semantics are lost [31]. For example, consider the model trained upon only the sentence "Jose likes chocolate and pasta, and Annabelle dislikes pizza and chocolate. The vector representing the text may be [1, 1, 2, 3, 1, 1, 1, 1] corresponding to the words ["Jose", "likes", "chocolate", "and", "pasta", "Annabelle", "dislikes", "pizza"]. From just this vector it's impossible to know who likes what, and who dislikes what. The context and location of words are not captured. In addition, BOW model vectors can be extremely long. Depending on the documents used to train, there can be thousands, millions, or billions of unique words in these documents. This can result in extremely long vectors for each document, many of which contain a majority of 0s, and this can result in longer computational time. Although sparse matrix operations can probably be used to decrease the computational time, other models have fixed length vectors with lengths between tens to hundreds instead of thousands to billions. In addition, these BOW models can't tell whether words are similar are not. For example, 'laptop' and 'computer' wouldn't be any more related to each other than 'laptop' and 'sky' [32]. And that is important in social media posts where the words and hashtags are always changing, but still representing similar concepts. However, BOW is a commonly used model, since it can do surprisingly well in certain circumstances. Because BOW can do surprisingly well in certain circumstances, we considered it for analysis in this project. Helen Lu, an Undergraduate Research Student, was interested in working on this project in Spring 2020. I provided general overarching guidance to Lu during the semester, and she implemented and assessed a BOW model for this dataset. Lu's work is discussed later in this chapter.

Word2Vec and FastText are models that takes into account contextual information unlike BOW. They are distributed word representations [33], [34]. Word2Vec and FastText are neural network embedding models that create a unique low-dimensional sized vector for each word. The model tries to predict a target word, $w_t$, given some range of the context words around it, $w_{t-k}, \dots, w_{t-1}$ and $w_{t+1}, \dots, w_{t+k}$ [31]. The model uses the context words to predict the target word. The model is trained in such a way that tries to maximize the average log probability of predicting the target word given the context words. In this equation, there are $T$ total words [31].

$$\frac{1}{T}\sum_{t=k}^{T-k} \log p(w_t | w_{t-k}, \dots, w_{t-1} \text{ and } w_{t+1}, \dots, w_{t+k})$$

According to Le (2014) prediction usually uses a multiclass classifier, like softmax [31]. The equation for the prediction in this case would be:

$$p(w_t | w_{t-k}, \dots, w_{t-1} \text{ and } w_{t+1}, \dots, w_{t+k}) = \frac{e^{y_{w_t}}}{\sum_i e^{y_i}}$$

Le (2014) explains that in this case, for each output word $i$, $y_i$ is an un-normalized log-probability using softmax [31].

Alternatively, one can think of Word2Vec and FastText as models that have a loss function that the neural network tries to minimize. That loss function can be thought of as

$$J = 1 - p(w_t |\, w_{t-k}, \dots, w_{t-1} \text{ and } w_{t+1}, \dots, w_{t+k})$$

where the model is trained to best predict the target word ($w_t$) given neighboring words $w_{t-k}, \dots, w_{t-1}$ and $w_{t+1}, \dots, w_{t+k}$ [32]. This architecture is called Continuous Bag-Of-Words (CBOW) where the surrounding context words predict the center target word [33], [34]. A diagram of this can be seen in Figure 5. The $k$ in $w_{t-k}, \dots, w_{t-1}$ and $w_{t+1}, \dots, w_{t+k}$ is equal to $k = 2$ in this diagram.

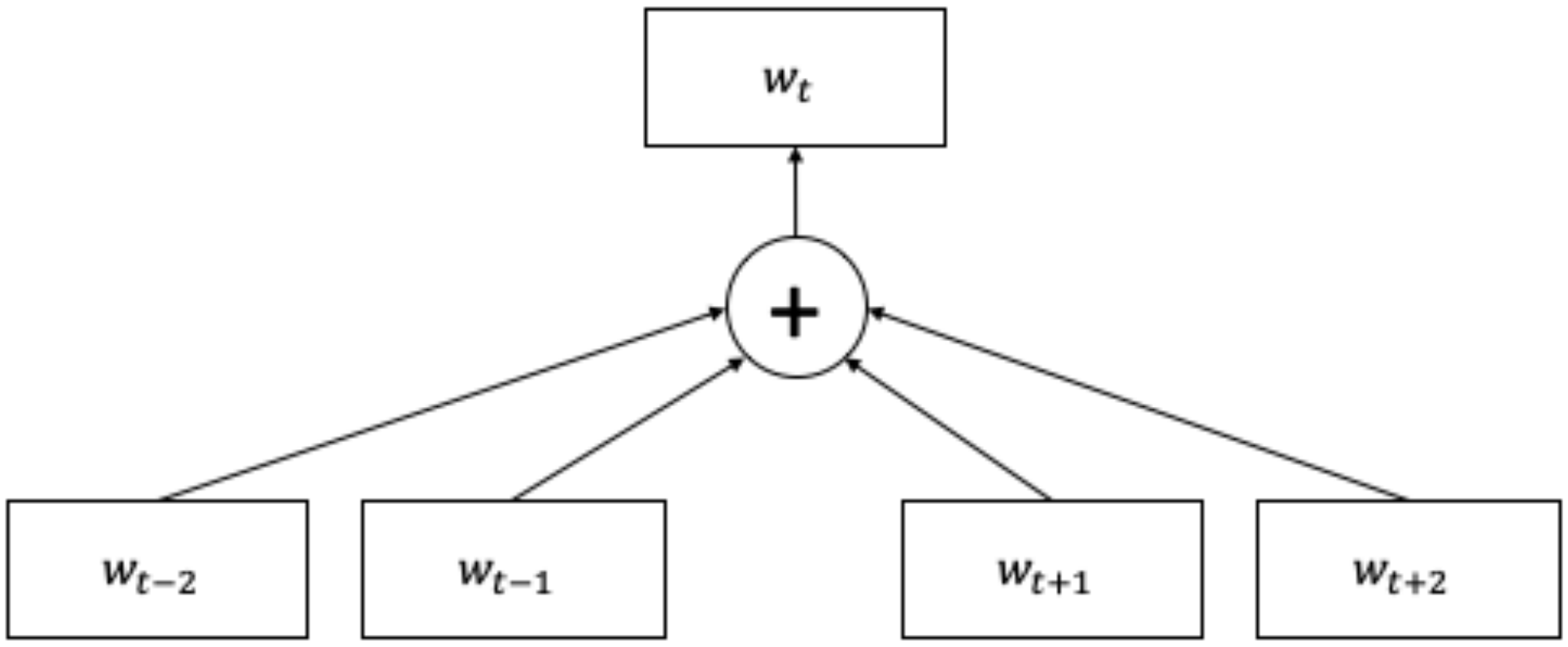

*Figure 5 - Diagram showing CBOW architecture where contextual words are used to predict the center target word [33].*

Alternatively, these models can use the center word to predict each surrounding context word one by one. In this case, the architecture is called Skip-gram, and is shown in Figure 6. Again, the $k$ in $w_{t-k}, \dots, w_{t-1}$ and $w_{t+1}, \dots, w_{t+k}$ is equal to $k = 2$ in this diagram.

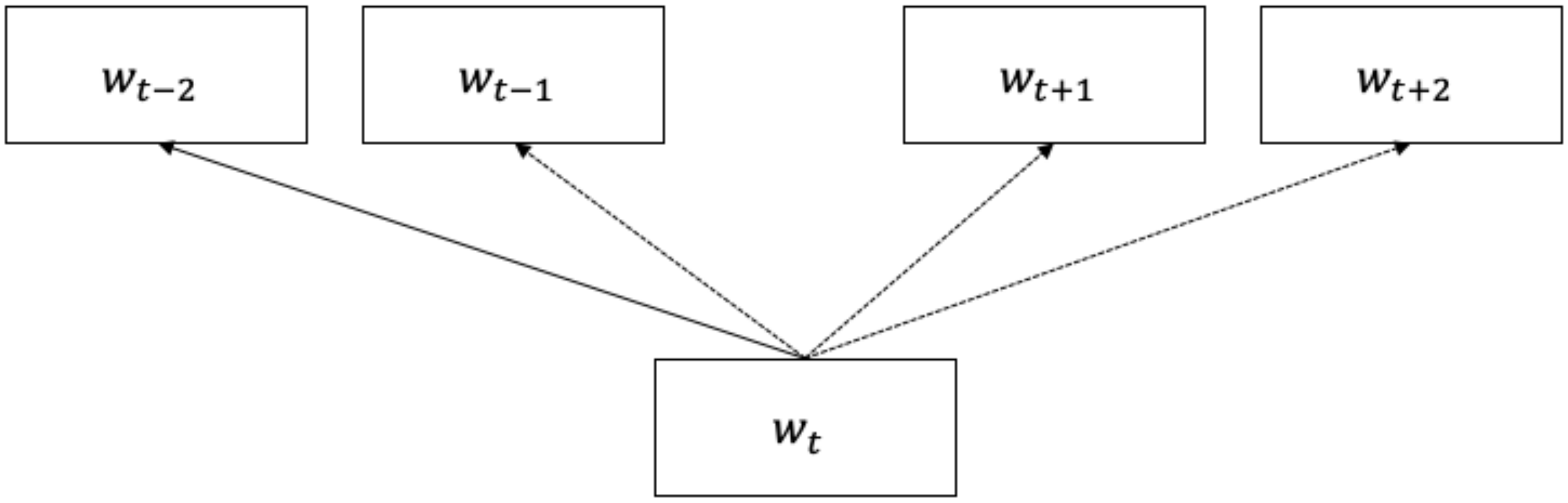

*Figure 6 – Diagram showing Skip-gram architecture where the center word predicts each surround context word one by one [33].*

The vectors for each word are varied throughout the neural network training as the loss function is minimized. These embeddings have become used in various situations, and Fréderic Godin applied them in the context of social media in his PhD thesis [33]. Fréderic Godin created Word2Vec and FastText embeddings for over 400 million Twitter tweets and compared the models. Had Godin used BOW he would have had vectors with a length of roughly 5 billion, which is clearly not practical. Therefore, FastText and Word2Vec in Godin's case were clearly better model choices for analysis.

In Godin's analysis, he mentions several important considerations that affect which model one would choose to use. One of the important differences between Word2Vec and FastText is that Word2Vec does not take into account character-level information. This would be extremely useful in the social media context. Word2Vec and FastText are often are set up to ignore words that occur below a certain threshold. For example, words occurring less than 5 times, may be ignored. Therefore, words that are misspelled, slang, or have different endings like -s, -ing, -ed, may not reach the minimum threshold for occurrences. This means that there is no vector in Word2Vec that corresponds to these words [33], [34]. This is especially important within the sex industry posts on social media. From my experience with the social media posts, the hashtags used to indicate a sex industry post have numerous small changes that represent essentially the same root word. In contrast to Word2Vec, FastText does take into account information at the character level. This means that if you had a post with words like 'runnin' and 'runnning', the FastText vectors that are generated by the model from character-level information will be very similar to the FastText model for the word 'running'. Depending on the number of occurrences of rare or misspelled words, this may or may not be a problem for Word2Vec. Since the dataset that I use is only roughly 50,000 posts, and there are many hashtags that have slight variants of each other, I chose to use FastText in my analysis instead of Word2Vec.

Paragraph Vector, implemented as Doc2Vec in Python's Gensim, is another distributed NLP model. It was introduced by Quoc Le and Tomas Mikolov in 2014 [31], [35]. Paragraph Vector is very similar to Word2Vec, but it outputs a vector for an entire document rather than a singular word. Paragraph Vector is trained upon multiple documents that each contain multiple words. Le and Mikolov refer to these documents as paragraphs in [31], but I will refer to them as documents since document better describes the posts used in the context of this thesis. During training, Paragraph Vector creates a vector for each word in the vocabulary in addition to creating a dense vector for each document that it is trained upon. During prediction, Paragraph Vector predicts the document vector instead of word vectors [31]. Figure 7 is useful for highlighting the differences between Word2Vec and Doc2Vec. In the Paragraph Vector Algorithm, there is an additional document token (a document id, $d_i$) for each document. As the target word, $w_t$, slides along a document, the contextual words ($w_{t-k}$ ... $w_{t+k}$) change, however the document id, $d_i$, remains the same for the entire document. In general, document ids different between different documents. Figure 7 is shown below.

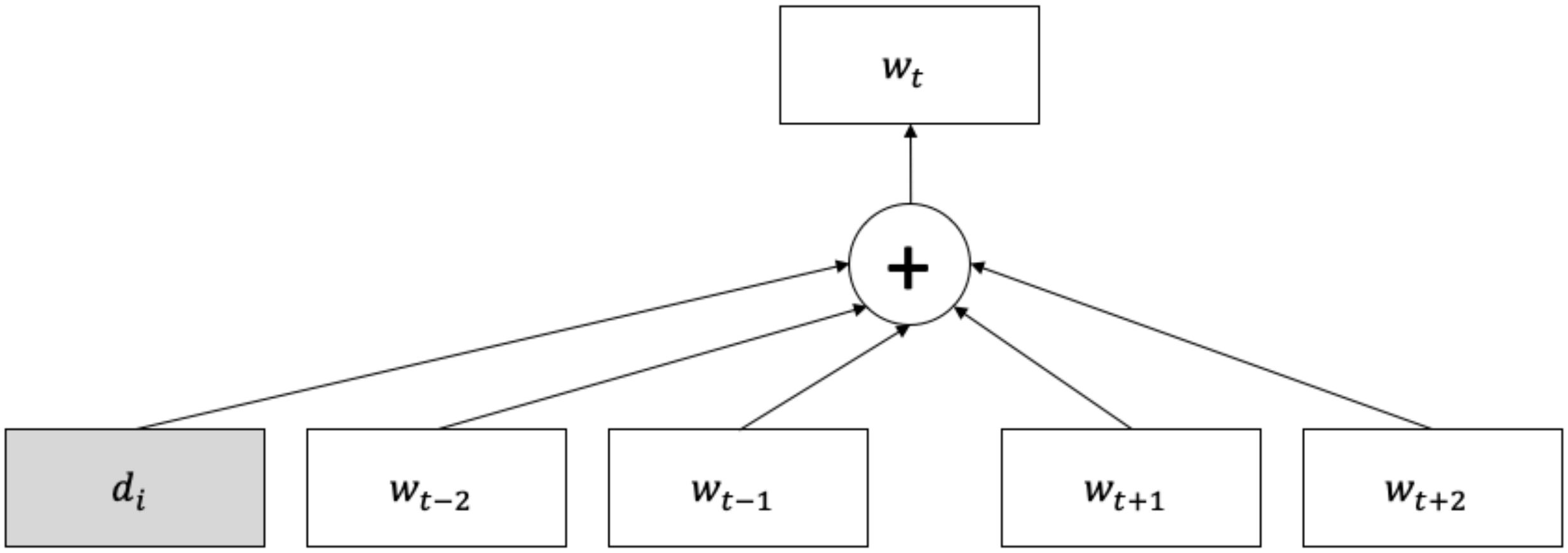

*Figure 7 - In the Paragraph Vector Algorithm, there is an additional document token (a document id, $d_i$) for each document.*

Le and Mikolov describe the document token as an extra 'word' that remains through the duration of the document. This gives a kind of memory that is shared among all word contexts within the document, but it is not shared between the different training documents [31]. Paragraph Vector has several advantages over BOW since it can be learned from unlabeled data and it takes into account word order. Paragraph Vector has the advantage over FastText and Word2Vec because it can give a vector to an entire document rather than giving vectors to single words [31]. Therefore, due to the advantages that Paragraph Vector has over Word2Vec, I also chose to use Paragraph Vector, implemented as Doc2Vec by Gensim, in my analysis.

## 11.2 Techniques Used to Cluster and Classify the Data

In this thesis, I use the silhouette score, k-means clustering, and cross-validation. I will briefly explain what these techniques are and why I chose to use them in this thesis.

The Silhouette score is used to classify how well data has been clustered. For a given point $x^{(i)}$, it takes into account two things. First it takes into account the average dissimilarity, $a(x^{(i)})$, between the point $x^{(i)}$ and the other points in its cluster. Second it takes into account the average dissimilarity, $b(x^{(i)})$, between the point $x^{(i)}$ and the cluster closest to it that it does not belong to. [36] Using these two dissimilarities, the Silhouette score, $S(x^{(i)})$, is calculated by:

$$S(x^{(i)}) = \frac{b(x^{(i)}) - a(x^{(i)})}{\max\left(a(x^{(i)}), b(x^{(i)})\right)}$$

The point is well clustered if $S(x^{(i)})$ is large (i.e. close to 1). The point is poorly clustered if $S(x^{(i)})$ is small (i.e. close to 0). And the point is in the wrong cluster if $S(x^{(i)})$ is negative [36]. To get the Silhouette score for the entire dataset that has been clustered, the Silhouette score is calculated for all points within the data and averaged. Note that the Silhouette score is increased when points of a cluster are close together. This is because $a(x^{(i)})$ would small, making $S(x^{(i)})$ large. The Silhouette score is decreased when two clusters are really close together. This is because $b(x^{(i)})$ would be small, making $S(x^{(i)})$ small. Thus, the Silhouette score can help determine the optimal number of clusters.

Consider the case where data is separated into very few clusters, or one large cluster. The Silhouette score makes $a(x^{(i)})$ large because there would be a lot of dissimilarity within that cluster, and the overall Silhouette score would be small. Thus, the silhouette score discourages making too few clusters. Now consider the case where the data is separated into many clusters, or every point is its own cluster. Here, the silhouette score is also decreased because $b(x^{(i)})$ will be small. It will be small because the points within one cluster will be very similar to points in neighboring clusters. Therefore, the Silhouette score is a good choice of metric to use when deciding how many clusters to have within the data because it balances having too few or too many clusters.

K-means clustering is used to cluster data into $K$ distinct clusters. The clusters are chosen based upon the minimization of a loss function. Specifically, the loss function is the within-group sum of squares (WGSS) [36]. The WGSS loss function is given below for $K$ clusters, where $x^{(i)}$ and $x^{(j)}$ are points within the data:

$$W(C) = \frac{1}{2}\sum_{k=1}^{K} \sum_{C(x^{(i)})=k} \sum_{C(x^{(j)})=k} d\left(x^{(i)}, x^{(j)}\right)^2$$

This equation is summing over the squared distance between every possible pair of points within a cluster, and then it is summed over all clusters. It checks how close items within the clusters are to each other [36]. $d(x^{(i)}, x^{(j)})$ is the distance equation between the two points $x^{(i)}$ and $x^{(j)}$ which are in cluster $k$ [36]. This exact solution to this loss function is not computationally feasible, therefore a greedy algorithm using random restarts can help to avoid local optima [36, pg. 18].

Lastly, cross-validation is used in this thesis to determine how accurately the labels are being applied [37]. Cross validation is a useful way to evaluate how well an algorithm is doing because it allows all the labeled data to be used. Normally, data is split into training and test sets. Cross validation allows for all the data to be used in situations where getting labeled data is costly either in terms of money or time [37]. The pseudocode, as described in [37], can be seen below.

1. Randomly separate the data, $D$, into divided into $k$ groups of equal size, $D_1, D_2, \dots, D_k$.
2. **For** j from 1 **to** k
3. Train the algorithm on all groups $D_1, D_2, \dots, D_k$ except $D_j$
4. Compute the "test error", $\varepsilon_j$ on the group $D_j$
5. **Return** the average error of all "test errors". $\frac{1}{k}\sum_{j=1}^{k} \varepsilon_j$

These techniques are used both in my analysis and in Lu's analysis.

### *11.3 Important Considerations of Unsupervised, Semi-Supervised and Supervised learning.*

**Notice to reader: In the two paragraphs, some readers might find hashtags related to the sex industry to be psychologically harmful, triggering or upsetting. In this case, the reader may skip to the paragraph right before 11.4.**

There are several important considerations that affect unsupervised, semi-supervised and supervised learning. It is important to note that in semi-supervised learning, the model will work well if most clusters within the data correspond well to the clusters that we are trying to label. In order for that to be the case, the vectors created by BOW (with dimensionality reduction), FastText, and Doc2Vec must be able to be clustered in ways that separate sex industry and non-sex industry posts. There are a few challenges that make this difficult given the dataset used in this thesis.

For one, the dataset consists of text from the post, but it does not indicate who made each comment/reply. We do not know whether it was the person posting or another user who made the comment/reply, which poses a challenge. When the expert and I labeled the data, we often used whether the original person posting or another user made the comment to label the data. For example, consider a post that only had one comment on it, and that comment was "I'm looking for a loyal sugarbaby, and I am willing to pay $1000 weekly #sugardaddy #sugarbaby". If the original poster made that comment, then the post would be classified as part of the sex industry because the original poster is clearly indicating that they would like to find someone to pay to have a sexual relationship with them. However, if the original poster did not post anything about the sex industry on this post or their profile, but another user posted that comment looking for a sugarbaby, then the post would be classified as not part of the sex industry. This kind of incident can happen when there is a post of an attractive individual, often a woman. Other users can then post looking for a sugarbaby, despite the fact that the original person posting may have had no intent of being in the sex industry. Therefore, depending on who created the comment, the post should be labeled differently even though it contains the exact same comment. Unfortunately, the dataset we currently have cannot be updated, so there is no way to improve the dataset in that way.

**End of paragraphs including potentially psychologically harmful, triggering or upsetting hashtags.**

In order to do Supervised learning however, we would need to have all the data labeled. It's not possible for us to label all the data, which contains over 50,000 posts. It takes roughly 1 hour to label 100 posts, meaning that to label all 50,000 posts would take roughly 500 hours of manual labor by individuals with specific knowledge. Given this information, it means that I expect some of the clusters presented within our data to contain both posts labeled as part of the sex industry and not part of the sex industry. The original hope was that only a few of these clusters would experience that problem, while other clusters would clearly fall within the sex industry or not sex industry categories.

### *11.4 Bag-Of-Words Model*

As mentioned earlier in this chapter, Helen Lu created a BOW model (with dimensionality reduction). This was done to determine how well a semi-supervised learning model with BOW would classify posts in this dataset. Although this is not part of the work that I did directly for my thesis, it is important work that is relevant to this project. Therefore, I briefly discuss it here. The BOW model vectors were long, as expected, so Lu did Principal Component Analysis (PCA) and T-distributed Stochastic Neighbor Embedding (t-SNE) to reduce the dimensionality of these vectors. PCA reduces dimensionality to a selected n-dimension while maximizing variance [38]. t-SNE is a non-linear dimensionality reduction that keeps points that are close to each other in the high-dimensionality data also close to each other in the low-dimensionality transformation [38]. Examples of the dimensionality reduction that Lu did can be seen in Figure 8 and Figure 9. In these figures, Lu plotted the 2 Component PCA of the posts and the 2 Component t-SNE of the 1,369 Component PCA. Lu also overlaid the original manual labels on the figures. Points labeled in red were labeled as not in the sex industry, and points labeled in green were labeled as in the sex industry.

Please note that during Spring 2020 when Lu worked on this research project, the manually labeled data had not yet been updated via the rubric rules that the expert Guevara gave. Nonetheless, this gives a good idea on whether or not the clusters can be separated into groups generally consisting of posts either in or not in the sex industry.

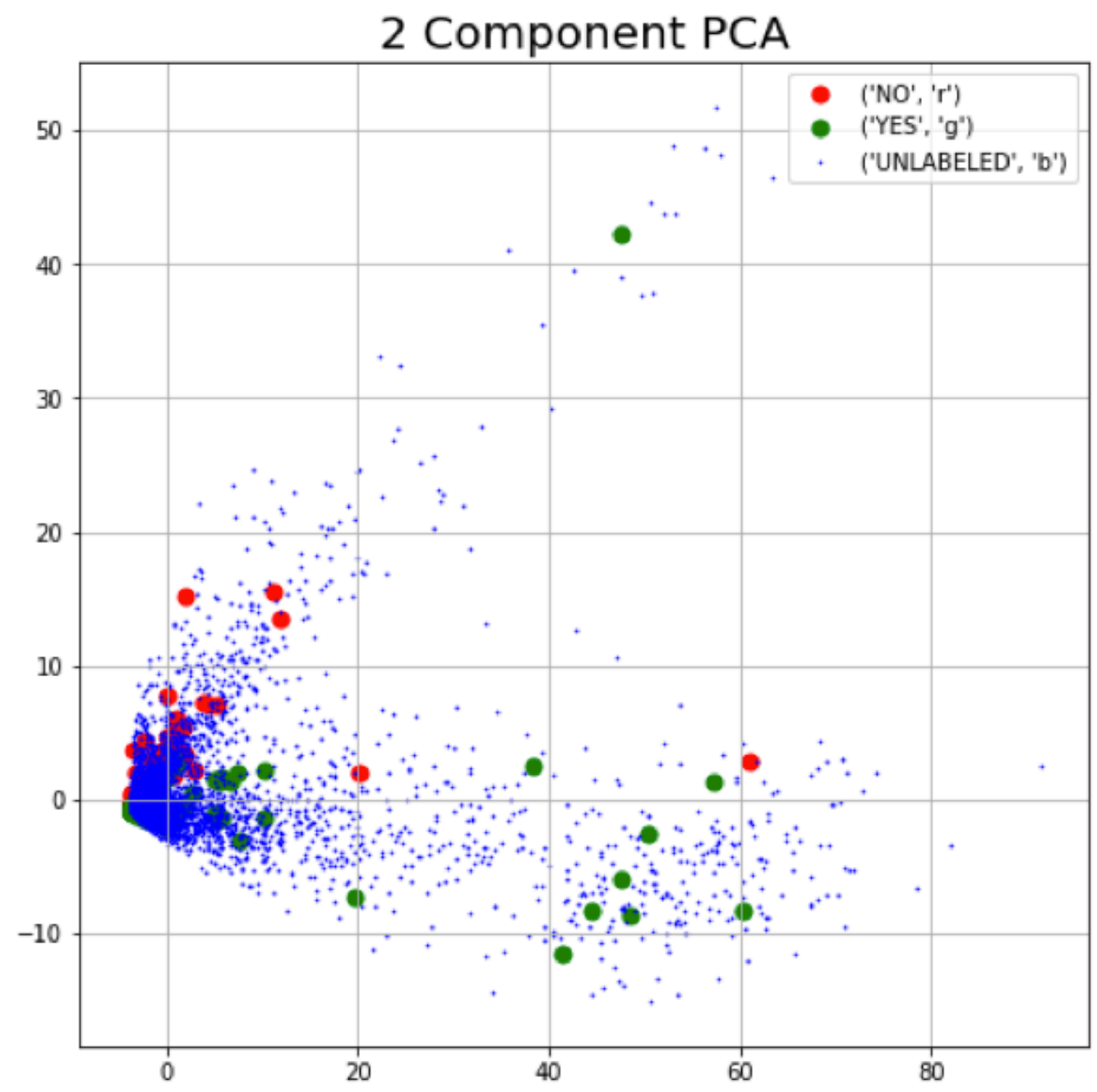

*Figure 8 - 2 Component PCA with Bag-Of-Words. Manually Labeled Posts are shown in Red for Not-Sex Industry Posts, and Green for Sex-Industry Posts. The posts labeled in the Sex Industry appear to generally go horizontally across the bottom of the figure, while posts that are not in the Sex Industry appear to generally trend diagonally upwards. Axes are first and second principal components of the data.*

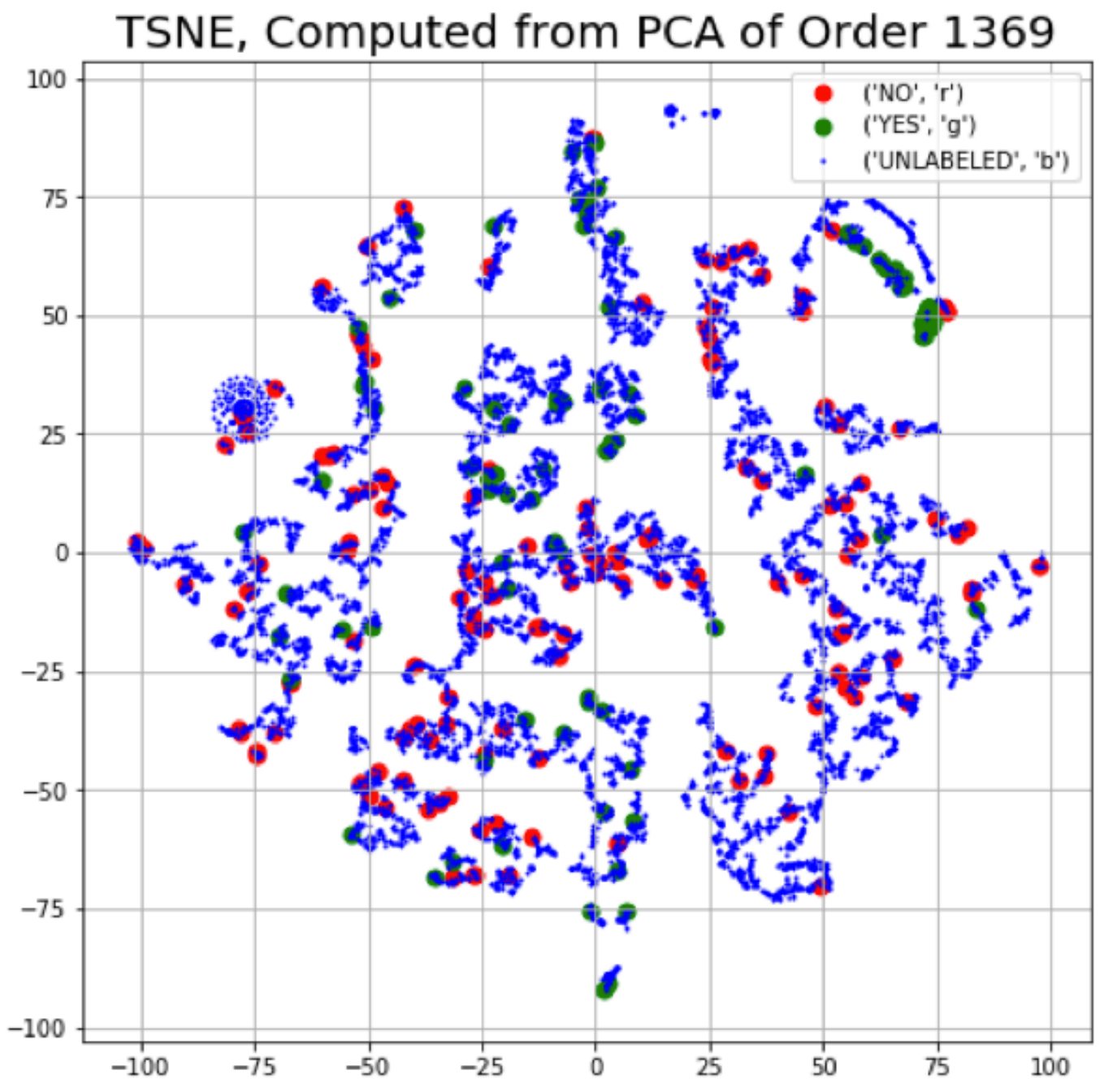

*Figure 9 – Plotted above is the result of first taking the 1,369 Component PCA of the BOW data points and then applying a 2 Component t-SNE to further reduce the dimensionality. Manually Labeled Posts are shown in Red for Not-Sex Industry Posts, and Green for Sex-Industry Posts. Here many clusters contain only not Sex Industry posts. Other clusters contain both, and a few clusters contain only Sex Industry Posts. Axes are first and second principal components of the data.*

From these diagrams, the clusters that occur do seem to correspond roughly to the manual labels at the time. The 2 Component PCA does suggest that the ‘yes’ and ‘no’ sex industry posts can be well spread out. However, there are many posts within the center that could be either yes or no. In the 2 Component T-SNE of the 1,369 Component PCA, the data is better separated into different clusters. Some clusters appear to contain only ‘no’ posts, others appear to contain only ‘yes’ posts, and others contain both. In order to determine how well these clusters can be used to label the unlabeled posts, Lu utilized several clustering and machine learning techniques including the Silhouette score, K-means, and cross-validation. Lu used the silhouette score and found for the 2-Component t-SNE of the 1,433-Component PCA that a cluster of about size 50 worked well. Lu used 50 clusters to cluster the data with K-means clustering. The next step in the semi-supervised learning was to propagate the labels within clusters. When Lu implemented the label propagation, she labeled the cluster with the label that occurred most frequently within that cluster. Then using cross-validation, Lu checked her accuracy and achieved an accuracy rate of 91.2% when applying labels to the clusters.

# Chapter 12: Results

## *12.1 FastText Skipgram, FastText CBOW, and Paragraph Vector (Doc2Vec) Models Semantically Displayed*

As discussed earlier in this chapter, I choose to use the FastText Skipgram, FastText CBOW, and Paragraph Vector (Doc2Vec) models to create word and document embeddings with the dataset. For the parameters for the FastText CBOW model and FastText Skipgram model, I choose a word embedding vector size of 100, a word context window of 5, a minimum word count of 5, and 5 iterations. The minimum word count of 5 means that words that occur less than 5 times will be ignored during training. For the Doc2Vec parameters, I used a vector size of 100, a minimum word count of 5, and 20 epochs.

In order to check to make sure that the models are classifying the words in an appropriate way, I predict the word vector given by each model for 52 words. This generates 52 word embeddings with a size of 100. In order to display these vectors on a 2-dimesional graph, I use t-distributed Stochastic Neighbor Embedding (t-SNE) to reduce these to a 2-dimensional graph. As I mentioned earlier, t-SNE is a non-linear dimensionality reduction that keeps points that are close to each other in the high-dimensionality data also close to each other in the low-dimensionality transformation [38].

**Notice to reader: In the next several pages, some readers might find hashtags related to the sex industry to be psychologically harmful, triggering or upsetting. In this case, the reader may skip to 12.2.**

Figure 10, shown below, contains the t-SNE Dimensionality Reduction of the Word Embeddings from the FastText Skipgram model.

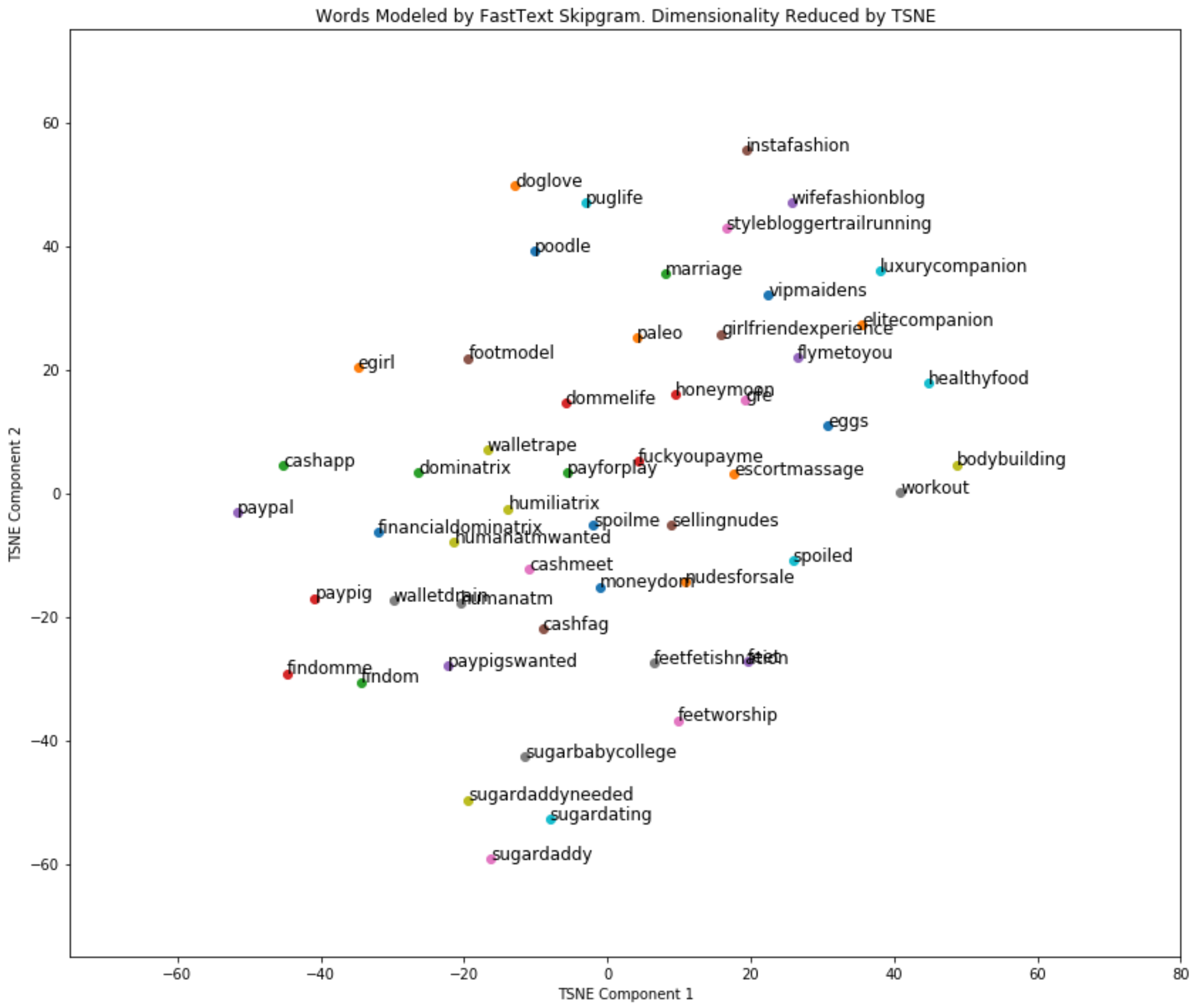

*Figure 10 - T-SNE Dimensionality Reduction of Word Embeddings from FastText Skipgram Model.*

Here we can see that many words with similar connotations are indeed close to each other. For example, ‘doglove’, ‘puglife’, and ‘poodle’ are close to each other. ‘instafashion’, ‘wifefashionblog’, and ‘stylebloggertrailrunning’ are close to each other. ‘healthyfood, ‘bodybuilding’, ‘workout’, and ‘eggs’ are close to each other. For hashtags used within the sex industry, hashtags often seen together on a comment are also close to each other on this diagram. Individuals who offer more expensive services in the sex industry use hashtags like ‘elitecompanion’, ‘flymetoyou’, ‘vipmaidens’, and ‘luxurycompanion’ and these are seen close to each other. Comments where someone is looking for a sugarbaby or sugardady use hashtags like ‘sugardaddy’, ‘sugardating’, ‘sugardaddyneeded’, and ‘sugarbabycollege’, which can all be seen next to each other. In the sex industry, and outside of the sex industry, apps used to transfer money like ‘paypal’ and ‘cashapp’ are used in similar contexts, and again, we can see those two words close to each other below. In addition, words like “paypig”, “findom”, “paypigswanted”, and “findomme” often occur in the same comments, and they are also seen close together. On the other hand, some words are near each other that are not expected. For example, “marriage” and “honeymoon” are close on the graph, but “paleo” and “girlfriendexperience” are located in between them, which is not expected. In addition, comments related to feet fetishes have hashtags like “feetfetishnation”, “feet”, and “feetworship”, can be seen together, but “footmodel” is far away. On the whole, this diagram suggests that the FastText Skipgram model is indeed creating appropriate vector embeddings. As a reminder, the Skipgram model uses the center word to predict the words around it [33].

Figure 11, below, shows the t-SNE Dimensionality Reduction of the Word Embeddings from the FastText CBOW model.

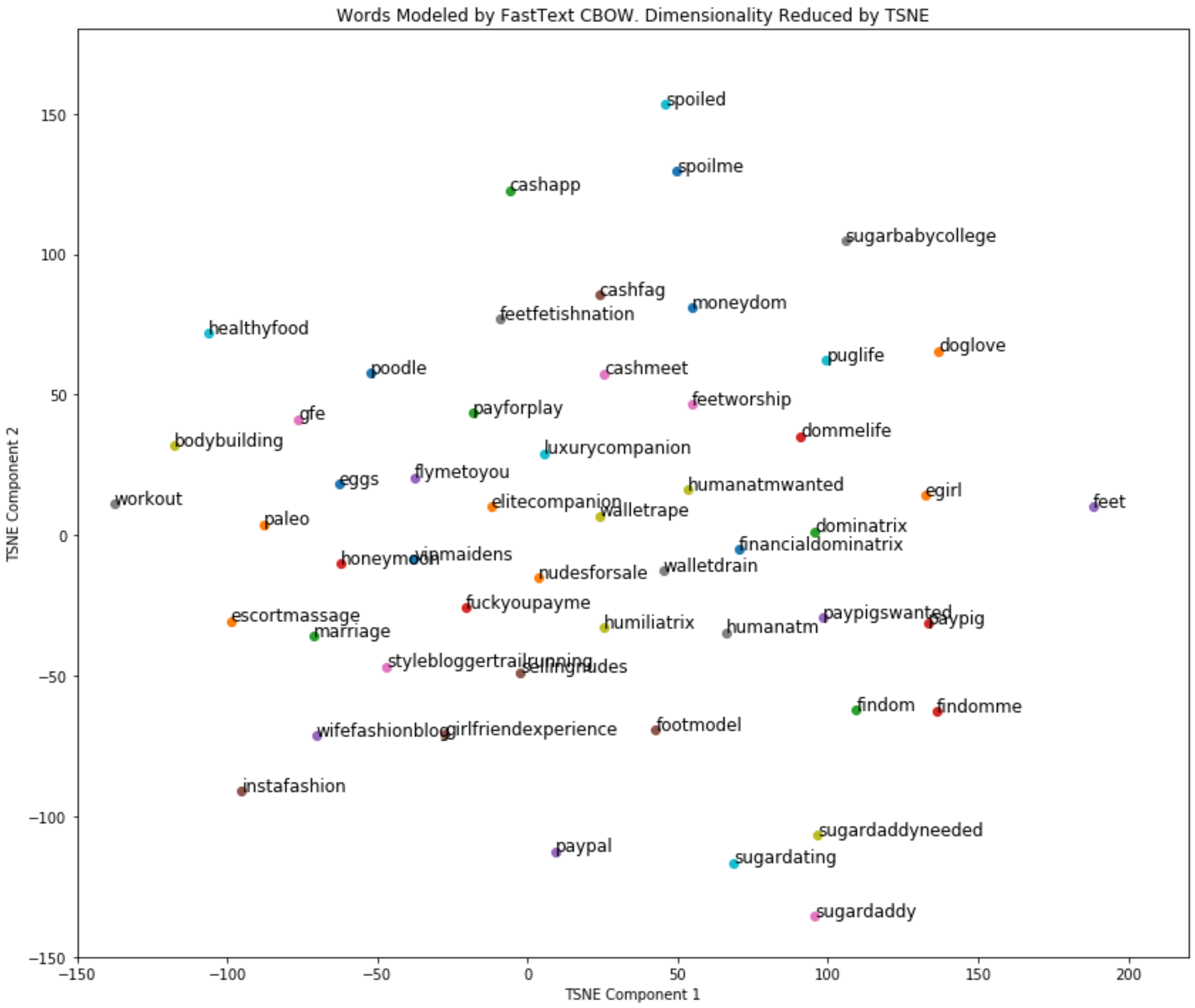

*Figure 11 - t-SNE Dimensionality Reduction of Word Embeddings from FastText CBOW Model.*

Here, the model did not do as good of a job putting words close together that are typically seen together. For example, “puglife” and “doglove” are close together, but they are both far away from “poodle”. “feet”, “feetworship”, and “footmodel” are all far apart from each other. “paypal” and “cashapp” are on opposite sides of the diagram. On the other hand, some words that are expect to be together are close together. “flymetoyou”, “luxurycompanion”, “elitecompanion”, and “vipmaidens” are all close together. “healthyfood”, “bodybuilding”, “workout”, “eggs”, and “paleo” are also all close together. “paypig”, “paypigswanted”, “findom”, and “findomme” are all close together. Overall, the FastText CBOW model does capture the semantics relatively well. However, it does not appear to capture the semantics as well as the FastText Skip-gram model. As a reminder, the FastText CBOW model uses the surrounding words to predict the center word [33].

Lastly, Figure 12 shows the t-SNE Dimensionality Reduction of the Word Embeddings from the Paragraph Vector (Doc2Vec) model.

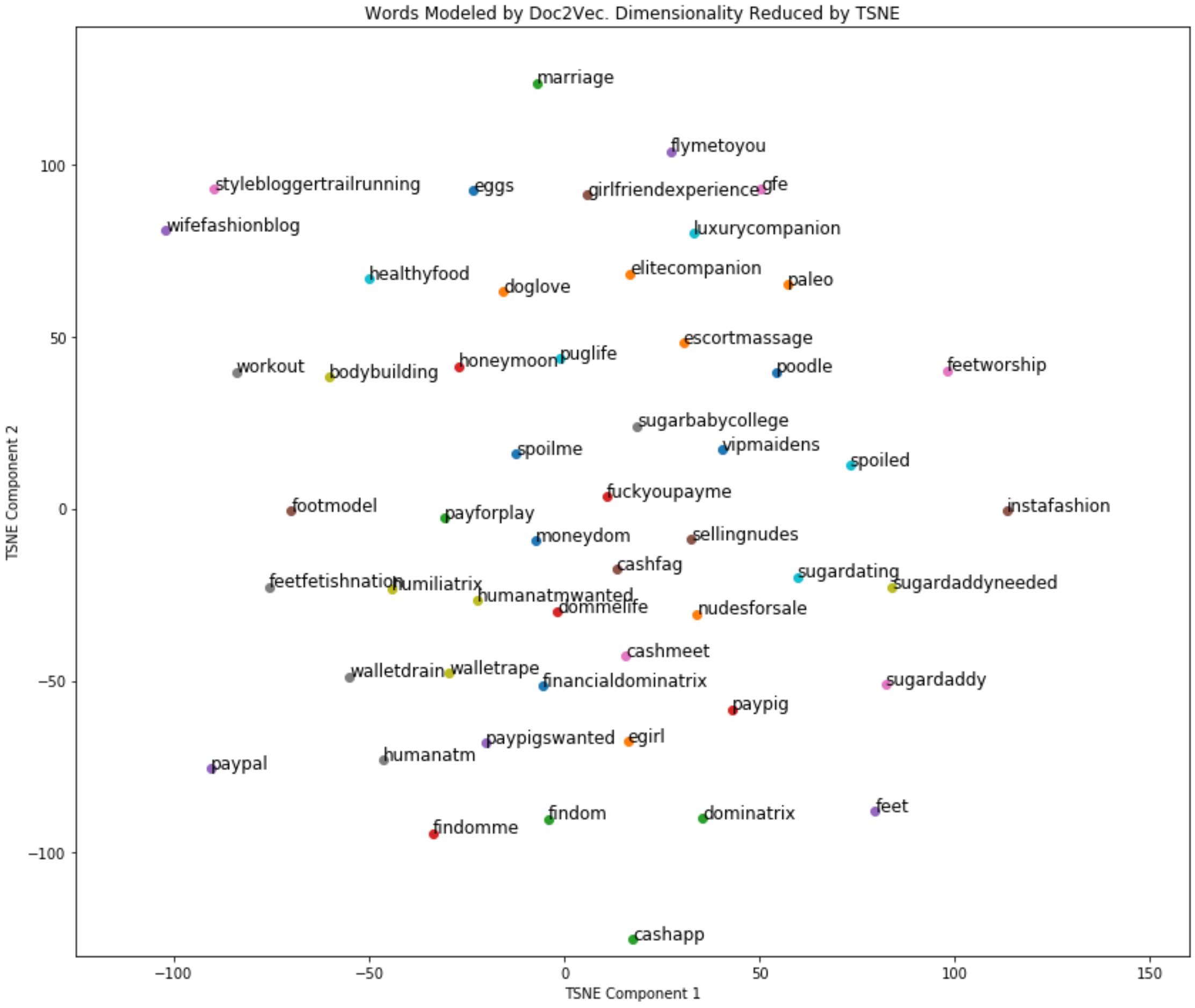

*Figure 12- T-SNE Dimensionality Reduction of Word Embeddings from Paragraph Vector (Doc2Vec) Model*

This model also appears to embed semantics worse than the FastText Skipgram model. For example, “feetworship”, and “feetfetishnation” and “feet” are all far apart from each other, even though “feetfetishnation” and “footmodel” are close together. “paypal” and “cashapp”, while on the same side of the diagram, are not close to each other. “instafashion” is on the other side of the diagram from “stylbloggertrailrunnign” and “wifefashionblog”. “flymetoyou”, “luxurycompanion”, “elitecompanion” are close together, but are not close to “vipmaidens”. On the other hand, “paypigswanted”, “paypig”, “findomme” and “findom” are all close together. In addition, “girlfriendexperience” and “gfe” are close together. (gfe stands for girlfriendexperience, so they are used in similar contexts.)

By looking at these plots, we might expect the FastText Skipgram model to outperform the FastText CBOW and Paragraph Vector (Doc2Vec) model.

**End of section including potentially psychologically harmful, triggering or upsetting hashtags.**

### *12.2 Kmeans Clustering and Silhouette Score Results*

After creating the FastText Skipgram, FastText CBOW, and Doc2Vec models, I transformed each post into an embedded vector. The FastText Skipgram model and FastText CBOW model only output vectors of words. They do not output an embedded vector for a whole document (post), which has multiple words in it. Therefore, in order to get the vector for the entire post, I sum the vector for each word in the post, and then divide by the number of words in the post to normalize the result so that it is comparable with other posts. This has its downsides because it is does not take into account placement of words. Doc2Vec on the other hand outputs a vector for an entire document which does take into account placement of words. After computing the vectors for each post in the dataset, I compute the Silhouette score for the vectors resulting from each model. Unfortunately, none of the models were able to get good Silhouette scores close to 1. This suggests that the data does not cluster well, and we will be able to see this later in this chapter. The Silhouette Score for each of the models can be seen below in Figure 13.

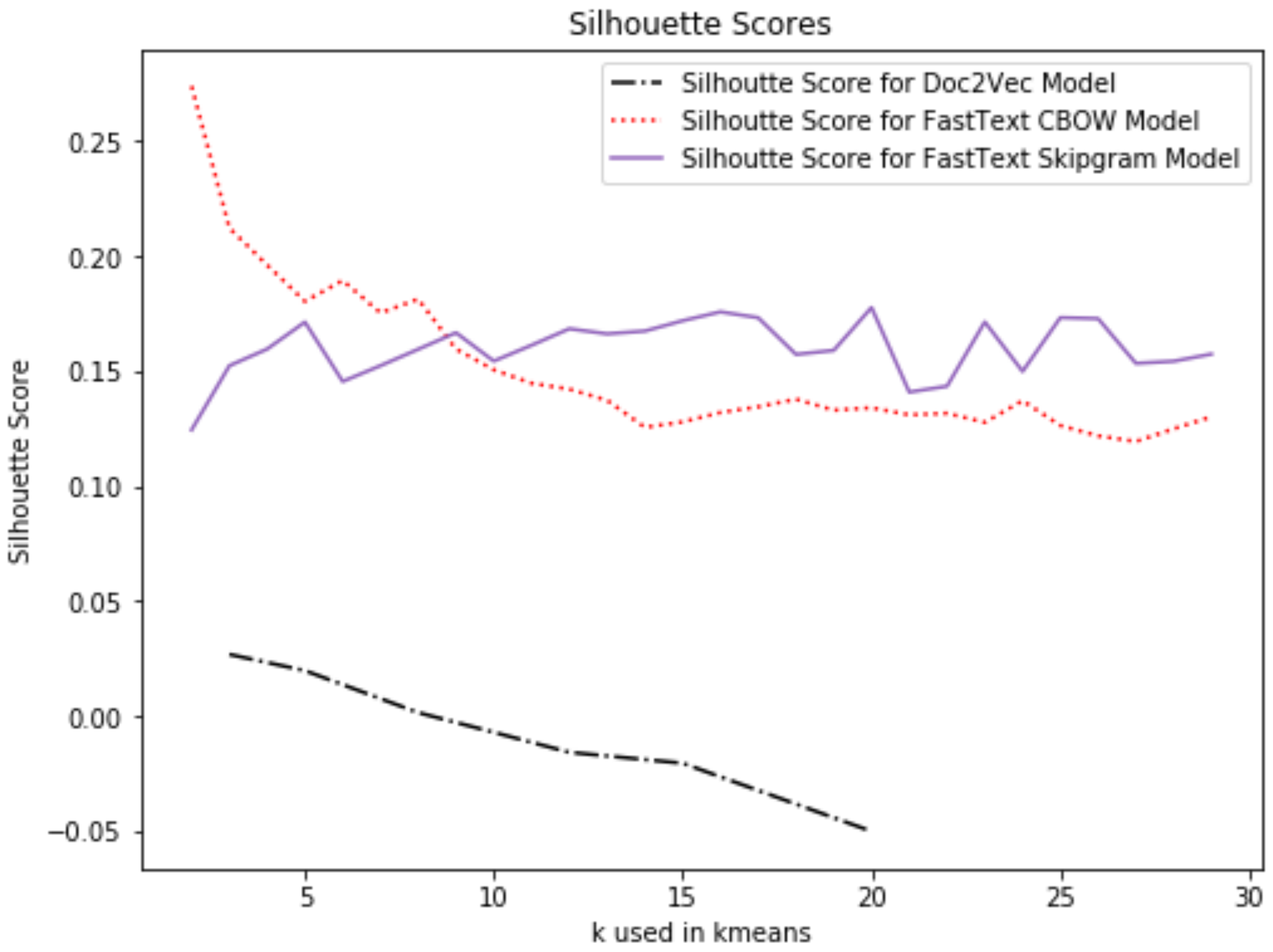

*Figure 13 - Silhouette Scores for each model show that none of the models cluster well. Between all of the models the FastText models appear to cluster better, so we might expect those to have better accuracies in the semi-supervised labeling.*

When deciding the number of clusters, there are several things to consider. First, we have 3 labels in the data. "Yes – post is part of sex industry", "No – post is not part of sex industry", and "Unclear – it is unclear whether the post is part of the sex industry". From here on, the label "No – post is not part of sex industry" will be referred to as "No". The label "Yes – post is part of sex industry" will be referred to as "Yes". The label "Unclear – it is unclear whether the post is part of the sex industry" will be referred to as "Unclear". Given that there are 3 label options for a given post, we cannot have 1 or 2 clusters, because there would not be enough clusters for the number of labels. Second, if there are too many clusters, there are not enough labeled posts within a cluster to classify that cluster as part of the sex industry or not. Using this knowledge, while also trying to maximize the Silhouette score, I decided to cluster the post vectors from the FastText Skipgram model into 16 and 20 clusters using Kmeans clustering. In the FastText CBOW model, I cluster the post vectors into 6 and 8 clusters using Kmeans clustering. In the Doc2Vev model, I cluster the post vectors into 5 and 8 clusters using Kmeans clustering.

### *12.3 UMAP Display of Clustered Data with Manually Labeled Data*

I use Uniform Manifold Approximation and Projection (UMAP) dimensionality reduction to visualize the clustered data. UMAP is similar to t-SNE in that it can reduce nonlinear high dimensional data into a low dimension, however its computational runtime is much faster than that of t-SNE [39]. This was particularly helpful given that I needed to display the data for 6 models each with 50,000+ posts. Please note that the UMAP dimensionality reduction done in this section is solely for the purposes of visualization. Using t-SNE, PCA or some other dimensionality reduction would not change the overall accuracy results, but would simply change how we visualize the data. The FastText and Doc2Vec models already sufficiently reduce the dimensionality of the posts to perform Kmeans clustering on the embedded document vectors.

Figure 14 and Figure 15 help show whether or not the clusters of the post data correspond well with the manually labeled posts. As we expected from the Silhouette score plots in the section above, the FastText Skipgram Models and FastText CBOW Models appear to cluster the data better than the Doc2Vec models. Within both FastText Skipgram Models, the clusters, shown by varying shades of blue, do appear to line up generally well with the labels corresponding to the posts. That is to say, that each cluster appears to predominantly contain either “No” posts or “Yes” posts. There are a few clusters that contain a fair amount of both labels, but this does not appear to be the majority. This will be tested in the following section when the labels are propagated to the clusters and cross-validation tests the accuracy of our labels.

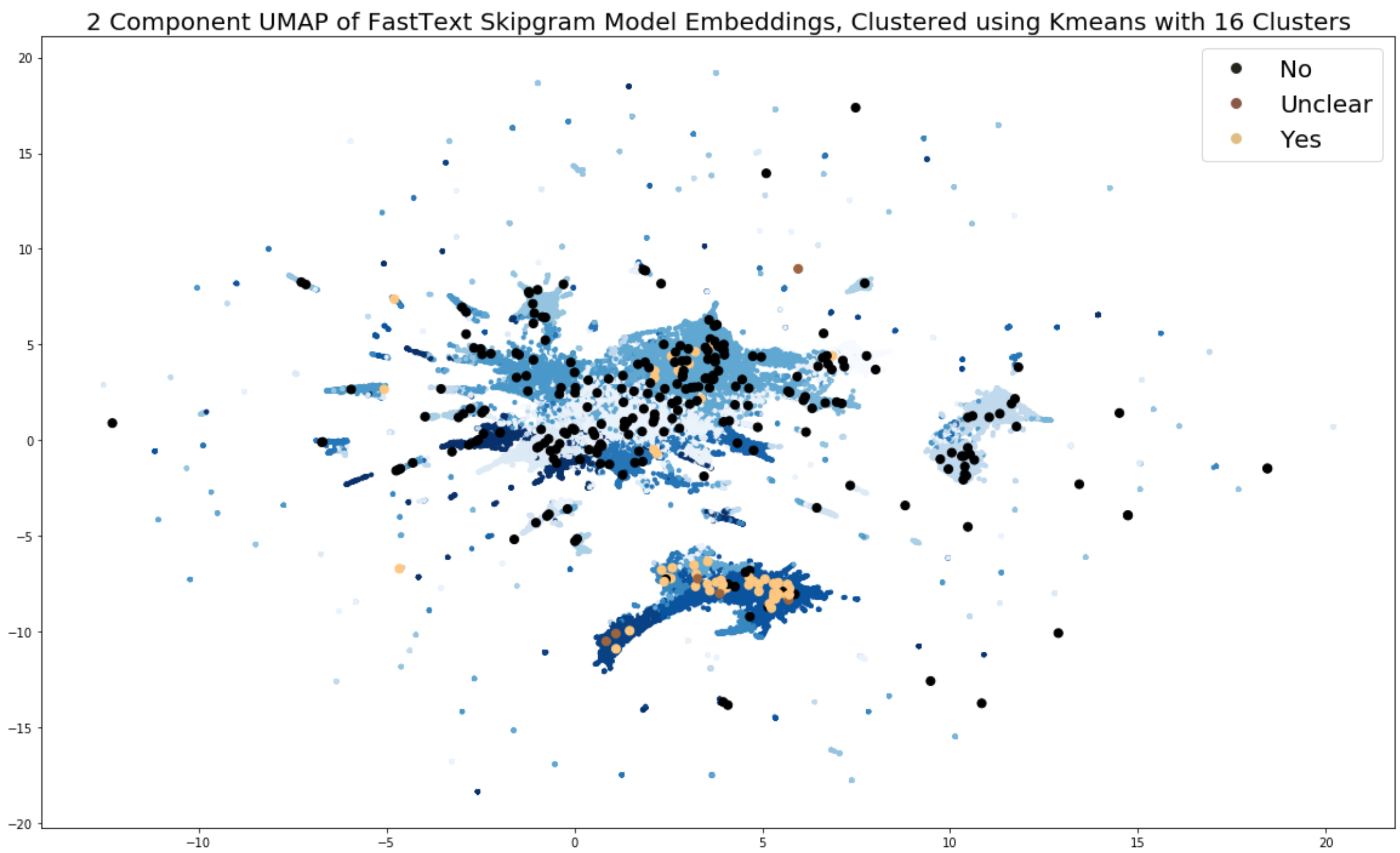

*Figure 14 – FastText Skipgram Model Embeddings are displayed with the labeled posts. There are 16 clusters in this model, and the clustering is done using Kmeans clustering. The clusters correspond to the varying shades of blue. The dimensionality reduction is done with UMAP to make the 100-dimensional word embeddings able to be plotted on a two-dimensional plot. Here the clusters do seem to correspond well to the labeled posts. Labels correspond to "Yes – Sex Industry Post", "Unclear – Post May be in Sex Industry", and "No – Not Sex Industry Post". Axes are 1st and 2nd UMAP Component.*

Chapter 12 continued on following page.

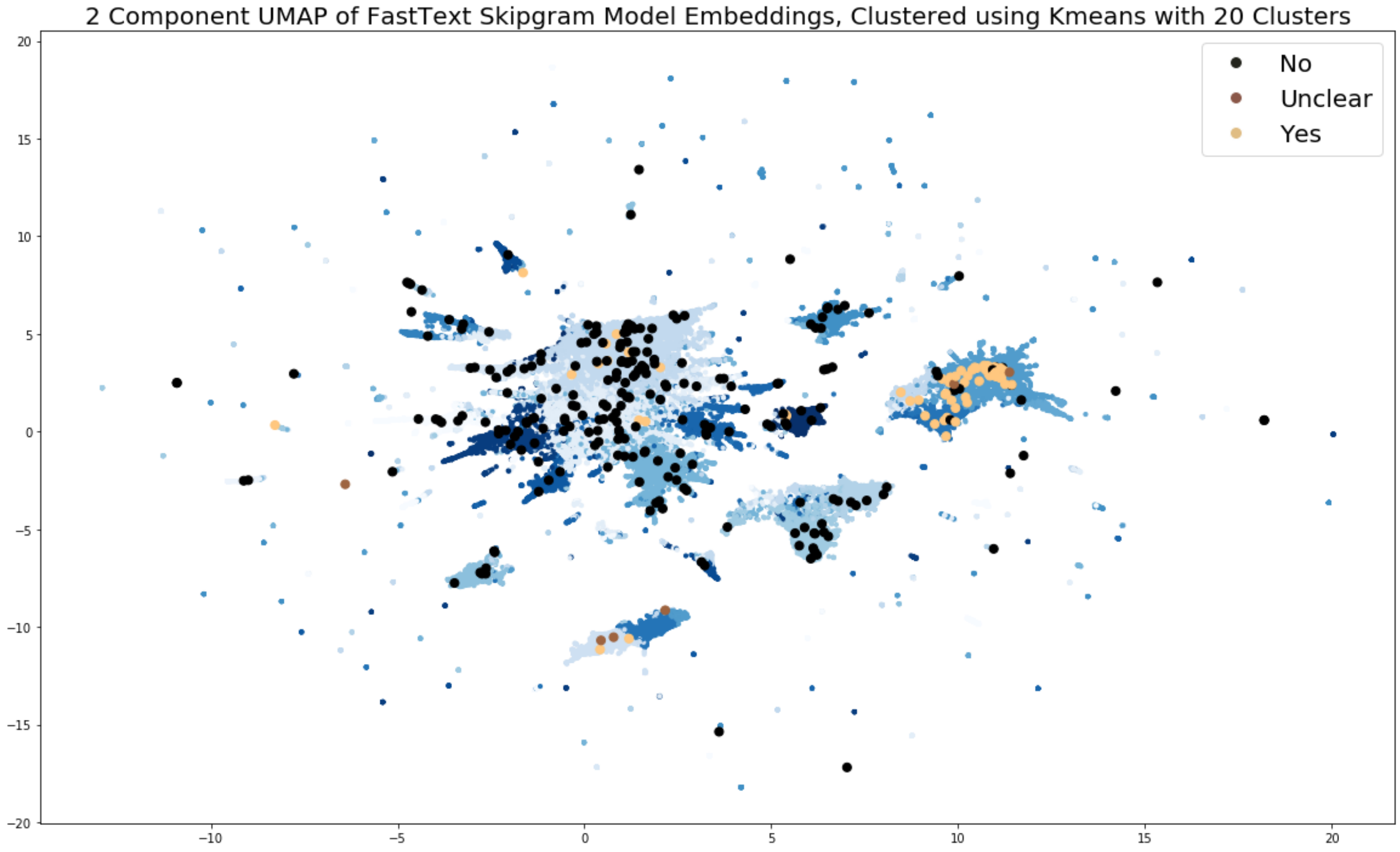

*Figure 15 - FastText Skipgram Model Embeddings are displayed with the labeled posts. There are 20 clusters in this model, and the clustering is done using Kmeans clustering. The clusters correspond to the varying shades of blue. The dimensionality reduction is done with UMAP to make the 100-dimensional word embeddings able to be plotted on a two-dimensional plot. Here the clusters do seem to correspond well to the labeled posts. Labels correspond to "Yes – Sex Industry Post", "Unclear – Post May be in Sex Industry", and "No – Not Sex Industry Post". Axes are 1st and 2nd UMAP Component.*

For the FastText CBOW Models, the Silhouette score started higher and decreased as the number of clusters increased. This suggests that the FastText CBOW Models are better suited to a few clusters, and led to the decision to use 6 and 8 clusters. Now, looking at the UMAP Projections of these FastText CBOW Model Embeddings in Figure 16 and Figure 17, it is easier to see that the embeddings do seem to be slightly less spread out than the FastText Skipgram Embeddings. In these FastText CBOW Models, the clusters also appear to correspond well with the post labels. Clusters appear to predominantly contain either "No – Not in Sex Industry" posts or "Yes – In Sex Industry" posts. There are a couple clusters that contain a fair amount of both labels, but this does not appear to be the majority.

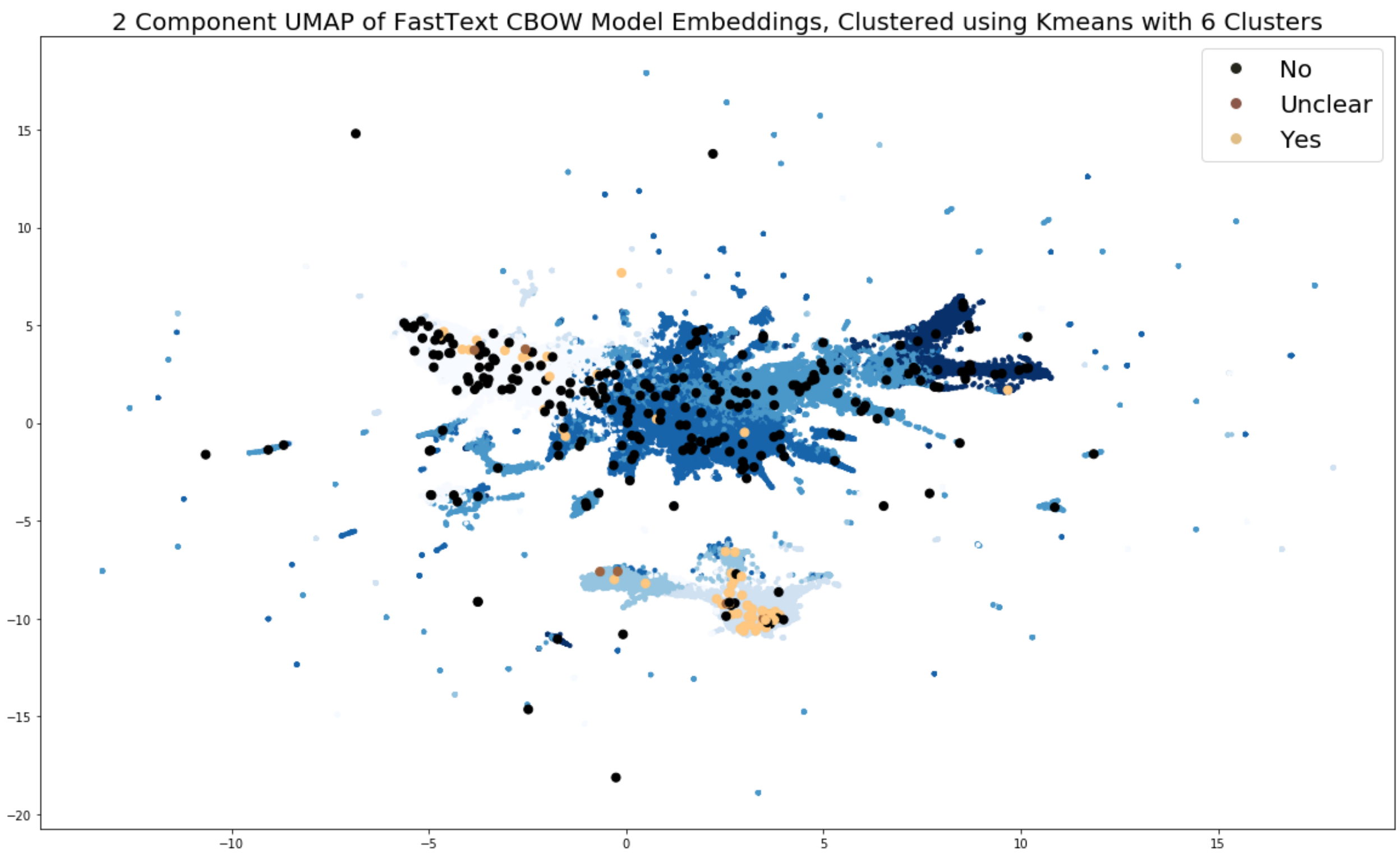

*Figure 16 - FastText CBOW Model Embeddings are displayed with the labeled posts. There are 6 clusters in this model, and the clustering is done using Kmeans clustering. The clusters correspond to the varying shades of blue. The dimensionality reduction is done with UMAP to make the 100-dimensional word embeddings able to be plotted on a two-dimensional plot. Here the clusters do seem to correspond well to the labeled posts. Labels correspond to "Yes – Sex Industry Post", "Unclear – Post May be in Sex Industry", and "No – Not Sex Industry Post". Axes are 1st and 2nd UMAP Component.*

Chapter 12 continued on following page.

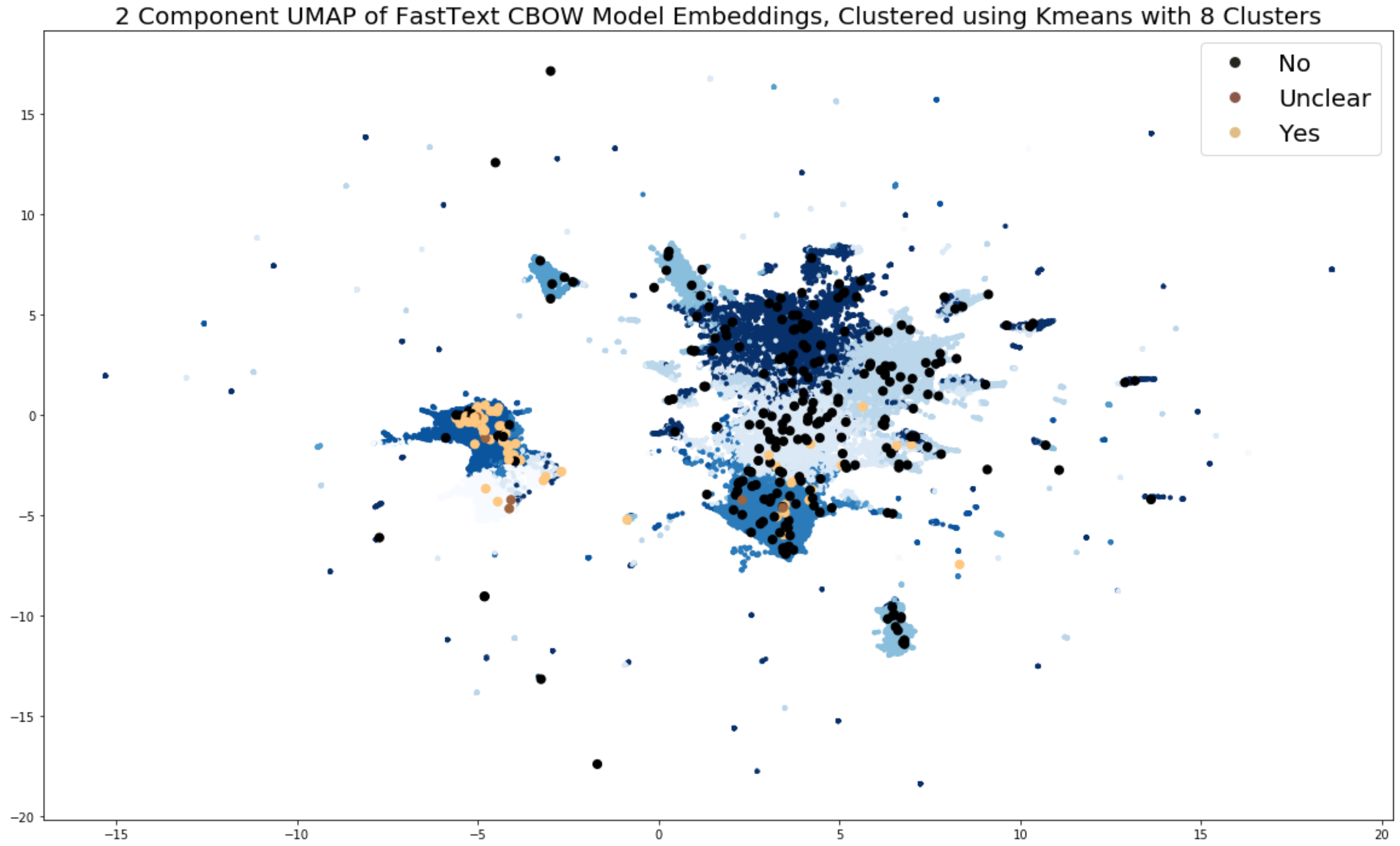

*Figure 17 - FastText CBOW Model Embeddings are displayed with the labeled posts. There are 8 clusters in this model, and the clustering is done using Kmeans clustering. The clusters correspond to the varying shades of blue. The dimensionality reduction is done with UMAP to make the 100-dimensional word embeddings able to be plotted on a two-dimensional plot. Here the clusters do seem to correspond well to the labeled posts. Labels correspond to "Yes – Sex Industry Post", "Unclear – Post May be in Sex Industry", and "No – Not Sex Industry Post". Axes are 1st and 2nd UMAP Component.*

The Silhouette score for the Doc2Vec (Paragraph Vector) Model Embeddings suggested that the Doc2Vec model does not separate into clusters very well. The Silhouette score hovered around 0, starting out positive and then going negative as the number of clusters increased. One might expect the data to all be uniformly spaced apart such that clusters cannot easily be formed. This is exactly what is seen in the UMAP Projection of the Doc2Vec Model Embeddings in Figure 18 and Figure 19. All of the data points are near each other and relatively evenly spaced out into the shape of an oval. The clusters chosen by Kmeans clustering appear to coincide with smaller sections, or sides of the oval shop. Earlier in this chapter, we saw that the Doc2Vec model does not embed words in a way that captures the semantics that the FastText models do. This can also be seen by the various "Unclear" and "Yes – Sex Industry" post labels scattered throughout the clusters. However, most of the "Yes – Sex Industry" posts do seem to be on one side of the Embeddings, so there is a chance that the generated semi-supervised cluster labels may match up well with the Doc2Vec clusters.

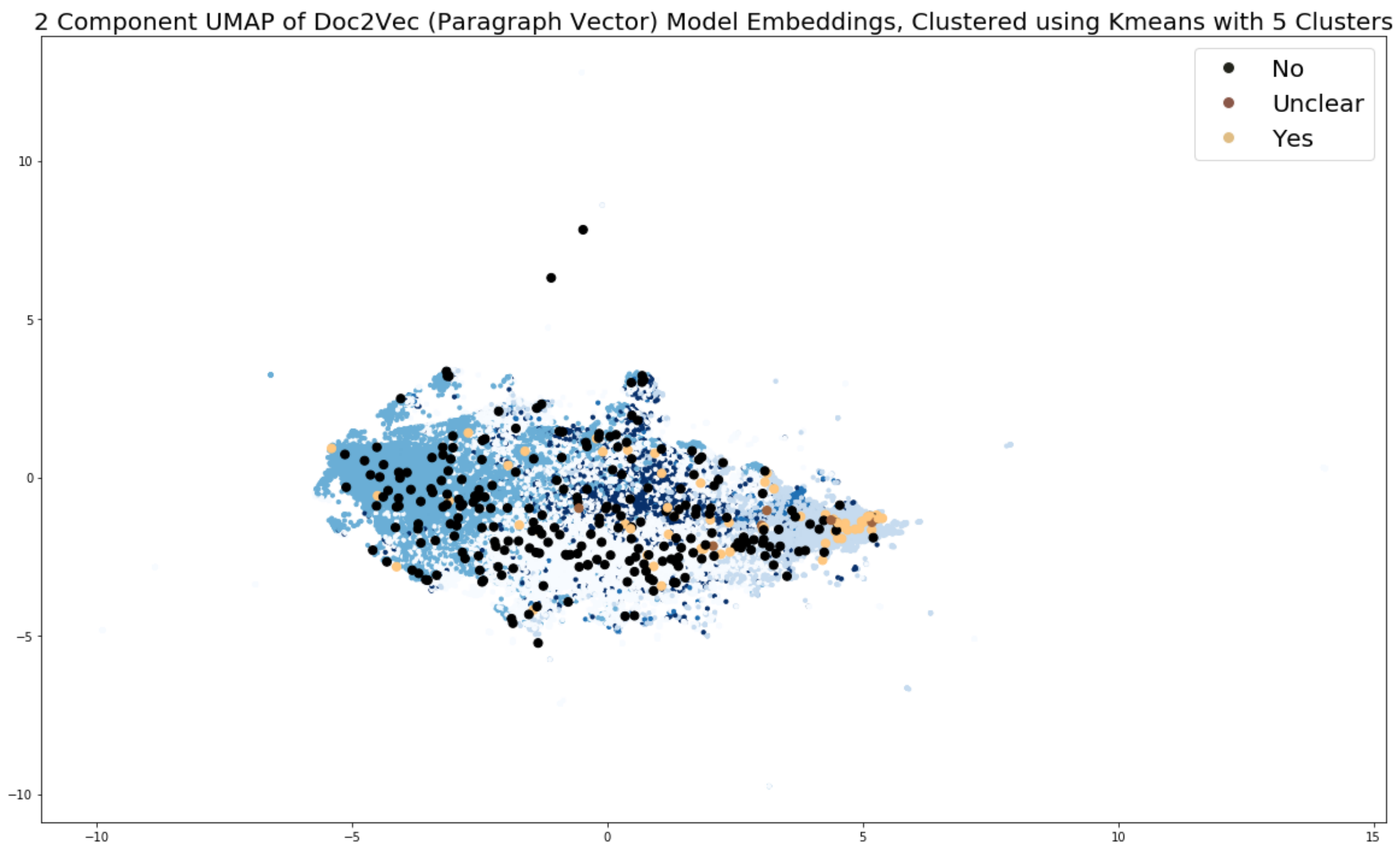

*Figure 18 - Doc2Vec (Paragraph Vector) Model Embeddings are displayed with the labeled posts. There are 5 clusters in this model, and the clustering is done using Kmeans clustering. The clusters correspond to the varying shades of blue. The dimensionality reduction is done with UMAP to make the 100-dimensional word embeddings able to be plotted on a two-dimensional plot. Here the clusters do not seem to correspond well with the labeled posts. Labels correspond to "Yes – Sex Industry Post", "Unclear – Post May be in Sex Industry", and "No – Not Sex Industry Post". Axes are 1st and 2nd UMAP Component.*

Chapter 12 continued on following page.

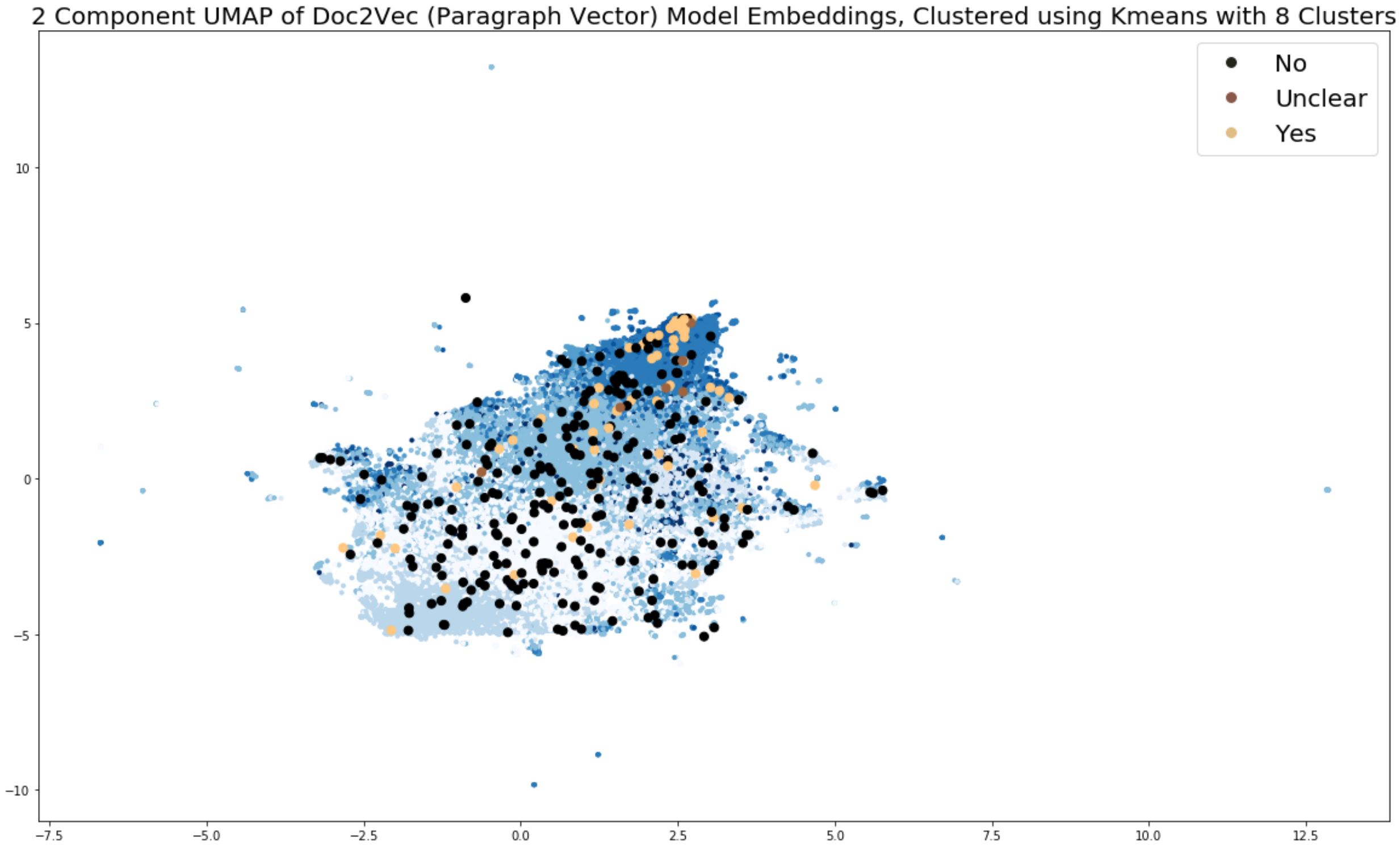

*Figure 19 - Doc2Vec (Paragraph Vector) Model Embeddings are displayed with the labeled posts. There are 8 clusters in this model, and the clustering is done using Kmeans clustering. The clusters correspond to the varying shades of blue. The dimensionality reduction is done with UMAP to make the 100-dimensional word embeddings able to be plotted on a two-dimensional plot. Here the clusters do not seem to correspond well with the labeled posts. Labels correspond to "Yes – Sex Industry Post", "Unclear – Post May be in Sex Industry", and "No – Not Sex Industry Post". Axes are 1st and 2nd UMAP Component.*

## 12.4 Cross-Validation Score for FastText CBOW, FastText Skipgram, and Doc2Vec

The posts were transformed into embedded vectors using FastText and Doc2Vec, and then these embedded vectors were clustered with Kmeans clustering. The next step in doing the semi-supervised machine learning is to propagate the labels to clusters where appropriate. The process of labeling unlabeled data in appropriate clusters is referred to as label propagation. Labels can only be propagated in certain circumstances. For example, if a cluster contains many posts labeled both "Yes" and "No", then there shouldn't be any label propagation in that cluster. If a cluster has many labeled posts within it that all have the same label, then it would make sense to label all those posts in that cluster with that respective label. Alternatively, if a cluster only has one or two posts labeled in it, even though both of them may have the same label, there aren't enough posts to suggest that the entire cluster should be labeled with that respective label. Therefore, during the label propagation step, I choose to apply labels to clusters only if there are 5 or more labeled posts within the cluster and those labels are 100% “yes”, “no”, or “unclear”. Then the respective label is propagated to that cluster.

I tested the label propagation with cross-validation, dividing the manually labeled data into 10 subsets. This was done for each model, and the results below are shown in Table 9.

*Table 9 - The table below shows the cross-validation accuracies of each word embedding model. The Doc2Vec Models performed very poorly and either did not classify any posts, or only had a 50% accuracy. The FastText Models performed much better overall. The FastText CBOW Models outperformed the FastText Skipgram Models.*

| Model | Cross-Validation Accuracy When Labels are Applied | Number of New Posts Labeled |
|---|---|---|
| FastText Skipgram Model With 20 Clusters | 93.6% | 10,492 |
| FastText Skipgram Model With 16 Clusters | 93.2 % | 9,785 |
| FastText CBOW Model With 6 Clusters | 98.6 % | 12,578 |
| FastText CBOW Model With 8 Clusters | 98.5 % | 12,141 |
| Doc2Vec Model With 5 Clusters | N/A | None |
| Doc2Vec Model With 8 Clusters | 50% | 0 |

As expected, both FastText CBOW Models and FastText Skipgram Models do fairly well. The UMAP projections of these models showed that many of the clusters contained predominantly one label, and so it makes sense that the label propagation did well here. Between those four models, the FastText CBOW Models did better than the FastText Skipgram Models. This is surprising because in the initial visualizations of how the models vectorized various hashtags, the FastText Skipgram Model appear to capture the semantics of the hashtags better. Please note that the Cross-Validation Accuracy is calculated over 10 cross-validation groups. On the other hand, the Number of New Posts Labeled are the number of posts labeled when the Label Propagation is applied to the whole dataset using all of the Manually Labeled Posts available. This helps explain why the Doc2Vec Model with 8 clusters couldn’t apply new labels to any posts. The FastText CBOW Model with 8 clusters had the highest cross-validation accuracy, and had the greatest number of new posts labeled.

There are several important things to note. For one, not having to manually label 12,578 posts saves roughly 125 hours of manual work. So, the fact that this can be used to manually label that many posts is a huge benefit. On the other hand, there are still 37,194 posts out of the entire dataset that cannot be labeled during this first round of label propagation. Recall that the entire dataset contains over 50,000 posts.

Another consideration, is that all of the clusters that were able to have Label Propagation accurately applied to them were labeled “no”. They were most likely control posts that are already known to most likely not be part of the sex industry. So, the new amount of knowledge that this gives us, is not a lot in terms of whether or not this semi-supervised machine learning model can distinguish posts that have a higher likelihood of being be a part of the sex industry. On the other hand, it seems that these posts were clustered separately from posts with a higher likelihood of being in the sex industry. Therefore, it seems that the FastText Models were indeed capturing some of this information. Given the fact that the dataset did not capture whether the original poster or a different user made a comment/reply, this makes sense.

When manually labeling the data, often just one or two words written specifically by the original poster on the post or on their profile indicated that it was part of the sex industry. If these words appear on the profile and do not appear on the post, it is much more challenging for our models to identify it as part of the sex industry because it is not encoded in the embedding. If these words do appear on the post itself, the way that our FastText model embeds documents by averaging the word vectors may minimize the importance of these words. The FastText model embeds documents by averaging the word vectors over the whole document. So, if only a few words indicate that a post is part of the sex industry, these sex industry word vectors may not contribute significantly to the entire document’s embedding. Ultimately, it would be much better if the dataset could be updated to describe whether the original poster or another user made a comment, and to include non-identifying information from the profile. I expect that this would allow the label propagations to be able to classify more posts.

### *12.5 Future Technical Work: Semi-Supervised Learning as a Cycle*

Ultimately, this is just the first round of the semi-supervised machine learning. After the appropriate clusters are labeled via the Label Propagation, then the Semi-Supervised Cycle must be repeated until all the posts in the dataset have been labeled. In Figure 20 below, the Semi-Supervised Cycle for this particular thesis project is shown. I have completed the first 5 steps of the first cycle of this work. This project is a work in progress, and ultimately, this cycle needs to be iterated upon several times before the dataset is fully labeled.

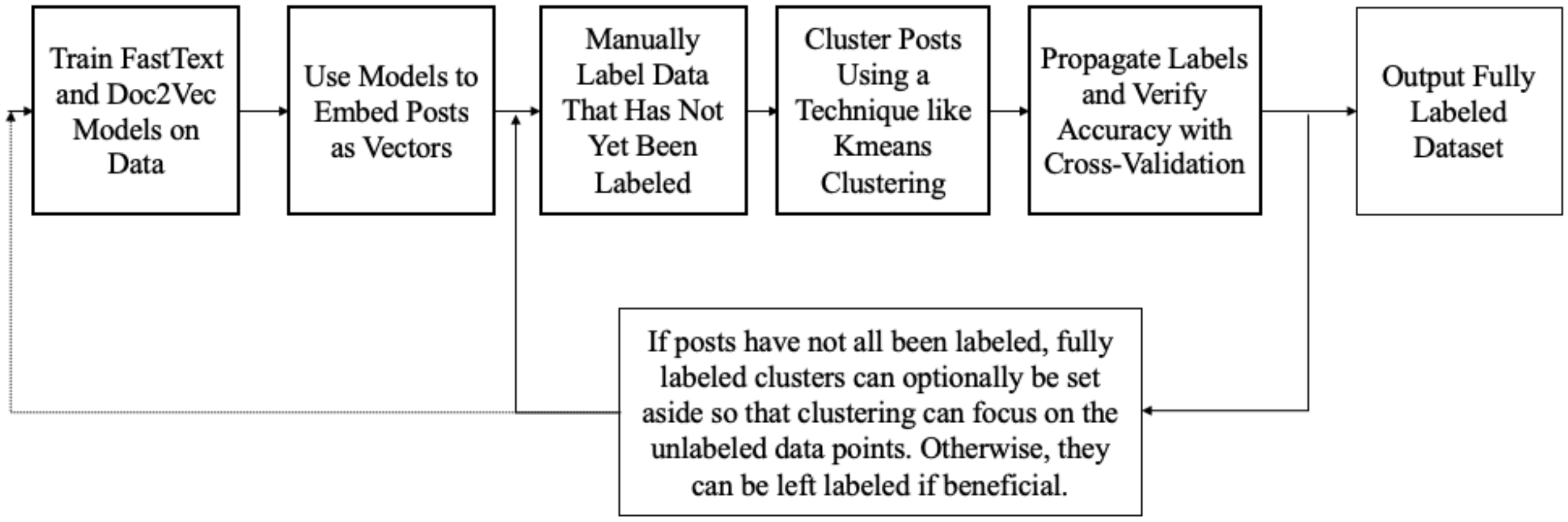

*Figure 20 - The Semi-Supervised Cycle for this particular thesis is shown above. When the cycle repeats, the researcher can choose to use a new model to capture different aspects as represented by the dashed arrow, or the researcher can reuse the same model and instead skip to manually labeling more data.*

Future researchers may also want to consider using word embeddings that have already been generated by models that other researchers have created. For example, Fréderic Godin created Twitter embeddings using FastText and Word2Vec on 400 million tweets [33], [34].

# Chapter 13: Future Work

This thesis sets up the possibility for a wide array of future work in this space, although much of it is filled with challenges from legal and ethical perspectives. This next paragraph briefly introduces ideas for future work, which will be discussed later in this chapter. Future work directly related to this project includes developing a supervised machine learning model that can identify Instagram posts in the sex industry. It could be used to create a spatial and/or temporal model of sex industry posts on Instagram, or in social media as a whole. Future work could track and monitor how hashtags used to indicate the sex industry change over time. There are several research areas that are tangentially related to this thesis that could be important or interesting to pursue. These include researching how SESTA-FOSTA has changed the online sex industry. It includes detecting advertisements that may be indicative of labor trafficking. Another area for possible future work is identifying communities that may have a high risk for being trafficked, but who have not yet been trafficked, and working with nonprofits or social media companies to increase awareness of resources and organizations that could provide support for those vulnerabilities. Lastly, generating, finding, or working with governments and organizations to generate high quality data would be extremely helpful to the research community.

## *13.1 Applications Directly Related to This Thesis*

### 13.1.1 Development of the Supervised Machine Learning Model for this Project

One of the most logical next steps in future work is to develop a supervised machine learning model to determine whether a post is in the sex industry or not. It is important that the results from that work, as well as from this thesis, remain in aggregate data form. Guevara, the expert in the field who helped label the data, and I usually used hashtags to help decide whether a post was in the sex industry or not. Hashtags are what users can search for to find posts, so by using a certain hashtag on Instagram, people in the sex industry (selling or buying) can find each other more easily. A supervised machine learning model could be useful to help people in the nonprofit space determine what other kinds of attributes make a post more likely to be in the sex industry. One of the most challenging attributes of the dataset used in this thesis is that it is unknown whether the comments and replies are made by the original poster or another individual. If the original poster uses hashtags clearly indicative of the sex industry, that post can be clearly identified as part of the sex industry. However, when another individual, not the original poster, uses this terminology and hashtags indicative of the sex industry, it does not necessarily mean the post is in the sex industry. This can occur when the other user thought the post was in the sex industry, but the original poster did not intend for it to be a sex industry post. The fact that this dataset does not have information on whether a comment/reply was made by the original poster may make it harder for the supervised machine learning model to work well. This means that someone working in this space may need to work in collaboration with the social media companies, for legal reasons mentioned in Chapter 6, to expand the dataset to include information on whether the original poster made the comment or reply. However, working with a social media company can make the work ethically challenging for reasons mentioned in Chapter 6. The challenges associated with this are discussed later in this chapter.

### 13.1.2 Spatial and/or Temporal Aggregate Analysis

If a researcher is able to build a supervised machine learning model that can detect whether a post is in the sex industry or not, they could then use this to build a spatial and/or temporal model of where and when social media is being used within the sex industry. Some research has been done in this space on websites designed to facilitate prostitution, that shows the number of advertisements on these websites increase during major sporting events like the World Cup or the Super Bowl, as well as during business conferences [40]. Other research has shown how advertisements on these websites respond to major weather events [41]. It would be interesting to see if the sex industry on social media follows similar trends. It is possible that social media is being used to target a different set of buyers in comparison to the buyers targeted by websites designed to facilitate prostitution.

In addition, by building a spatial and temporal model, researchers could create an aggregate map of areas where Instagram is used within the sex industry. Cities with high usages of the sex industry on social media could put up lists of resources in public spaces, like bathroom stalls, or hold awareness and educational sessions in the community. These resources, which could include information on where people experiencing trafficking can access support, may be able to have high impact if posted in areas that are simultaneous public and private. For example, bathrooms in parks, restaurants, schools, or other public spaces are open to the public, however they are also private in the sense that a trafficker likely will not be inside the bathroom stall with the person experiencing trafficking. This means that people experiencing trafficking could view the resources while reducing the risk that their trafficker knows that they looked at the resources. If a trafficker knows that someone they are controlling is accessing and viewing resources, they may punish them, possibly quite violently, for viewing those resources [2]. This is why locations that are simultaneously public and private, can be good places to post resources.

### 13.1.3 Modeling How Hashtags Indicative of The Sex Industry Evolve Over Time

Another area for possible future research is to look at how hashtags that indicate a post within the sex industry change over time. What people post in the sex industry, just like what most everyone posts on social media, will likely change over time. What hashtags are commonly used one year, may not be commonly used the next year. In addition, people in the sex industry on social media often have to filter their posts so that their accounts are not removed from the social media companies. Company's algorithms that are used to detect posts that don't fall within 'community guidelines' will likely change over time to reflect the hashtags being used within the sex industry. Then people in the sex industry may develop new hashtags so that users can keep using social media to participate in the sex industry.

### 13.1.4 Challenges Anticipated with This Future Work

In order to do any of the research above, it may be necessary to improve the dataset that was used in this thesis. It might, as mentioned earlier, be critical to know whether the original poster wrote a comment/reply or if another user wrote the comment/reply. This means future researchers may have to work with social media companies, for legal reasons, to get a better dataset. However, as mentioned in Chapter 6, this may have negative ethical outcomes. If a researcher partnered with or got permission from a social media company, that social media company may want access to the research. Then, that company may simply use the research to delete any account related to the sex industry for fear of facilitating prostitution. It is illegal for a website to facilitate prostitution of another person. This became illegal when SESTA-FOSTA passed in 2018 [1]. However, many of the posts in the sex industry may not be prostitution. For example, they could be sexual photos or videos of adults or they could be photos and videos of feet supplied to people with foot fetishes. The definition of the sex industry that Guevara, the expert labeler for this project, used to identify sex industry posts was if someone was using their body in some way for the sexual enjoyment of others in exchange for money or something of value [28]. This definition includes much more than prostitution. Therefore, many people who are in the sex industry may be forced out of a website that they view as a safer alternative. Then these individuals, if they are participating in survival sex work or if they are experiencing sex trafficking, may have no other option but to resort to working on the streets or on more dangerous websites. Therefore, by working in collaboration with a social media company, a researcher may cause unintended harm.

Additionally, there may be serious legal and ethical aspects to consider if the future researcher wants to do any work with images on social media. This is discussed in Chapter 4 – Scope of Thesis and Problem Definition. Transgender and people of color make up a disproportionate percentage of people in the sex industry due to oppression and discrimination in society [11], [12]. By creating a machine learning algorithm that uses images to classify posts as sex industry or not sex industry, there is a big risk for bias that could cause further discrimination against transgender individuals and people of color. In fact, this is already a problem in anti-sex trafficking machine learning models that have looked at images. People who are women of color or transgender individuals have been falsely accused of posting content in the sex industry and have been banned from sites [14]. In addition, there are legal considerations to consider when working with images in an anti-sex trafficking context. Researchers may accidentally possess or store images of people under 18 engaging in sexual acts. This means that they could be liable for possession, intent to view, or storage of child pornography. This is a severe crime that must be taken seriously. All known images of child pornography must be immediately reported to the National Center for Missing and Exploited Children, and the data must immediately be removed from the dataset that the researcher is using. Chapter 6 goes into greater detail of this, and future researchers, who have the intention to use images in their dataset, may want to read through that chapter.

## *13.2 Applications Indirectly Related to This Thesis*

### 13.2.1 Detecting Advertisements Posted by Labor Traffickers or Indicative of Labor Trafficking

One area that is tangentially related to this thesis would be doing machine learning to detect posts or ads that may be indicative of labor trafficking. I expect that research in this area would have less ethical and legal considerations that result in conflicting best practices. This means that a researcher may be able to partner or get permission from a social media company to try to identify ads or posts on social media that may be indicative of labor trafficking with potentially fewer ethical concerns. In addition, there may be attributes that allow experts to identify advertisements that are indicative of labor trafficking vs legitimate job postings. In contrast, in the sex industry, even experts in the field find it impossible or extremely challenging to identify sex trafficking vs survival sex work vs freely chosen sex work via an online posting [2]. In addition, researchers are unlikely to run into child pornography issues, like researchers might with anti-sex trafficking research. Depending on the images used, they may have more to do with large sums of cash or other kinds of photos as opposed to sexual or nonsexual images of people in the sex industry. This may reduce some of the risk associated with creating bias within a machine learning model that can further discrimination. In addition, this is an area that, in my experience, has been significantly less researched than images potentially indicative of sex trafficking. However, it goes without saying, that researchers would nonetheless need to analyze all this for themselves to make sure that their model does not have bias, legal concerns, ethical concerns, or other unintended harm that could result from the project. I have not looked deeply into the project, interviewed nonprofits nor interviewed people who have experienced labor trafficking to know what they would think of the project proposal. This would be an important first step if a future researcher wanted to work on this idea.

### 13.2.2 SESTA-FOSTA Impact on Sex Workers Finding a Safe Place to Advertise

Another area for potential future research is to use data to look at how SESTA-FOSTA changed the online sex industry. SESTA-FOSTA was passed in 2018, and the goal of this law was to decrease sex trafficking online. As I mentioned in Chapter 3 – The Role of Technology in Human Trafficking, this law has had some unintended negative consequences. Because of this law, people who are still in the sex industry have had to move to more dangerous platforms or to street sex work, which is considered more dangerous for people in the sex industry [14]. Looking at how SESTA-FOSTA changed the online sex-industry from a data science perspective could help congress and lawmakers know what the positive and negative results were from the SESTA-FOSTA law [2], [14]. This project could be challenging because it is hard to get good data in this field, and if interviewing people, it may be hard to get unbiased data that represents the entire group of individuals in the sex industry. There are several reasons why interviewing individuals and having those individuals represent the community as a whole could be hard. For one, people who choose to be in the sex industry may not want to have conversations with researchers because their work is criminalized in the USA. Second, if they do want to have conversations, they are seen as consultants and request significant monetary compensation. This is fair given that they are experts in the field, they deserve to be compensated for their time, they are not paid a salary wage by a company, and they are taking a risk by having a call with a researcher. As far as they know, a researcher could actually be an undercover law enforcement officer, which means having a call with a researcher could be a big risk for them. I was not able to get an estimate for prices, but I would suspect a budget of around $300 per hour per call with sex workers in advocacy groups may be sufficient to cover the costs. This may make the research cost prohibitive. Another option could be a written anonymous interview that would take less time and thereby require less compensation, allowing for more data points and opinions to be collected. For individuals who are experiencing sex trafficking, there would be no clearly safe way to communicate with them to get their feedback. It is unknown what percentage of people in the sex industry experience sex trafficking, so it is hard to know how many individuals would need to be interviewed to allow for results to be unbiased by the reason someone is in the sex industry. Therefore, if pursuing this research topic, it may be useful for researchers to look at aggregate trends. What websites closed down? How many users in the sex industry had those websites been hosting? What location/website did those advertisements change to? Has there been an increase in street prostitution? Was there an overall decrease in sex work on the whole? And from those aggregate numbers, it may be possible to see how the online sex industry changed after SESTA-FOSTA. Then researchers could interview people who are in advocacy groups that represent the different reasons for entering the sex industry. They could ask them which of those websites/means of connecting with ‘clients’ / ‘dates’ they felt safer using while in the sex industry. This could allow for a better picture to be painted on how the SESTA-FOSTA has change the online sex industry and what positive or negative results it has led to.

### 13.2.3 Identifying Communities That Have High Risk of Experiencing Trafficking

There are many things that make someone more likely to experience human trafficking than other people. By providing effective resources to people with vulnerabilities, it may reduce their risk of experiencing human trafficking. Social media could be used as a site to disseminate these resources when someone experiences vulnerabilities. For example, recent migration is listed as one of the factors associated with a higher risk of trafficking [13]. Individuals who have recently migrated may benefit from advertisements on social media for local resources that provide support and resources to immigrants. Or individuals who are experiencing a lack of income may benefit from advertisements on social media for local resources that help with job-training or connections to good-paying jobs. Of course, there may be vulnerabilities where it is not appropriate to provide advertisements on social media. For example, if someone is transgender, they are at a higher risk for being trafficked. However, they may not have come out to their family or friends. By listing any targeted advertisements on social media for resources that support the LBGTQAI+ community, this could have serious negative consequences for that individual. If not done well, this could result in potential microaggression or discrimination against individuals. People may feel like they are being treated a certain way simply for attributes about themselves that they cannot change. If not handled appropriately, it could result in lower self-esteem or individuals feeling like there is something wrong with themselves. Therefore, I discuss this potential research idea with hesitation. Just like this thesis, there would need to be significant reading and consulting done to ensure the project is implemented in an ethical way.

#### 13.2.4 Getting High Quality Human Trafficking Data

As I mentioned in Chapter 3 – The Role of Technology in Human Trafficking, getting good data has always been one of the biggest challenges for individuals working in the anti-trafficking space. Data is often very poor, and not consistent across countries. The estimates for the levels of trafficking have wide ranges. There are many confounding factors when people try to measure the levels of human trafficking. For example, whenever a community does an awareness or educational campaign on human trafficking, the levels of reported human trafficking generally go up in that community. It is unlikely that human trafficking suddenly increased when the campaign was done. It is more likely that people in the community can identify trafficking better, so there are more cases being identified and reported. Or, when prostitution is legalized, the way that data is collected might change. This means reported levels of sex trafficking could change, but it may not be clear how the actual levels of sex trafficking have changed. Figuring out ways to collect good data for labor, sex, and other types of trafficking and implementing those data collecting techniques would be extremely valuable for the research against human trafficking.

Overall, there are many areas for future work to be done. They all have challenges that require the researcher to consult various experts in order to make sure it is done in an ethical and legal way.

# 14. Conclusion

The analysis discussed in this thesis leads to various conclusions including technical conclusions, project future work and application conclusions, and future work conclusion and recommendations.

### *14.1 Technical Conclusions*

From this thesis there are multiple technical conclusions reached in this thesis.

First, this thesis concludes that FastText worked better than Doc2Vec for this project. FastText and Doc2Vec were used to embed the social media posts into 100-dimensional vectors. Between the two, FastText was better at capturing the semantics of words and it performed significantly better during the label propagation with 98.6% accuracy compared to Doc2Vec's 50% accuracy. The 12,587 posts that were labeled by the FastText Continuous Bag-Of-Words algorithm with 6 clusters had a 98.6% cross-validation accuracy. It is important to note that I designed the label propagation to only label the posts that the model was most confident about. This means that even though the dataset had over 50,000 posts, only 12,587 of them were labeled during this first iteration of the semi-supervised classification cycle.

Second, this thesis concludes that future iterations of semi-supervised learning may be able to label many of the posts in this dataset. Future work on this project will continue to iterate throughout the semi-supervised learning cycle, but this thesis suggests promising results that indicate many of the posts in this dataset can be labeled using semi-supervised learning. However, it is important to note that all of the posts that were labeled using label propagation were posts that are labeled "No - not in the sex industry". There is a good chance that these posts were part of the control group, and thus are very different semantically from the posts containing hashtags indicative of the sex industry. Therefore, this research cannot make a conclusive determination on whether or not semi-supervised learning can be used to classify and label the posts that contain many words semantically similar to posts in the sex industry. It does, however, suggest that semi-supervised classification can separate posts that are clearly not in the sex industry from posts that are more similar to sex industry posts.

Third, this thesis concludes that the algorithm's ability to label new clusters may have been significantly reduced because the original dataset did not include whether or not the original poster made a comment. One challenge that this thesis faced is that the dataset does not take into account whether the original poster or another user made the comments on the post. Throughout this thesis both the expert in the field and I found that knowing whether or not the original poster or another user made a comment was crucial to determining whether or not to manually label a post as is the sex industry. Without an updated dataset, there is a good chance that the semi-supervised learning will continue to struggle with posts that contain words semantically associated with the sex industry. This is because there is no way for the algorithm to distinguish who wrote the comment, and thus it is not clear whether or not that word should be indicative of the sex industry for a given post.

Fourth, this thesis concludes that non-identifying public profile information and other publicly available posts may help the algorithm label clusters better. The expert in the field and I also found the profile bios and other posts from the original poster to be integral in manually labeling posts. Future researchers may want a database that includes non-identifying public profile bios and a sampling of the user's other publicly available posts.

Fifth, there are several techniques that may help the model cluster the data better. The first technique researchers may want to consider is using transformations upon the vector embeddings. It is important to note that the FastText Model embeddings were able to be clustered much better than the Doc2Vec Model embeddings. However, there is still a lot of room for improvement for the FastText model embeddings, as the Silhouette score was only around 0.18 for the various FastText models used. It is possible that doing a transformation upon these vector embeddings, such as t-distributed Stochastic Neighbor Embedding (t-SNE), may be able to increase the ability for these embeddings to be clustered during future iterations of the algorithm. The second technique researchers may want to consider is using word embeddings that have been generated on large social media datasets may help the algorithm cluster data better. It is possible that improvements could also be derived from the work of Fréderic Godin, who created FastText and Word2Vec word embeddings on 400 million Tweets [33], [34]. Future iterations of labeling could consider using those word embedding models and updating them with the post data from this thesis's dataset. The third technique researchers may want to consider is that a Bag-Of-Words (BOW) model may do surprisingly well. There are some words that are generally only used by individuals who are selling sex in the sex industry on their own posts. BOW might help to cluster those kinds of posts. The fourth technique researchers may want to consider is doing future iterations of semi-supervised learning on only the posts that are in the database because they contain hashtags indicative of the sex industry. This may allow for greater separation and clustering within this specific subset of posts. Those are several techniques future researchers may want to consider for this work.

Lastly, this thesis found that manually labeled data can be highly variable depending on an individual's definition of the sex industry which may depend on the project application. Work in and related to the sex industry is on a spectrum. Most individuals would agree that selling sex is part of the sex industry. However, is selling sexual photos and videos of oneself part of the sex industry? What if those photos and videos are of one's feet sold to individuals with foot fetishes? When deciding the answers to these questions for this thesis, one must consider the applications of this work. The intended application of this research involves determining at an aggregate level where and how the sex industry on social media is used so resources can be better distributed. This application would allow nonprofits and governments to know which locations may be ideal for providing public resources or funding. Because this intended application is designed to provide resources of support, it is better to have potentially false positives than false negatives. Therefore, the expert in the field chose to including a post as part of the sex industry whenever someone was using their body for the sexual enjoyment of others in exchange for anything of value (finances, shelter, food, presents, etc.) [28], [29]. On the other hand, if research was being done in a way that punished individuals in the sex industry, such as removing social media accounts, then having false positives would be worse than having false negatives.

### *14.2 Project Future Work and Application Use Conclusions*

Throughout this thesis, I interviewed individuals at anti-trafficking nonprofits and read documents from sex worker advocacy groups and anti-trafficking organizations. These led to several important project future work and use application conclusions.

The first conclusion to take from this is that working with a social media company poses many potential risks due to the new SESTA-FOSTA law that was passed in 2018. The law was intended to hold companies hosting websites accountable for facilitating sex trafficking [1]. However, because it also makes it illegal for a company to knowingly facilitate or promote the prostitution of another person online, it has resulted in companies removing profiles that they suspect or know are participating in sex work. Instead of targeting specifically the traffickers, this law has resulted in companies also targeting individuals selling sex. It is not likely going to reduce the sales quota an individual experiencing trafficking has to make for their trafficker. However, it may make that individual have to work on the streets instead of online. Working on the streets is generally more dangerous because individuals in the sex industry cannot verify their clients before meeting them [14]. Therefore, future work on this project should be careful how they work with social media companies in the future if there is any collaboration at all.

The second conclusion is that it is extremely important that real world applications of this project only use data in an aggregate form. Traffickers can and do monitor the social media accounts of the individuals they are trafficking [1]. Therefore, by attempting to provide resources at an individual level over social media, researchers and organizations would be putting people experiencing trafficking at risk of being violently punished by their traffickers for accessing or viewing those resources [2]. In my understanding, communication between social workers and people experiencing trafficking done over social media is generally led by the person experiencing trafficking so they can decide when it is right for them to send a message. Clever techniques like sending disappearing messages or location “check-ins” can be used to send communication between someone experiencing sex trafficking and social workers [1]. However, it is not appropriate for researchers to attempt to send messages to someone suspected of experiencing sex trafficking, even if they have good intentions of providing resources. This is because it could put that individual in immediate serious danger.

Therefore, it is recommended that future work of this thesis only look at aggregate data and that resources are disseminated at a community or city level. For example, in areas with higher sex industry posts on social media, advertisements of local resources could be put up in public bathrooms where people experiencing trafficking may have a chance to view those resources in private. In addition, there could be public awareness campaigns intended to help community members know how to identify the signs of sex trafficking. These campaigns could also show them how to enter a suspected case of human trafficking into the Polaris Project's National Human Trafficking Hotline or into the National Center for Missing and Exploited Children.

Lastly, I want to caution future researchers against using images in any machine learning for semi-supervised learning on anti-sex trafficking work. Using images in this way has led to an increased bias and discrimination against transgender individuals and women of color [14]. While it may be appropriate in very specific situations, researchers must be extremely cautious in how they proceed in order to guarantee that their machine learning algorithms are not causing biased discrimination, especially against individuals who are already so heavily discriminated against in our society.

### *14.3 Future Work Conclusions and Recommendations*

This research has led to the idea of two new research projects that are separate from this thesis that are described below.

The SESTA-FOSTA law passed in 2018 was designed to reduce sex trafficking online [1]. It was passed in response Backpage.com knowingly altering advertisements of minors for commercial sex. Backpage.com removed information indicating the individuals selling sex were minors [1]. Backpage.com was explicitly altering advertisements of children experiencing sex trafficking to keep those advertisements on their site rather than reporting them to the National Center for Missing and Exploited Children. The SESTA-FOSTA law prohibits online websites from knowingly facilitating sex trafficking. In addition, it prohibits online websites from facilitating the prostitution of another individual [1]. Depending on the organization asked, this law has positive impact, negative impacts, or both. For one, what Backpage.com did was unethical. They should have reported advertisements suspected of advertising a minor to the National Center for Missing and Exploited Children. On the other hand, it means that some websites have shut down that used to be a safer location for people in the sex industry to find "clients". On those previous websites, people in the sex industry could have the chance to vet clients via their phone numbers or emails before meeting them in person [14]. Because many of these sites shut down, sex workers may have to resort to more dangerous websites or working on the street. Nevertheless, individuals experiencing sex trafficking may still have to meet the same financial quota required by their traffickers [14].

Therefore, the first related research project I suggest is to genuinely determine both the harm and the benefits that have resulted from the SESTA-FOSTA law. This could be used to help make more effective laws in the future [2]. That project would require getting input and feedback from all members of the sex industry and the individuals who support people in the sex industry. This includes, but is not limited to, consensual sex workers, sex workers experiencing survival sex, people who have or are currently experiencing sex trafficking, and social workers who support these individuals. Of course, some of the individuals listed may be in very vulnerable positions, and it may be dangerous for them to be contacted by a researcher. Care should be taken to work alongside social workers who can help the researcher gain the needed insights and data in a way that is not harmful to people in the sex industry.

The second related research project I suggest is related to high quality data. Having high quality data on human trafficking is one of the biggest challenges faced by lawmakers and researchers alike. Without high quality data, it is hard, if not impossible, for lawmakers to make effective laws that support the individuals within the sex industry. Future research projects dedicated to the development of high quality data on human trafficking would be enormously impactful.